\RequirePackage[svgnames,table]{xcolor}

\documentclass[11pt,letterpaper]{arxiv}

\usepackage{mathtools}   %
\usepackage{dsfont}      %

\usepackage{natbib}
\usepackage{hologo}
\setcitestyle{square}

\usepackage{subcaption}  %
\usepackage{float}       %
\usepackage{longtable}
\usepackage{algorithm}
\usepackage{algpseudocode}

\usepackage{listings}
\usepackage{iftex}
\iftutex
\fi
\lstdefinestyle{promptstyle}{
  basicstyle=\ttfamily\footnotesize,
  breaklines=true,
  breakautoindent=false,
  breakindent=0pt,
  postbreak=\mbox{\textcolor{gray}{$\hookrightarrow$}\space},
  columns=fullflexible,
  keepspaces=true,
  frame=single,
  framesep=5pt,
  xleftmargin=6pt,
  xrightmargin=6pt,
  aboveskip=8pt,
  belowskip=8pt,
  showstringspaces=false,
}

\makeatletter
\def\munderbar#1{\underline{\sbox\tw@{$#1$}\dp\tw@\z@\box\tw@}}
\makeatother

\AddToHook{cmd/appendix/before}{%
  \setcounter{axiom}{0}%
}

\newcommand{\be}{\begin{equation}}
\newcommand{\ee}{\end{equation}}
\newcommand{\bea}{\begin{equation*}\begin{aligned}}
\newcommand{\eea}{\end{aligned}\end{equation*}}

\newcommand{\x}{x}

\usepackage{fontawesome5}   %
\usepackage{tikz}
\usetikzlibrary{arrows.meta,shapes.geometric,shapes.callouts}
\usepackage[nameinlink]{cleveref}

\newlength{\coverlogoheight}
\newcommand{\coverlogos}{%
    \raisebox{-0.04\coverlogoheight}{%
        \includegraphics[height=\coverlogoheight]{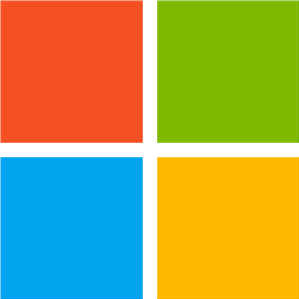}}%
    \hspace{5pt}%
    \textcolor{ghpurplemid}{\rule[-0.1\coverlogoheight]{0.5pt}{1.1\coverlogoheight}}%
    \hspace{5pt}%
    \raisebox{-0.04\coverlogoheight}{\resizebox{!}{\coverlogoheight}{\faGithub}}%
}

\makeatletter
\apptocmd{\abscontent}{%
    \par\vspace{2mm}\noindent\makebox[\linewidth][r]{\coverlogos}\par
}{}{}
\makeatother

\fancypagestyle{firststyle}{%
    \fancyhf{}%
    \fancyhead[L]{}%
    \fancyhead[R]{%
        \ifdefined\paperurl
            \if\relax\the\paperurl\relax\else
                \href{\the\paperurl}{\urlheaderfont\itshape\the\paperurl}\fi
        \fi
    }%
    \fancyhead[C]{}%
    \fancyfoot[C]{\footerfont\bfseries\relax}%
}

\newcounter{insight}

\newcommand{\namelist}[1]{%
  \begingroup\raggedright
  \def\namesep{\def\namesep{, }}%
  \renewcommand*{\do}[1]{\namesep\mbox{##1}}%
  \docsvlist{#1}\par
  \endgroup}

\title{Building a Unified Inline Code Editing\\ Model for GitHub Copilot}

\runningtitle{Building a Unified Inline Code Editing Model for GitHub Copilot}

\newcommand{\reportauthors}{Microsoft Code|AI and VS Code Teams}
\newcommand{\aff}{%
  \textsuperscript{\textcolor{ghpurple}{\normalfont\faGithub}}%
}

\author{\reportauthors\aff\\
}

\hypersetup{colorlinks=true, linkcolor=ghpurple, citecolor=ghpurple,
            urlcolor=ghpurple}

\makeatletter
\def\UrlBreakChars{\do\.\do\@\do\/\do\:\do\_\do\|\do\;\do\>\do\-}
\g@addto@macro{\UrlBreaks}{\UrlBreakChars}
\makeatother

\begin{document}

\begin{abstract}
AI-assisted coding ranges from fully agentic coding to inline code suggestions in the editor. Inline suggestions require predicting a developer's most likely next edit from an editing session. The next edit may lie at the cursor, in nearby lines, or farther away, and one editing intent can span several of these. The preceding system queries separate completion, nearby edit, and long-distance edit models, each with its own output format, in a fixed sequence; it thus cannot express intents spanning distant locations in one response, can require several sequential model calls per suggestion, and prioritizes completion even when a correction would serve better. We present a unified model that predicts all three edit types in one output space: a sequence of anchored diff patches specifying locations, deleted lines, and added lines. This representation permits multiple edits per response, so the client can serve later edits from a cache without further model calls. We train this model in two phases, first unifying next edit and long-distance edit prediction, then adding code completion. We synthesize multi-edit supervision from single-edit data via model pseudolabeling and edit history playback. We use a binary LLM-graded reward, a patch validity taxonomy shared by data filtering, reward, and evaluation, and grader boosting for targeted failure modes. Phase~2 also adapts how the client processes and presents model outputs. In controlled online A/B experiments in GitHub Copilot, Phase~1 significantly reduces time to display by 6.77\% and output tokens by 58.93\% relative to the three-model baseline. In Phase~2, the 3-in-1 model and client adaptations reduce dismissed suggestions per user by 10.06\% relative to the 2-in-1 model paired with the completion model. Throughout this report, we also share our methodology and learnings in evaluation, training, and end-to-end model and system design.
\end{abstract}

\maketitle
\footnotetext[0]{See the \hyperref[sec:contributors]{Contributors and Acknowledgments} section for the full list of contributors. Correspondence should be sent to \href{mailto:codeai@microsoft.com}{\texttt{codeai@microsoft.com}}. Please cite as shown in Appendix~\ref{app:citation-reference}.}

\clearpage
\setcounter{tocdepth}{2}
\tableofcontents
\clearpage

\section{Introduction}
\label{sec:introduction}

\begin{figure}[t]
  \centering
  \begin{subfigure}[b]{\textwidth}
    \centering
    \includegraphics[width=0.57\textwidth]{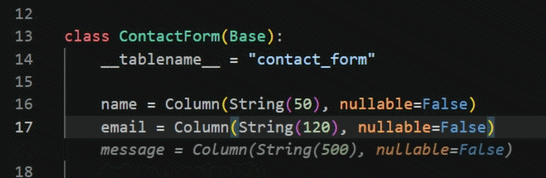}
    \caption{Completion at the cursor, rendered as ghost text in the VS Code client.}
    \label{fig:teaser-ghost-text}
  \end{subfigure}

  \vspace{6pt}
  \begin{subfigure}[b]{\textwidth}
    \centering
    \includegraphics[width=0.88\textwidth]{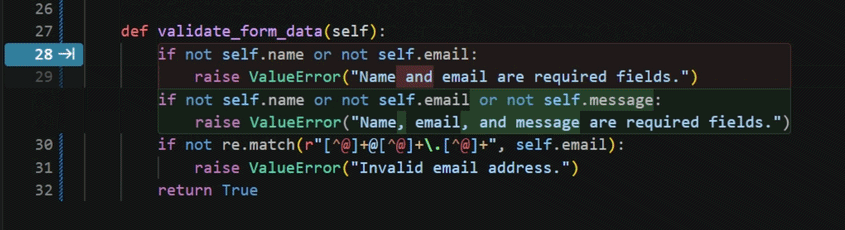}
    \caption{Follow-up edit 10 lines below the cursor, rendered as a side-by-side
    rewrite and served from the client cache.}
    \label{fig:teaser-long-distance-nes}
  \end{subfigure}
  \caption{A single unified model response spanning two edits that completes a \texttt{message} field at
  the cursor and updates the validation logic a few lines below. Predicting both edits together in a single response
  allows the client to show the second edit from its cache without another model call.}
  \label{fig:teaser}

\end{figure}

Inline code suggestion requires a model to predict, from the state of an editing
session, the edit a developer is most likely to make next, and to do so within the
latency budget of interactive typing. The best next edit may extend the current line, revise the surrounding lines (including characters already typed), or propagate a change to
a distant location in the file, and a single editing intent may span
multiple such edits. Code completion and prediction of subsequent edits have been extensively studied using statistical and neural language models and learning-based edit models \citep{hindle2012naturalness,raychev2014completion,bhoopchand2016suggestion,gupta2017deepfix,tufano2019bugfix,chen2021sequencer,yin2019edits,zhao2019sourceedits,brody2020contextual,chakraborty2022codit,chen2021codex,bavarian2022fim,fried2023incoder,reid2022editing,zhang2022coditt5,zhang2022overwatch,gupta2023grace,wei2024coeditor,liu2024coedpilot,li2024instructcoder,aggarwal2025nextcoder,lu2025nextedit}.

In this work, we introduce a unified model for inline code editing
that we post-train for GitHub Copilot Inline Suggestions. This model spans three key code
editing scenarios: code completion at the cursor, next edit suggestion (NES) near the cursor,
and long-distance NES farther from the cursor.
Unifying these scenarios within one model enables it to propagate intent across an editing context by composing multiple edits in a single response, as shown in
Figure~\ref{fig:teaser}.
Our approach spans data curation techniques, supervised fine-tuning (SFT) \citep{wei2022finetuned,ouyang2022training} and
reinforcement learning (RL) \citep{ziegler2019preferences,ouyang2022training}, along with carefully designed
offline and online evaluation methods for model selection.

Three principles guide our design:

\textbf{Earn each intervention.}
Because a correct edit can still be unwanted in the developer's current context, the best response is sometimes no edit.
The system must therefore decide both whether to make a suggestion and what to suggest.
Any suggestion the system makes must meet a quality standard for relevance and utility.
This requirement reflects human-AI interaction principles that emphasize contextually appropriate assistance \citep{amershi2019guidelines}.
We encode this requirement in our data curation and evaluations: each model suggestion must make a meaningful contribution to the developer's current task (Sections~\ref{sec:offline-benchmarks} and~\ref{sec:phase1-data-quality-filtering}).

\textbf{Optimize the end-to-end system, not just the model.}
Developers experience a suggestion through its timing, presentation, and behavior across successive edits.
We therefore co-design the model and client, including the output representation, invocation policy, rendering, and patch caching.
As an example, with the checkpoint held fixed, suppressing the reappearance of ignored suggestions reduces dismissed suggestions per user by 22.40\% relative to leaving suppression disabled (Table~\ref{tab:phase2-suppression-ablation}).

\textbf{Improve the system and its measures together.}
No single benchmark, metric, or scorecard captures the full editing experience.
We combine offline evaluation, end-to-end testing, self-testing, and online A/B experiments because each exposes different aspects of system behavior (Section~\ref{sec:evaluation-methodology}).
In a complex end-to-end system, we need both quantitative metrics and qualitative testing to assess and improve the end-to-end experience.
Finally, we carry learnings forward across iterations, using evaluation outcomes to improve the system and development flywheel and refine our evaluation protocol (Section~\ref{sec:phase1-phase2-offline-online-correlation}).

The model unification described in this report builds on the existing inline
suggestion system in GitHub Copilot. This system decomposes the problem by using
dedicated models for each coding task (completion~\citep{copilotcompletions},
NES~\citep{copilotnes}, and long-distance NES~\citep{vscodelongdistance}) that the code editing client then orchestrates
using a fixed policy.
After careful experimentation to mature the quality of each of these individual models, we take the
natural next step of unifying these models into a single model that can handle all code editing tasks,
with the goal of preserving suggestion quality while reducing latency and simplifying coordination across edit types.

There are several challenges when moving from such a decomposed system to a
unified one. For instance, since the prior standalone models do not produce
multiple edits, we develop a method to bootstrap multi-edit data
to supervise the unified model. We also conduct a careful assessment of 
output format validity and ablate the tradeoffs of RL grader strictness when enforcing
format requirements. In addition to malformed outputs, we address other undesirable behaviors,
such as code duplication and poor insertion patterns, by experimenting with the data distribution and reward design.
The RL grader design is a challenge, as it must balance several
objectives, including correctness, necessity, ordering, validity, and
completeness. We hypothesize that requiring a single grader to enforce all objectives may weaken
its judgments on some objectives. We introduce a technique called RL grader boosting
to improve the enforcement of multiple objectives while keeping the main grader focused on common scenarios.
Offline benchmarks do not perfectly predict online outcomes, so we carefully design the model selection
process to expedite learning and convergence to a strong model candidate.
Finally, and critically, we evolve the end-to-end system as a unit, not merely the model;
the user-facing client is a critical component we optimize in tandem with the model
to deliver a strong user experience.

We train the unified 3-in-1 model in two phases, with the second building upon the first to methodically enhance the model's capabilities.
In each phase, we evaluate the model offline and in controlled online A/B
experiments in GitHub Copilot.
Phase~1 produces a 2-in-1 model that combines next edit suggestion (NES) and
long-distance NES, which the client orchestrates with the completion model. Relative to the original standalone three-model system, it reduces time to
display by 6.77\% and output tokens by 58.93\% with no statistically significant
regression in shown, accepted, dismissed, or ignored suggestions, recurring engagement, or accepted and retained characters.
In Phase~2, we continue reinforcement learning from
the Phase~1 checkpoint to further add code completion capabilities. This produces the
\emph{3-in-1} model, which, together with its client configuration, reduces dismissed suggestions per user by 10.06\% relative to the 2-in-1 model paired with the completion model, although this result is not statistically significant after Bonferroni correction (see Section~\ref{sec:limitations}).
Despite lower latency, output tokens increase by 20.98\% in this Phase~2 comparison, while accepted and retained characters have negative point estimates at both retention windows, neither statistically significant.
These phase-specific results use different controls and do not establish the final system's cumulative effect relative to the original three-model baseline.

In developing these initial unified inline suggestions models for GitHub Copilot,
we produce not only the models and client experience themselves, but also an evolving recipe for a model development
flywheel that spans data pipelines, model training, offline and online evaluation, and more.
We aim to continue refining this flywheel while sharing our learnings
with the community as we advance our shared mission of
supporting every builder in the workflow that best meets their needs.

\section{Unified Code Editing Problem Formulation}
\subsection{Task Definition}
\label{sec:task-definition}

We consider three distinct types of code edits in the unified code editing problem: 
completion, next edit suggestion, and long-distance next edit suggestion.

\paragraph{Completion.}
Completion is an inline suffix prediction task.
Given an editing context, the model generates a strict continuation of the code prefix
currently at the cursor. The client inserts the code at the cursor's position.

\paragraph{Next Edit Suggestion (NES).}
Next edit suggestion (NES) predicts the next code edit to make within a few lines
of the cursor. In our formulation, we consider 2 lines before and 5 lines
after the cursor.
Given an editing context, the model rewrites the code within this specified window.

\begin{figure}
  \centering
  \input{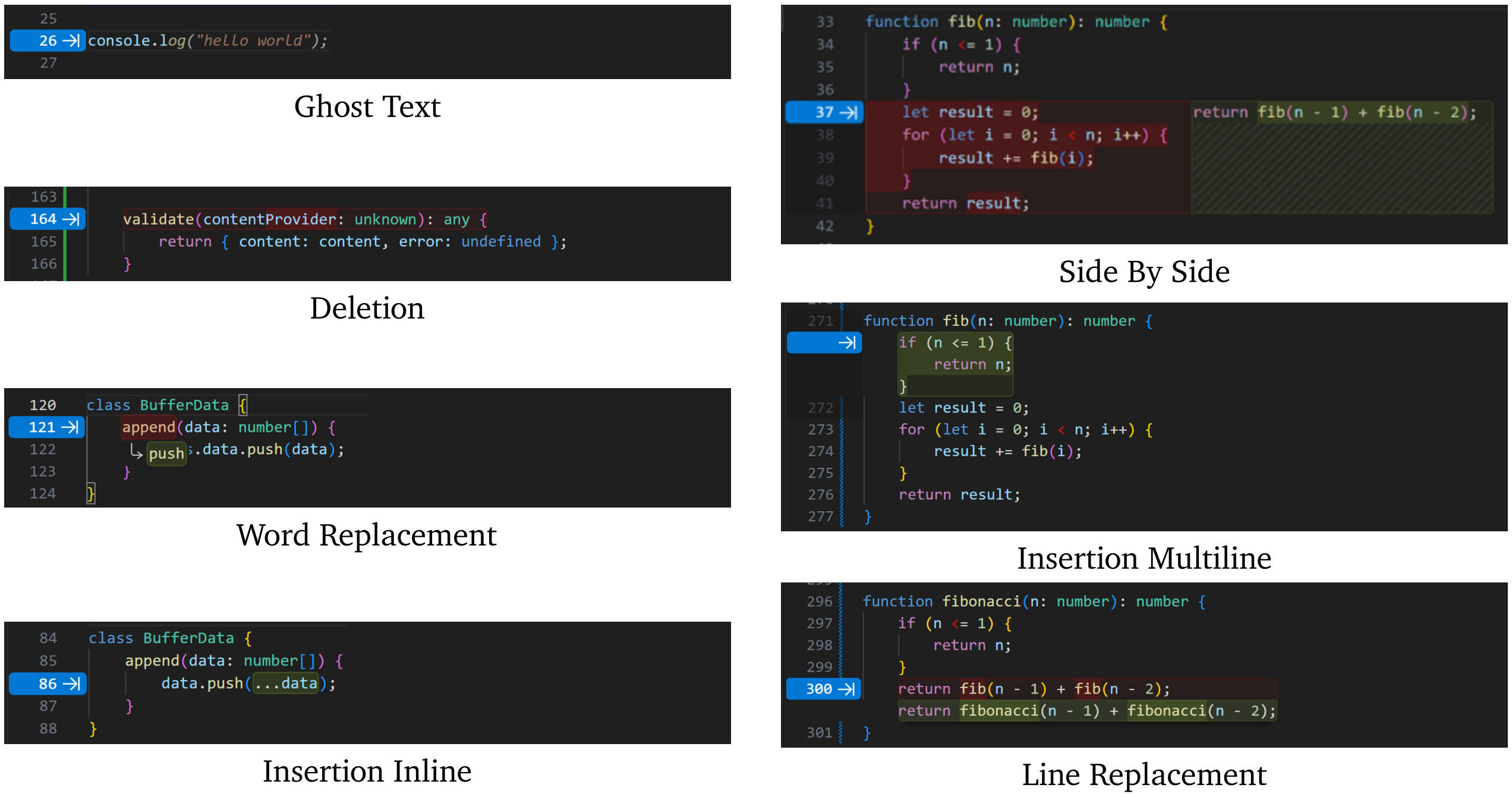}
  \caption{The seven possible NES view kinds as rendered by the VS Code client. The client chooses
  the rendering based on the diff between the suggested edit and the original code.}
  \label{fig:view-kinds}
\end{figure}

The model's suggested edit replaces the code in this window, and the client selects a rendering
based on the diff between the suggested edit and the original code.
We refer to this rendered view as the \textbf{view kind}, which includes ghost text,
side by side, insertion inline, insertion multiline, deletion, word replacement,
and line replacement, as shown in Figure~\ref{fig:view-kinds}.
While completion suggestions are always rendered as \textbf{ghost text at the cursor} (directly after the cursor's current position), NES suggestions can use any of the seven view kinds. NES ghost text suggestions can be at the cursor, just like completion suggestions, or \textbf{away from the cursor}, displayed in the same inline manner at a different location.

\paragraph{Long-Distance NES.}
Long-distance NES is the prediction of the next code edit that is far from the
current cursor position. The edit may lie elsewhere in the current file or in another file.
We refer to the latter as a cross-file edit.

\begin{figure}[t]
  \centering
  \begin{subfigure}[t]{0.48\textwidth}
    \centering
    \includegraphics[width=\linewidth]{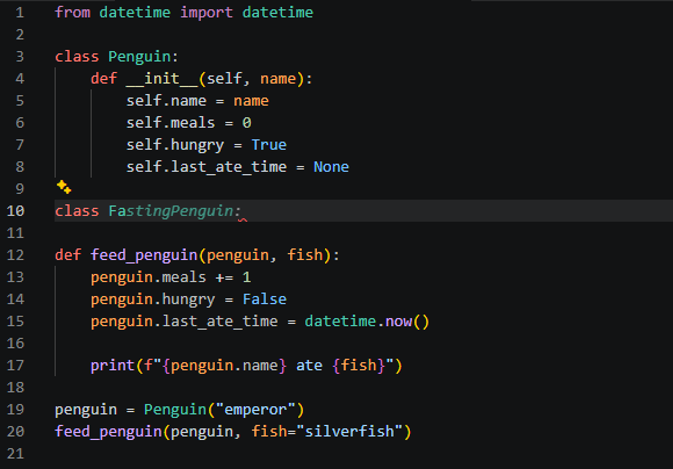}
    \caption{Baseline with three standalone models.}
    \label{fig:quality-before}
  \end{subfigure}
  \hfill
  \begin{subfigure}[t]{0.48\textwidth}
    \centering
    \includegraphics[width=\linewidth]{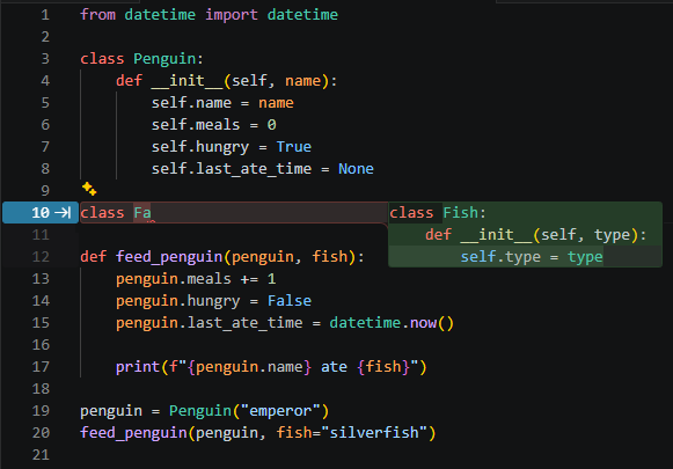}
    \caption{A single unified model.}
    \label{fig:quality-after}
  \end{subfigure}
  \caption{Suggestions predicted by the (a) standalone three-model baseline and (b) unified model
  for the same editing context. The baseline deterministically queries the completion model first, which completes \texttt{Fa} to
  \texttt{FastingPenguin}. The unified model rewrites \texttt{Fa} to \texttt{Fish}.
  Unification allows a single model to choose a context-appropriate correction end-to-end rather than always prioritizing completion.}
  \label{fig:quality-example}
\end{figure}

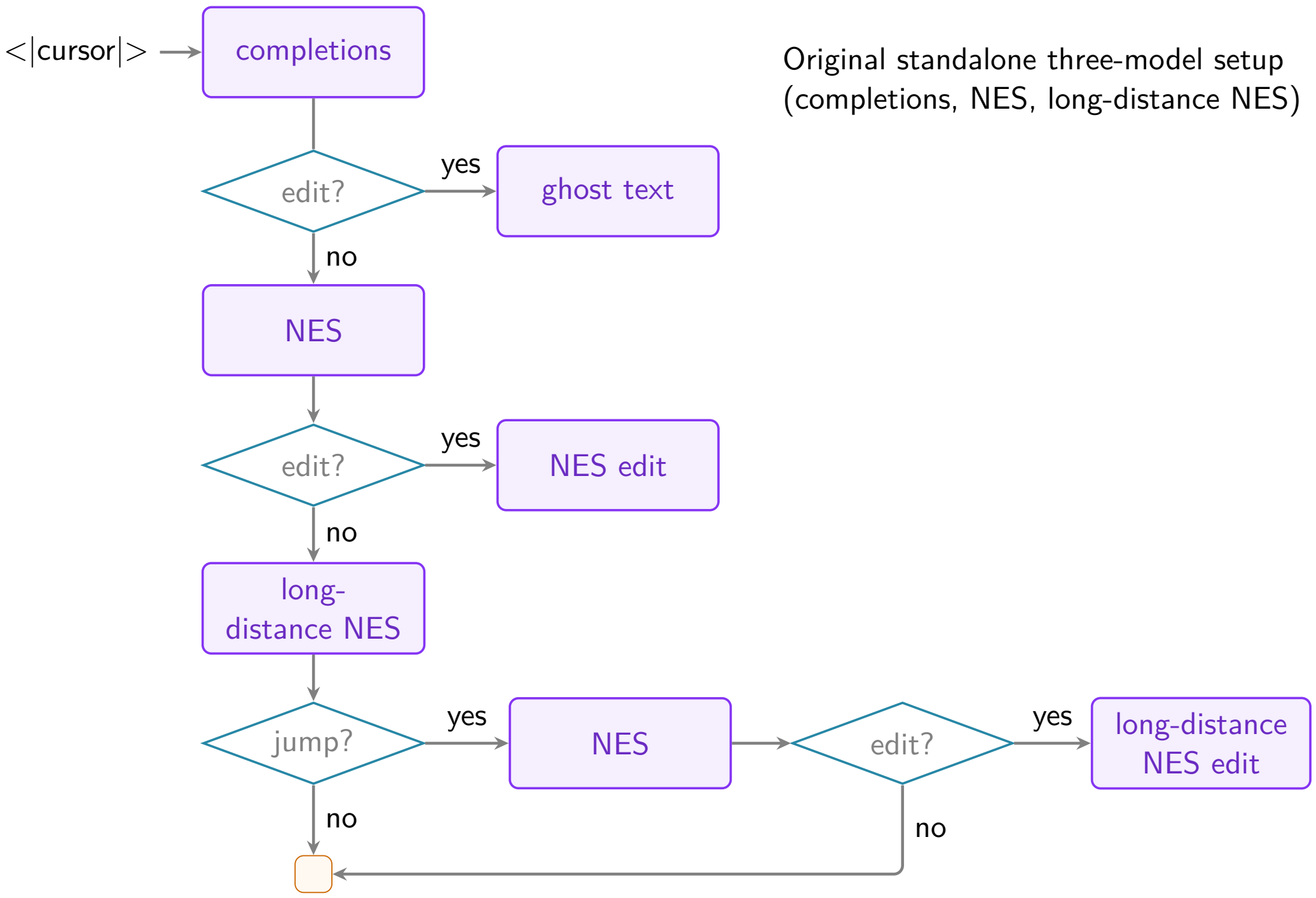
\begin{figure}[t]
  \centering
  \resizebox{0.82\textwidth}{!}{\begin{tikzpicture}[
  font=\sffamily\normalsize,
  model/.style={draw=ghpurple, fill=ghpurple!8, text=ghpurple!80!black, rounded corners=3pt, line width=0.8pt, minimum width=2.7cm, minimum height=1.1cm, inner xsep=8pt, inner ysep=5pt, align=center},
  completion/.style={model},
  nes/.style={model},
  distant/.style={model},
  decision/.style={diamond, aspect=2.6, draw=cyan!60!black, fill=white, text=gray, minimum width=2.7cm, minimum height=1cm, inner sep=0pt, line width=0.8pt},
  output/.style={model, minimum width=2.35cm},
  terminal/.style={draw=orange!80!black, fill=orange!5},
  line/.style={draw=gray, line width=1pt, rounded corners=3pt, >={Stealth[length=5pt, width=4.5pt]}}
]
  \node (cursor) at (0.8,10.5) {<|cursor|>};
  \node[model, completion] (completion) at (3.7,10.5) {completions};
  \node[align=left, anchor=north west] at (9.3,10.7) {Original standalone three-model setup\\(completions, NES, long-distance NES)};

  \node[decision] (completion-edit) at (3.7,8.8) {edit?};
  \node[output, completion] (ghost) at (7.3,8.8) {ghost text};
  \node[model, nes] (nes) at (3.7,7.1) {NES};
  \node[decision] (nes-edit) at (3.7,5.45) {edit?};
  \node[output, nes] (local-output) at (7.3,5.45) {NES edit};
  \node[model, distant] (distant) at (3.7,3.7) {long-\\distance NES};
  \node[decision] (jump) at (3.7,2.05) {jump?};
  \node[model, nes] (jump-nes) at (7.45,2.05) {NES};
  \node[decision] (jump-edit) at (10.9,2.05) {edit?};
  \node[output, distant, minimum width=2.6cm] (distant-output) at (14.55,2.05) {long-distance\\NES edit};
  \node[terminal, rounded corners=3pt, minimum size=0.45cm, inner sep=0pt] (stop) at (3.7,0.45) {};

  \draw[line, ->] (cursor.east) -- (completion.west);
  \draw[line] (completion.south) -- (completion-edit.north);
  \draw[line, ->] (completion-edit.east) -- node[above, text=black] {yes} (ghost.west);
  \draw[line, ->] (completion-edit.south) -- node[right, text=black] {no} (nes.north);
  \draw[line, ->] (nes.south) -- (nes-edit.north);
  \draw[line, ->] (nes-edit.east) -- node[above, text=black] {yes} (local-output.west);
  \draw[line, ->] (nes-edit.south) -- node[right, text=black] {no} (distant.north);
  \draw[line, ->] (distant.south) -- (jump.north);
  \draw[line, ->] (jump.east) -- node[above, text=black] {yes} (jump-nes.west);
  \draw[line, ->] (jump.south) -- node[right, text=black] {no} (stop.north);
  \draw[line, ->] (jump-nes.east) -- (jump-edit.west);
  \draw[line, ->] (jump-edit.east) -- node[above, text=black] {yes} (distant-output.west);
  \draw[line, ->] (jump-edit.south) |- node[pos=0.25, right, text=black] {no} (stop.east);
\end{tikzpicture}}
  \caption{Client-side orchestration for the three-model baseline. A single
  suggestion trigger can require up to four sequential model calls before the client
  displays an edit, adding latency and orchestration complexity.}
  \label{fig:three-model-baseline}
\end{figure}

One way to design a system that handles all three types of edits is to develop a
dedicated model for each edit type and orchestrate them using a model orchestration 
policy. Prior work in GitHub Copilot uses this strategy. However, several limitations
arise when we decompose the problem this way: fixed prioritization of edit types,
as exemplified in Figure~\ref{fig:quality-example}, and latency and system complexity
of multiple-model orchestration, as shown in the example logic of the
standalone three-model system shown in Figure~\ref{fig:three-model-baseline}.

This motivates the development of a unified model that can handle all three types 
of code edits within a single framework. The unified model design
must satisfy the following conditions:
\begin{enumerate}
  \item \textbf{Unification.} Handle completion, next edit suggestion, and long-distance next edit suggestion
  with the goal of maintaining suggestion quality relative to the three-model baseline.
  \item \textbf{Single-model.} Make all edit predictions in a single model.
  \item \textbf{Multi-edit.} Predict zero or more appropriate suggestions given the current editing context.
  \item \textbf{Consistency.} Provide edit sequences that are consistent with user intent, user flow, and logical flow.
  \item \textbf{Extensibility.} Support easy extension to additional code editing tasks and scenarios.
\end{enumerate}

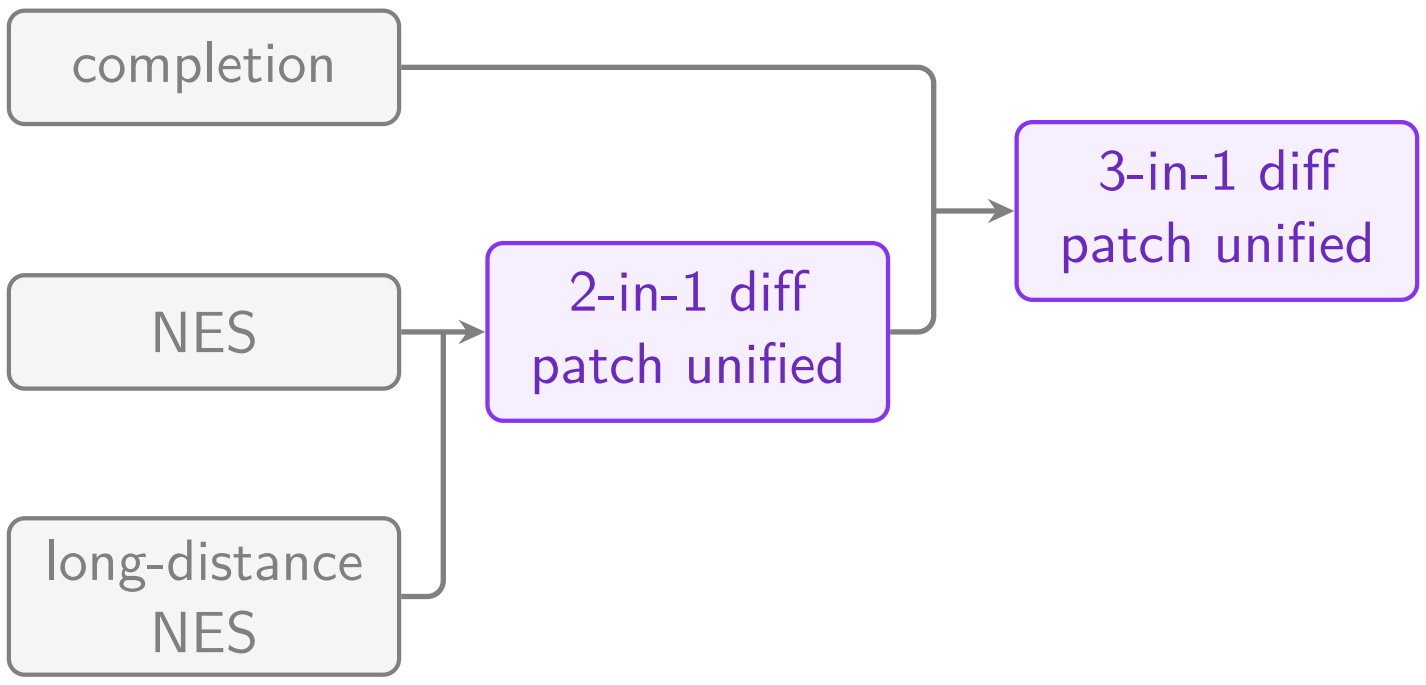
\begin{figure}[t]
  \centering
  \begin{tikzpicture}[
  font=\sffamily\normalsize,
  legacy/.style={draw=gray, fill=gray!8, text=gray, rounded corners=3pt, line width=0.8pt, text width=2.3cm, minimum height=0.75cm, inner sep=4pt, align=center},
  unified/.style={draw=ghpurple, fill=ghpurple!8, text=ghpurple!80!black, rounded corners=3pt, line width=0.8pt, minimum height=0.75cm, inner xsep=8pt, inner ysep=5pt, align=center},
  completed/.style={unified},
  line/.style={line width=1pt, rounded corners=3pt, >={Stealth[length=5pt, width=4.5pt]}}
]
  \node[legacy] (completion) at (1.3,4.25) {completion};
  \node[legacy] (nes) at (1.3,2.5) {NES};
  \node[legacy] (distant) at (1.3,0.75) {long-distance NES};
  \node[unified] (two) at (4.5,2.5) {2-in-1 diff\\patch unified};
  \node[completed] (three) at (8,3.3) {3-in-1 diff\\patch unified};

  \path (nes.east) -- (two.west) coordinate[midway] (two-merge);
  \path (two.east) -- (three.west) coordinate[pos=0.35] (three-merge);

  \draw[line, ->, gray] (nes.east) -- (two.west);
  \draw[line, gray] (distant.east) -| (two-merge);
  \draw[line, gray] (completion.east) -| (three-merge |- three.west);
  \draw[line, gray] (two.east) -| (three-merge |- three.west);
  \draw[line, ->, gray] (three-merge |- three.west) -- (three.west);
\end{tikzpicture}
  \caption{Two-phase model unification strategy. Phase~1 combines NES and long-distance
  NES and Phase~2 adds completion. This allows completion training to build on the
  patch-format editing capabilities developed in Phase~1.}
  \label{fig:phase1-phase2-unification-strategy}
\end{figure}

To systematically develop the unified model's capabilities, we take a two-phase
approach toward unification. We first train a 2-in-1 model combining next edit and long-distance edit prediction, followed by the 3-in-1 model that additionally includes completion.
We intentionally first unify NES and long-distance NES before code completion because completion accounts for more than 70\% of edit opportunities, and experiments preceding this work show NES suggestions are more prone to dismissal than ghost text completion at the cursor. Deferring completion allows Phase~1 to develop the diff-patch data and training recipe for the two editing tasks, which Phase~2 can then build on.
Figure~\ref{fig:phase1-phase2-unification-strategy} illustrates this process.

\subsection{Diff Patch Output Format}
\label{sec:diff-patch-output-format}
To develop a unified model that satisfies the unification requirements, we introduce the
diff patch model output format, which represents code edits as a series of changes
anchored to specific lines in the source files. All unified models we discuss in this report use
this format. Each model response can contain either the
empty string, meaning no suggested edits, or one or more edits represented as anchored patches. Each patch specifies the absolute file path and
line number, followed by at least one deletion line, and optionally one or more addition lines.
We represent a deletion as a patch containing only deletion lines.
We represent an insertion as a patch that deletes an existing anchoring line and
adds it back together with the inserted content. Preliminary experiments show that requiring an anchoring deletion reduces
off-by-one errors in predicted line numbers.

Figure~\ref{fig:diff-patch-format} illustrates how each of the completion, NES, long-distance NES, and unified models expresses suggested edits in the same
editing context. In this example, the diff patch response has multiple edits: it
completes a parameter rename at the cursor and propagates it to two later lines of the same file and to a call
site in a related file.

Note how the output format represents
all three types of code edits while being less verbose on average for NES edits
since it no longer requires a full edit window rewrite. 
In addition, the multi-edit capabilities of the diff patch format allow it to 
naturally and compactly express changes spanning multiple locations in the code
without limitations on distance from the cursor. The multi-edit nature of the
model response format also affords a user experience improvement: \textbf{patch caching}.
The client can cache subsequent patches and show the next edit immediately after the developer
accepts the first, without additional model calls. This leads to a smooth tab-tab-tab client behavior that supports in-flow productivity.

\begin{figure}[!ht]
  \centering
  \footnotesize
  \setlength{\tabcolsep}{5pt}
  \begin{tabular}{@{}p{0.40\textwidth}p{0.54\textwidth}@{}}
    \toprule
    \multicolumn{2}{@{}l}{\textbf{Code to edit}, cursor marked \texttt{<|cursor|>}} \\
    \multicolumn{2}{@{}l}{%
      \begin{minipage}[t]{\dimexpr\textwidth-2\tabcolsep}
      \ttfamily\vspace{2pt}
      0|def feed\_penguin(penguin, fo<|cursor|>):\\
      1|\ \ \ \ penguin.meals += 1\\
      2|\ \ \ \ penguin.hungry = False\\
      3|\ \ \ \ penguin.last\_ate\_time = datetime.now()\\
      4|\\
      5|\ \ \ \ print(f"\{penguin.name\} ate \{fish.type\}")\\
      6|\\
      7|penguin = Penguin("emperor")\\
      8|feed\_penguin(penguin, fish="silverfish")\\
      \vspace{2pt}
      \end{minipage}} \\
    \midrule
    \textbf{Completion:} a suffix at the cursor
      & \texttt{od} \\
    \addlinespace
    \textbf{NES:} the rewritten window, including unchanged lines
      & \begin{tabular}[t]{@{}l@{}}
          \texttt{def feed\_penguin(penguin, food):} \\
          \texttt{\ \ \ \ penguin.meals += 1} \\
          \texttt{\ \ \ \ penguin.hungry = False} \\
          \texttt{\ \ \ \ penguin.last\_ate\_time = datetime.now()} \\
          \texttt{} \\
          \texttt{\ \ \ \ print(f"\{penguin.name\} ate \{food.type\}")}
        \end{tabular} \\
    \addlinespace
    \textbf{Long-distance NES:} the line number of the next edit, followed by a
    second NES call at that line
      & \texttt{5} \\
    \addlinespace
    \textbf{Diff patch:} one or more anchored patches, at any distance and in any
    file
      & \begin{tabular}[t]{@{}l@{}}
          \texttt{/path/to/penguins.py:0} \\
          \texttt{-def feed\_penguin(penguin, fo):} \\
          \texttt{+def feed\_penguin(penguin, food):} \\[0.25em]
          \texttt{/path/to/penguins.py:5} \\
          \texttt{-\ \ \ \ print(f"\{penguin.name\} ate \{fish.type\}")} \\
          \texttt{+\ \ \ \ print(f"\{penguin.name\} ate \{food.type\}")} \\[0.25em]
          \texttt{/path/to/penguins.py:8} \\
          \texttt{-feed\_penguin(penguin, fish="silverfish")} \\
          \texttt{+feed\_penguin(penguin, food="silverfish")} \\[0.25em]
          \texttt{/path/to/antarctica.py:42} \\
          \texttt{-feed\_penguin(penguin2, fish="krill")} \\
          \texttt{+feed\_penguin(penguin2, food="krill")}
        \end{tabular} \\
    \bottomrule
  \end{tabular}
  \caption{The three standalone response formats and the unified diff patch
  format for the same editing context. Diff patches express the parameter rename
  at the cursor and its propagation within and across files in one response
  without rewriting unchanged lines.}
  \label{fig:diff-patch-format}
\end{figure}

We also consider more general formats, including tool calls, which pretraining data
represents better and which support actions beyond editing. For code edits
specifically, however, we find that the patch format is more compact at equal expressiveness.

\subsection{Input Context}

\paragraph{Changes Relative to NES Prompt.}
To support the unified task, we modify the input context used by the original NES model.
The original NES model prompt context contains five components:
recently viewed code snippets, the current file,
recent edit history as a unified diff, a padded region around the edit
target, and the rewrite window. The most notable prompt context change concerns the edit
target and rewrite window: because the unified task can express an edit anywhere
in the file, we replace the last two sections with
a three-line block specifying the line number and content at the cursor. We also
add line numbers to the current file and recently viewed code snippets sections accordingly.

Table~\ref{tab:input-context} summarizes the changes relative to the original NES model prompt.
In addition to these two changes, we remove the backtick fences around the prompt
for simplicity and rewrite the closing instruction to describe the patch task and ordering
requirement (see Appendix~\ref{app:input-context} for an example of the unified model prompt).

\begin{table}[t]
  \caption{Changes to input context sections between the NES and unified prompts. The unified prompt replaces the padded target region and rewrite window with a cursor block, and adds line numbers to the current file and recently viewed code snippets.}
  \label{tab:input-context}
  \centering
  \begin{tabular}{lll}
    \toprule
    Section & NES prompt & Unified prompt \\
    \midrule
    Backtick fences around the prompt & yes & no \\
    Recently viewed code snippets      & yes, without line numbers & yes, with line numbers \\
    Current file & yes, without line numbers & yes, with line numbers \\
    Edit history, unified diff     & yes & yes \\
    Padded region around target    & yes & no \\
    Code to edit, rewrite window   & yes & no \\
    Cursor line block              & no  & yes \\
    \bottomrule
  \end{tabular}
\end{table}

\paragraph{Key-Value Cache Optimization.}
We place the cursor block at the end of the prompt rather than inline in the file
content. Because the cursor moves on nearly every keystroke, an inline marker
would invalidate the shared prefix of consecutive prompts. Placing it at the end
keeps the file body and edit history as a stable prefix, which preserves
key-value cache hits across requests within an editing session.

\section{Evaluation Methodology}
\label{sec:evaluation-methodology}

\subsection{Offline Benchmarks}
\label{sec:offline-benchmarks}

\paragraph{Simulation Tests.}
Simulation Tests (STests) cover common NES editing scenarios and undesirable suggestions identified by the team.
The benchmark consists of workspace recordings. A client simulator replays recordings to reconstruct the workspace state and construct prompts for the candidate model.
The benchmark has two versions, v1 and v2, each organized into three subsets: MustHave, NiceToHave, and MustAvoid.
Each example includes one or more human-labeled reference suggestions with scores of 0 (bad), 0.1 (acceptable), or 1 (good).
If a candidate suggestion matches a reference suggestion, it receives the corresponding score; otherwise, an LLM judge \citep{zheng2023judge} scores it using the labeled suggestions as in-context examples.
Throughout this report, LLM judges in offline benchmarks, data filtering, and RL grading use a general-purpose frontier reasoning model.
We report three metrics: MustHave, the sum of scores on the MustHave subset; Total, the sum of scores across the MustHave and NiceToHave subsets; and MustAvoid, the percentage of examples in the MustAvoid subset that receive a score of 0.1 or 1, indicating successful avoidance of an undesirable suggestion.
Higher values are better for all three metrics.

\paragraph{Augmented HumanEval.}
Augmented HumanEval measures code completion capability by extending HumanEval \citep{chen2021codex} to completion tasks within existing code.
We remove a code span and ask the model to complete the missing code.
The shape of the removed span defines the benchmark's three subsets: Single Line, Multi-Line, and Random Span, abbreviated as Single, Multi, and Span.
We evaluate the completed code using unit tests and report pass@1, the percentage of tasks for which a single generated completion passes the tests, separately for each subset.

\paragraph{Output View Kind.}
\label{sec:output-view-kind}
The Output View Kind benchmark shows what kinds of suggestions the model tends to produce over a curated set of recorded editing contexts.
It reports the distribution of view kinds, which we define in Section~\ref{sec:task-definition}, and the fraction of responses that contain no edit.
These measurements describe output characteristics rather than assign an overall suggestion quality score.
Version 1 distinguishes the seven named view kinds shown in Figure~\ref{fig:view-kinds}, together with no edit, whitespace only, and unknown.
We compute the no-edit ratio by dividing the no-edit count by the total response count.
We compute the ghost text ratio by dividing the ghost text count by the total response count excluding no edit, whitespace-only, and unknown responses.
Version 2 uses a simplified taxonomy for diff-patch models: ghost text, NES, no edit, and invalid.
We classify any valid suggestion that is not ghost text as NES.
The benchmark reports patch validity and average patches per response.
It also reports the average size of insertions and deletions per patch, which we measure in characters and lines.

\paragraph{Pseudo-Online Evaluation (POE).}
POE provides an offline model-selection signal that aims to predict online outcomes
more accurately than the preceding benchmarks.
The benchmark organizes recorded editing contexts into three equally sized
sets based on the outcome of a prior model suggestion: acceptance, rejection,
or no edit.
An LLM judge evaluates each patch from a candidate response in the context of
the prior suggestion and its recorded outcome.
We summarize the judge outcomes as POE acceptance, rejection, and no-edit rates.
These rates reflect offline assessments of candidate suggestions rather than outcomes
measured in online controlled experiments.
We also report the acceptance-to-rejection rate ratio as a key metric.
Two conversion rates measure abstention relative to the recorded outcome: the
fractions of previously accepted and previously rejected suggestions for which
the candidate produces no edit.
These rates distinguish between likely desirable abstention from previously rejected suggestions and
likely less desirable abstention from previously accepted suggestions. We thus also report the rejection-to-no-edit to acceptance-to-no-edit ratio as a key metric, for which higher values are better.
POE also reports the average number of patches per response, the average line
distance between patches, and patch validity error counts (both critical and suboptimal, two classes we discuss further in Section~\ref{sec:phase1-data-quality-filtering}).
Unless otherwise noted, the patch validity error rate refers to the rate of total errors (both critical and suboptimal errors).
POE also breaks down patch-level statistics by patch index.

\paragraph{Qualitative Must-Haves (QMH).}
We use the QMH benchmark to efficiently filter out checkpoints that regress on previously observed failure modes.
We curate its cases from failures that we identify during internal testing.
During Phase~1 model development, the team identifies insertion placement errors in internal testing, but the available offline benchmarks and online experiments do not capture them. This gap motivates QMH as a compact canary benchmark for recurring high-priority failures. We discuss these errors further in Section~\ref{sec:phase1-targeted-behavior-curation}.
The cases in this benchmark cover a small set of high-priority behaviors that candidate models
must perform reliably.
The cases include ghost text completions, nearby NES, long-distance NES, and cross-file examples.
QMH therefore serves as a quality guardrail for recurring regressions rather
than as an estimate of overall model quality.
For each case, an LLM judge compares the candidate response with desired and
undesired reference responses and assigns a pass or fail outcome.

\subsection{Online Evaluation Metrics}
\label{sec:online-metrics}
Table~\ref{tab:online-metric-definitions} defines the key online metrics.
No single metric establishes model acceptability because a gain on one metric
can mask a regression on another.
For example, showing more suggestions may increase acceptances but also cause
more interruptions, while shorter responses may reduce latency but contribute
less retained content.
We therefore assess each candidate jointly across suggestion quality, latency,
and inference efficiency.
We accept a candidate only when it achieves its intended improvement without
introducing a statistically significant regression in the designated guardrail
metrics. In the case of model unification, our primary aim is to unify an existing set of models while maintaining performance on par with the control, rather than to outperform it.

The absence of statistically significant guardrail regressions, however, is necessary but not sufficient for acceptance.
We also assess the direction and width of confidence intervals for each metric and the results of qualitative testing.
To maintain a strong standard for the editing experience, we delay deployment when metrics show a meaningful risk of regressions, experiment results are too uncertain to rule out regressions, or internal testing identifies qualitative regressions that we judge unacceptable for deployment.

\begin{table}[t]
  \caption{Online metric definitions. We use multiple metrics to assess various dimensions of the user experience, including suggestion quality, engagement, latency, and inference efficiency.}
  \label{tab:online-metric-definitions}
  \centering
  \small
  \begin{tabularx}{\linewidth}{
    >{\raggedright\arraybackslash\hangindent=1em\hangafter=1}p{3.7cm}
    >{\raggedright\arraybackslash}X
  }
    \toprule
    Metric & Definition \\
    \midrule
    Shown suggestions & Mean suggestions displayed per user during the experiment \\
    Accepted suggestions & Mean suggestions accepted per user during the experiment \\
    Dismissed suggestions & Mean suggestions explicitly dismissed per user during the experiment \\
    Ignored suggestions & Mean suggestions neither accepted nor dismissed per user during the experiment \\
    Accepted and Retained Characters (ARC) &
      Mean characters accepted and not reverted per user at each retention window (30 seconds or 5 minutes after acceptance) \\
    Recurring engagement &
      Proportion of users who accept an inline suggestion on $\geq 2$ different days \\
    Time to display &
      Mean per-user latency from suggestion trigger to suggestion display \\
    Time to produce &
      Mean per-user latency from suggestion trigger to model response \\
    Output tokens & Mean tokens emitted per response per user, excluding the standalone completion model \\
    Prompt tokens & Mean tokens submitted per request per user, excluding the standalone completion model \\
    Cached suggestion share & Percentage of shown suggestion events served from the client cache, excluding the standalone completion model \\
    \bottomrule
  \end{tabularx}
\end{table}

For ARC, we cap each user's total at 35{,}000 characters at each retention window before averaging across users.
Token metrics first average token counts across each user's requests or responses, then average across users, excluding the standalone completion model in both phases.
In Phase~1, completion remains a separate component in both arms.
In Phase~2, token metrics compare the 3-in-1 model with the 2-in-1 model alone, excluding the control's separate completion model.
These Phase~2 metrics therefore compare models with different task coverage.

\subsection{A/B Experiment Design}
\label{sec:online-experiment-design}
We obtain results using controlled online A/B experiments, which are randomized controlled
experiments that run over a fixed exposure period. Most reported experiments run for at least 7 days, with some running for 14 days and the shortest running for 5 days. We choose these durations so that results reflect developers' experience over sustained use and reduce sensitivity to day-of-week effects.
For each experiment, we randomly select a small fraction of eligible VS Code stable users and then randomly assign them in equal proportions to the experiment's conditions.
None of the reported experiments shows a sample ratio mismatch (SRM), which would indicate that an arm does not receive its intended share of users.
In this work, the treatment is the candidate unified model and client configuration and the control is the then-production baseline model and client configuration unless otherwise specified.
We hold the configuration constant between groups except for the model or
client change under evaluation, allowing us to attribute metric differences to
that intervention.

\paragraph{Target Metrics and Hypotheses.}
We specify target metrics before each experiment based on the intervention we evaluate.
In Phase~1, the goal is to unify NES and long-distance NES while maintaining suggestion quality relative to the three-model baseline.
We also hypothesize that edit caching with the multi-patch representation leads to fewer sequential model calls, and that the more compact diff-based format reduces time to display, time to produce, and output tokens.
Accepted suggestions, dismissed suggestions, ignored suggestions, ARC, and recurring engagement serve as quality and engagement guardrails that detect whether the model achieves these reductions through undesirable means, such as suggesting less useful content or interrupting developers more often.
In Phase~2, the goal is to replace the 2-in-1 model paired with the completion model while maintaining performance across the guardrail metrics.
Results from earlier experiments also motivate new hypotheses for subsequent experiments.
For instance, the initial experiment using the 2-in-1 model without the separate completion model reveals regressions in accepted suggestions and ARC; completion-focused training targets these regressions by improving completion coverage and completeness (Section~\ref{sec:phase2-unifying-2-in-1-with-code-completion}).
Client experiments also test other hypotheses.
For example, we hypothesize that preventing an ignored ghost text suggestion from reappearing as a different view kind reduces repeated interruptions without reducing useful suggestions, thereby reducing dismissed suggestions (Section~\ref{sec:phase2-client-ablations}).

\paragraph{Estimation and Candidate Selection.}
Each experiment uses a new independent random sample of eligible users.
We aggregate most metrics per user and report the treatment-control comparison for each metric in an A/B experiment scorecard.
The only exception to per-user aggregation is cached suggestion share, which is computed at the event level.
We report relative changes as the estimate $\pm$ the 95\% confidence interval half-width, expressing both quantities as percentages.
For each metric, the scorecard reports a \(p\)-value for the null hypothesis
that the treatment and control do not differ.
We describe \(0.001 < p < 0.05\) as statistically significant and
\(p \leq 0.001\) as highly statistically significant.
In result tables, bold entries indicate statistical significance, and we display \(p\)-values below 0.00005 as \(\approx 0\).
We primarily rely on statistically significant metric changes to guide our decision-making.
Following statistical reasoning, it is important to note that a lack of statistical significance does not establish equivalence between the candidate and control models, only that the evidence does not meet the chosen significance threshold.
We calculate $p$-values separately for each experiment, without adjustment for multiple metrics or candidates.
The reported significance levels therefore do not control false positives across the full candidate search.
We interpret each result against its stated target, the full guardrail scorecard, and qualitative testing, as opposed to relying solely on an isolated favorable metric change.

\paragraph{Cross-Phase Comparisons.}
We do not directly compare the final 3-in-1 model and client configuration with the original three-model baseline.
The phase-specific experiments use different controls, exposure periods, and sampled users, so we do not multiply their relative changes across phases to estimate a cumulative effect relative to the original baseline.

\subsubsection{Client Configuration in A/B Experiments}
Each A/B experiment evaluates a model checkpoint together with a client configuration.
The client determines when it invokes a model, how it renders patches, whether
it preserves ignored suggestions, and when it shows cached suggestions.
These choices can change suggestion outcomes and latency even when we hold the
model checkpoint fixed.
Section~\ref{sec:phase2-client-ablations} examines the effects of individual client
settings.

\subsection{Model Selection Process}
\label{sec:model-selection-process}

Our model selection process consists of several stages that efficiently narrow
the candidate checkpoint sets by progressing from lower-effort to more involved evaluation methods.
We use the process described here from late in Phase~1 RL stage~1 onward, which covers the final Phase~1 candidate and all of Phase~2.
Earlier Phase~1 checkpoints, including the SFT checkpoint, predate the 19-case QMH set, so we screen them with the other offline benchmarks.

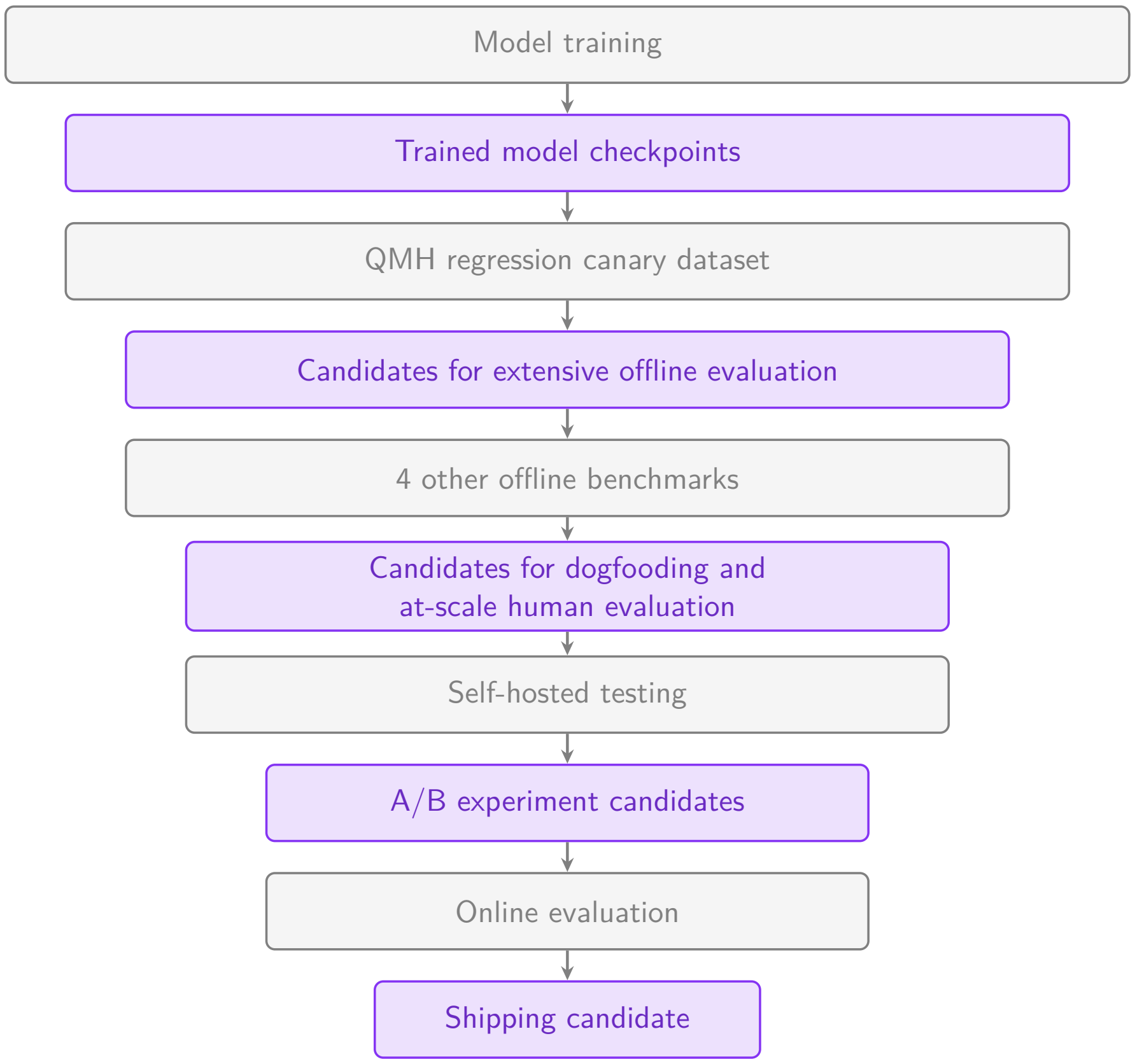
\begin{figure}[!ht]
  \centering
  \resizebox{0.70\linewidth}{!}{\begin{tikzpicture}[
  font=\sffamily\normalsize,
  stage/.style={rounded corners=3pt, line width=0.8pt, minimum height=0.95cm, inner xsep=8pt, inner ysep=5pt, align=center},
  evaluation/.style={draw=gray, fill=gray!8, text=gray},
  candidates/.style={draw=ghpurple, fill=ghpurple!15, text=ghpurple!80!black},
  line/.style={draw=gray, line width=1pt, >={Stealth[length=5pt, width=4.5pt]}}
]
  \node[stage, evaluation, minimum width=14cm] (training) at (0,12.15) {Model training};
  \node[stage, candidates, minimum width=12.5cm] (checkpoints) at (0,10.8) {Trained model checkpoints};
  \node[stage, evaluation, minimum width=12.5cm] (qmh) at (0,9.45) {QMH regression canary dataset};
  \node[stage, candidates, minimum width=11cm] (offline-candidates) at (0,8.1) {Candidates for extensive offline evaluation};
  \node[stage, evaluation, minimum width=11cm] (benchmarks) at (0,6.75) {4 other offline benchmarks};
  \node[stage, candidates, minimum width=9.5cm] (testing-candidates) at (0,5.4) {Candidates for dogfooding and\\at-scale human evaluation};
  \node[stage, evaluation, minimum width=9.5cm] (testing) at (0,4.05) {Self-hosted testing};
    \node[stage, candidates, minimum width=7.5cm] (online-experiment-candidates) at (0,2.7) {A/B experiment candidates};
  \node[stage, evaluation, minimum width=7.5cm] (online) at (0,1.35) {Online evaluation};
  \node[stage, candidates, minimum width=4.8cm] (shipping) at (0,0) {Shipping candidate};

  \draw[line, ->] (training.south) -- (checkpoints.north);
  \draw[line, ->] (checkpoints.south) -- (qmh.north);
  \draw[line, ->] (qmh.south) -- (offline-candidates.north);
  \draw[line, ->] (offline-candidates.south) -- (benchmarks.north);
  \draw[line, ->] (benchmarks.south) -- (testing-candidates.north);
  \draw[line, ->] (testing-candidates.south) -- (testing.north);
  \draw[line, ->] (testing.south) -- (online-experiment-candidates.north);
  \draw[line, ->] (online-experiment-candidates.south) -- (online.north);
  \draw[line, ->] (online.south) -- (shipping.north);
\end{tikzpicture}}
  \caption{Progressive model selection process. Lightweight canary checks narrow the
  candidate set before broader and more extensive offline, internal, and online evaluation.
  If no candidate meets a stage's criteria, we return to an earlier stage to iterate.}
  \label{fig:phase1-phase2-model-selection-process}
\end{figure}

Figure~\ref{fig:phase1-phase2-model-selection-process} illustrates these progressive stages of model selection.
We first cast a wide net by sampling model checkpoints every 100 training steps and
evaluating each checkpoint on the lightweight QMH dataset of 19 examples.
To ensure QMH serves effectively as a canary set, we carefully curate QMH
to adequately represent the key editing contexts that define a high-quality user
experience. Without this careful design, using it in this important proxy role would lead to a higher rate of inclusion of poor candidates
or exclusion of strong candidates for further testing.
In more mature stages of training, we further expand the pool of candidates by considering additional checkpoints
around loci of promising checkpoints in the first QMH evaluation round; for instance,
if the 400th step has a strong score, we may also consider evaluating the 
350th step through the 450th step with a higher granularity (say, every 10th step) to ensure we do not miss nearby promising candidates.

Once QMH identifies promising candidates,
we evaluate them on the four other more extensive offline benchmarks.
Based on the composite scores across these benchmarks, we identify a smaller set of
strong candidates for internal testing via team dogfooding and at-scale human evaluation studies.
The feedback from the end-to-end testing in the client further narrows down the
candidates for larger-scale online A/B testing via the aforementioned randomized
controlled online A/B experiments.

After each experiment concludes, we carefully analyze
the target and guardrail metrics to determine whether any candidate meets the
deployment criteria. If no candidate meets these criteria, we may return
to any prior step in the process to iterate on the appropriate methods, including
data, model training, client settings, and other factors. This also applies to all
intermediate steps; if no promising candidates exist, we return to prior steps to iterate.

\section{Phase 1: Unifying NES and Long-Distance NES}
\label{sec:phase1-nes-long-distance-unification}

\subsection{Model Architecture}
We fine-tune a causal large language model \citep{brown2020language,chowdhery2022palm,anil2023palm2,touvron2023llama,touvron2023llama2,jiang2023mistral} based on the Transformer architecture \citep{vaswani2017attention} to generate code edits in the diff patch format described in Section~\ref{sec:diff-patch-output-format}.
We use the \texttt{o200k\_base} tokenizer \citep{openai2024o200k}.

\subsection{Supervised Fine-Tuning}
\label{sec:phase1-sft}

\subsubsection{Data Format Conversion}
\label{sec:phase1-data-format-conversion}
We use two pipelines to produce patch-format samples from existing NES data.
The first pipeline operates on serialized prompt-response samples in the window-rewrite format and therefore
requires format-specific conversion tooling.
It parses each NES prompt and response and expresses the ground truth edit as one or
more patches.
The same tooling converts the five offline benchmarks and preserves their
underlying samples across the output-format reformulation.

The second pipeline operates on workspace recordings and does not depend on a
particular prompt or output format.
During playback, a simulator in the client reconstructs the workspace state,
transforms it into the requested prompt format, and uses the recorded action as
the ground truth label.
Because the client simulator handles sample formatting, we can convert the same
recording again after client-side prompt changes without a new source-format
converter.

\subsubsection{Data Bootstrapping}
\label{sec:phase1-data-bootstrapping}

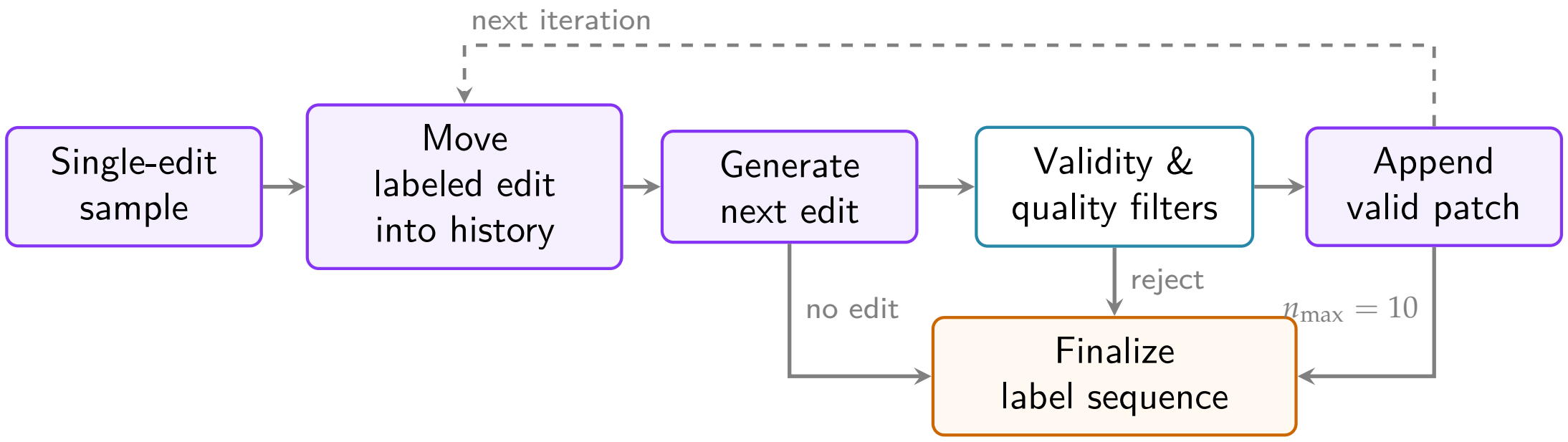
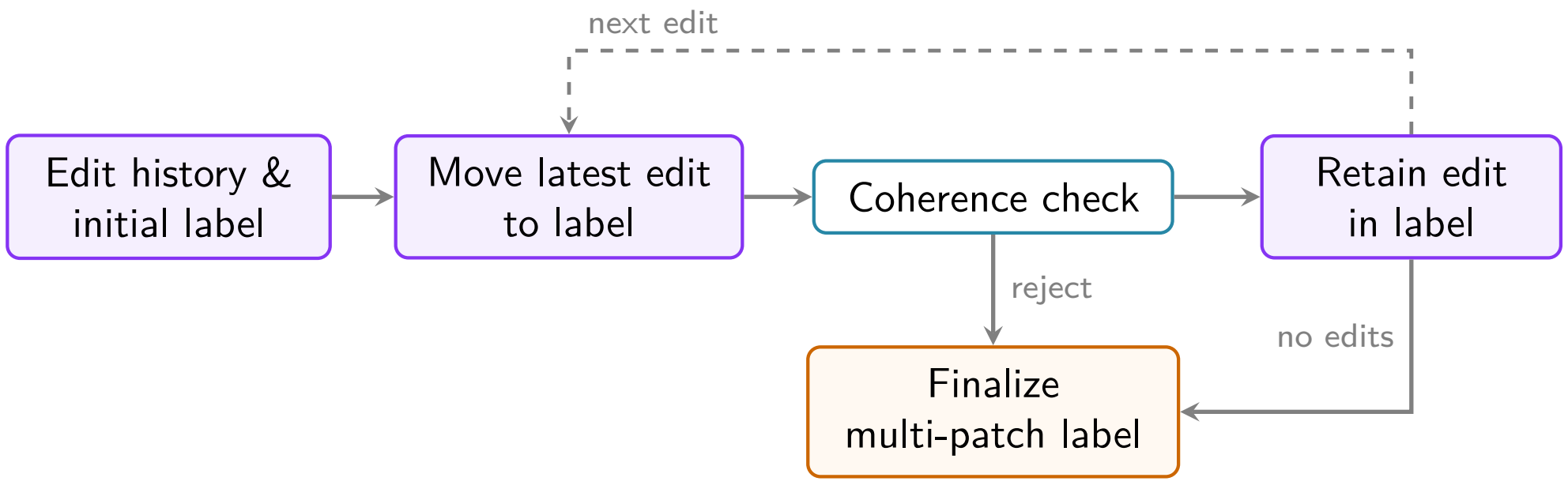
\begin{figure}[!ht]
  \centering
  \captionsetup{justification=centering}
  \captionsetup[subfigure]{font=small,justification=centering}
  \begin{subfigure}{\linewidth}
  \centering
\begin{tikzpicture}[
  font={\sffamily\small\hyphenpenalty=10000\exhyphenpenalty=10000\relax},
  box/.style={
    draw=ghpurple,
    fill=ghpurple!8,
    rounded corners=3pt,
    line width=0.8pt,
    align=center,
    inner xsep=5pt,
    inner ysep=5pt
  },
  check/.style={
    draw=cyan!60!black,
    fill=white,
    rounded corners=3pt,
    line width=0.8pt,
    align=center,
    inner xsep=5pt,
    inner ysep=5pt
  },
  stop/.style={
    draw=orange!80!black,
    fill=orange!5,
    rounded corners=3pt,
    line width=0.8pt,
    align=center,
    inner xsep=5pt,
    inner ysep=5pt
  },
  flow/.style={->, draw=gray, line width=1pt, >=stealth},
  iterate/.style={->, dashed, draw=gray, line width=0.9pt, >=stealth},
  note/.style={font=\sffamily\scriptsize, text=gray}
]
  \node[box, text width=2.0cm] (single) at (1.1,2.0)
    {Single-edit\\sample};
  \node[box, text width=2.55cm] (advance) at (4.15,2.0)
    {Move labeled edit\\into history};
  \node[box, text width=2.0cm] (model) at (7.15,2.0)
    {Generate\\next edit};
  \node[check, text width=2.2cm] (filter) at (10.15,2.0)
    {Validity \&\\quality filters};
  \node[box, text width=2.0cm] (append) at (13.1,2.0)
    {Append\\valid patch};
  \node[stop, text width=3.0cm] (terminate) at (10.15,0.25)
    {Finalize\\label sequence};

  \draw[flow] (single.east) -- (advance.west);
  \draw[flow] (advance.east) -- (model.west);
  \draw[flow] (model.east) -- (filter.west);
  \draw[flow] (filter.east) -- (append.west);
  \draw[flow] (model.south) |- (terminate.west)
    node[pos=0.24, right, note] {no edit};
  \draw[flow] (filter.south) -- (terminate.north)
    node[midway, right, note] {reject};
  \draw[flow] (append.south) |- (terminate.east)
    node[pos=0.24, left, note] {\(n_{\max}=10\)};
  \draw[iterate] (append.north) -- ++(0,0.75) -| (advance.north)
    node[pos=0.45, above, note] {next iteration};
\end{tikzpicture}
  \caption{Model pseudolabeling extends a label until generation or filtering
  determines that the sequence should end.}
\end{subfigure}

\medskip
\begin{subfigure}{\linewidth}
  \centering
\begin{tikzpicture}[
  font={\sffamily\small\hyphenpenalty=10000\exhyphenpenalty=10000\relax},
  box/.style={
    draw=ghpurple,
    fill=ghpurple!8,
    rounded corners=3pt,
    line width=0.8pt,
    align=center,
    inner xsep=5pt,
    inner ysep=5pt
  },
  check/.style={
    draw=cyan!60!black,
    fill=white,
    rounded corners=3pt,
    line width=0.8pt,
    align=center,
    inner xsep=5pt,
    inner ysep=5pt
  },
  output/.style={
    draw=orange!80!black,
    fill=orange!5,
    rounded corners=3pt,
    line width=0.8pt,
    align=center,
    inner xsep=5pt,
    inner ysep=5pt
  },
  flow/.style={->, draw=gray, line width=1pt, >=stealth},
  iterate/.style={->, dashed, draw=gray, line width=0.9pt, >=stealth},
  note/.style={font=\sffamily\scriptsize, text=gray}
]
  \node[box, text width=2.35cm] (history) at (1.25,1.75)
    {Edit history \&\\initial label};
  \node[box, text width=2.55cm] (move) at (4.6,1.75)
    {Move latest edit\\to label};
  \node[check, text width=2.65cm] (intent) at (8.15,1.75)
    {Coherence check};
  \node[box, text width=2.15cm] (retain) at (11.65,1.75)
    {Retain edit\\in label};
  \node[output, text width=2.75cm] (labeloutput) at (8.15,-0.05)
    {Finalize\\multi-patch label};

  \draw[flow] (history.east) -- (move.west);
  \draw[flow] (move.east) -- (intent.west);
  \draw[flow] (intent.east) -- (retain.west);
  \draw[flow] (intent.south) -- (labeloutput.north)
    node[midway, right, note] {reject};
  \draw[flow] (retain.south) |- (labeloutput.east)
    node[pos=0.25, left, note] {no edits};
  \draw[iterate] (retain.north) -- ++(0,0.7) -| (move.north)
    node[pos=0.45, above, note] {next edit};
\end{tikzpicture}
  \caption{Edit history playback iteratively transfers coherent edits from the
  history into the patch sequence in the label.}
\end{subfigure}
  \caption{Multi-patch synthesis from single-edit data through iterative model
  pseudolabeling and edit history playback. Both pipelines bootstrap multi-edit samples from samples with single-edit supervision to teach the model when and how to continue edits beyond the first patch.}
  \label{fig:phase1-data-bootstrapping}
\end{figure}

Moving from a fixed edit window to a sequence of patches introduces a stopping
decision: after each patch, the model must decide whether another edit should
follow or the response should end.
The existing datasets provide insufficient multi-edit supervision for the
model to learn when to continue a patch sequence, biasing it toward terminating
after the first patch.
We therefore synthesize multi-edit samples through model pseudolabeling and
edit history playback to provide explicit supervision for continuation
(Figure~\ref{fig:phase1-data-bootstrapping}).

\paragraph{Model Pseudolabeling.}
We advance a single-edit sample by moving its ground truth edit into the edit history
and running our strongest model checkpoint at the time on the resulting context.
If the predicted edit passes all patch validity checks, we append it to the
patch sequence in the ground truth label and repeat the process.
We stop generation when the model produces no edit, a prediction fails a
validity check, or the sequence reaches \(n_{\max}=10\) patches.
Each successful iteration therefore extends an \(n\)-patch sample to
\(n+1\) patches.

\paragraph{Edit History Playback.}
We iteratively move the latest edit from the recorded edit history into the
patch sequence in the ground truth label.
An LLM judge retains the edit only if it shares the ground truth label's intent or forms a
logical next step given the remaining history, after which the process repeats.
For each sample, we transfer between 1 and \(n\) edits in this way, with
\(n_{\max}=10\).

\subsubsection{Data Quality Filtering and Curation}
\label{sec:phase1-data-quality-filtering}
We view data quality through three lenses: patch validity, suggestion quality, and targeted behavior curation.
We filter all curated samples, whether raw samples or those we generate through iterative model pseudolabeling or edit history playback.
Our overall process is to first filter samples by suggestion quality and then use
an iterative LLM feedback loop to ensure patch validity. In the case of addressing
specific problematic behaviors, we employ dedicated curation pipelines that identify and rewrite problematic examples
before these two steps.

\paragraph{Suggestion Quality.}
We first filter raw data for quality using an LLM-as-a-judge \citep{zheng2023judge}.
We supply 14 criteria in this filter that represent developer preferences we identify while developing our preceding code editing models.
These include correctness, necessity, and alignment with intent and edit history,
as well as other criteria that high-quality code edits should satisfy.
The filter also includes guidance on patch ordering, requiring patches to follow
a logical order, both structurally and semantically.
We also adhere to this logic for bootstrapped samples as we discuss in Section~\ref{sec:phase1-data-bootstrapping}.

We also have separate filters for samples whose responses have no edits
and those that have at least one edit. In our experience, deciding whether to make
an edit is a nuanced decision that has a separate decision boundary from edit quality. In practice, we find that leniency toward abstention in ambiguous cases leads to models that perform better online.
Separating these filters allows us to intentionally curate data that adequately
characterizes both distributions.

We use the same grader for suggestion quality filtering and RL training. This maintains a shared standard for suggestion quality
that benefits both our data pipelines and model training recipes.

\paragraph{Patch Validity.} As we define in Section~\ref{sec:diff-patch-output-format},
a patch consists of a file path, line number, and a sequence of deletions
and additions. A model suggestion contains zero or more patches. However, there are many
nuances to the definition of a valid patch that we consider. First, we develop a set
of seventeen rules that define ideal patch structure, which we discuss in Appendix~\ref{app:patch-validity-taxonomy}.
The rules cover various aspects such as valid file paths and line numbers, deletion and addition sequencing and redundancy, and code duplication.
As we discuss in later sections, we ultimately divide these rules into two
groups: critical and suboptimal, as Tables~\ref{tab:patch-validity-critical} and \ref{tab:patch-validity-suboptimal} detail.

During RL training, as we discuss in Section~\ref{sec:phase1-rl-ablations}, we find that while it is important to penalize critical issues,
penalizing suboptimal issues may be too strict and thus negatively impact model quality.
However, for data filtering, we can curate desired behavior more strictly. We first identify every example that violates any rule, then use an
iterative LLM feedback loop to rewrite the offending patch or response until it satisfies all rules while preserving the original edit's intent.
This ensures that every training sample has an ideal patch representation as its ground truth.

\paragraph{Targeted Behavior Curation Pipelines.}
\label{sec:phase1-targeted-behavior-curation}
We also construct dedicated data curation pipelines to target specific model behaviors. One particular class of behaviors we target is recurring problematic
behaviors from earlier model iterations. We address these behaviors through the preceding patch validity
pipeline and a curation pipeline targeting insertion pattern propagation.

Insertion pattern propagation is a failure mode in which the model repeats an insertion
pattern across multiple edits in the same response. The model generates the correct
content for each insertion but places it incorrectly, often one or a few lines from its
intended location.
For example, the editing context may show a user adding a new form field,
\texttt{password}, after \texttt{email}, while the model places \texttt{password}
before \texttt{email} or after the field that follows \texttt{email}.

\begin{figure}[htbp]
  \centering
  \begin{tikzpicture}[
  font={\sffamily\small\hyphenpenalty=10000\exhyphenpenalty=10000\relax},
  box/.style={
    draw=ghpurple,
    fill=ghpurple!8,
    rounded corners=3pt,
    line width=0.8pt,
    align=center,
    inner xsep=5pt,
    inner ysep=5pt
  },
  check/.style={
    draw=cyan!60!black,
    fill=white,
    rounded corners=3pt,
    line width=0.8pt,
    align=center,
    inner xsep=5pt,
    inner ysep=5pt
  },
  flow/.style={->, draw=gray, line width=1pt, >=stealth},
  iterate/.style={<->, dashed, draw=gray, line width=0.9pt, >=stealth},
  note/.style={font=\sffamily\scriptsize, text=gray}
]
  \node[box, text width=3.1cm, minimum height=1.45cm] (identify) at (1.9,2.2)
    {Identify samples exhibiting target failure mode};

  \node[box, text width=3.0cm, minimum height=1.45cm] (rewrite) at (7.3,2.2)
    {LLM rewrites response};
  \node[check, text width=3.0cm, minimum height=1.05cm] (judge) at (7.3,0.1)
    {LLM judge: passes target-mode filter?};

  \node[box, text width=3.2cm, minimum height=1.45cm] (final) at (12.8,2.2)
    {Apply quality filter (and patch validity filter, if not yet run)};

  \node[font=\sffamily\bfseries\small, text=ghpurple] at (1.9,3.45) {Step 1};
  \node[font=\sffamily\bfseries\small, text=ghpurple] at (7.3,3.45) {Step 2. Iterative rewrite};
  \node[font=\sffamily\bfseries\small, text=ghpurple] at (12.8,3.45) {Step 3};

  \draw[flow] (identify.east) -- (rewrite.west);
  \draw[iterate] (rewrite.south) -- node[left, note, align=right] {no: rewrite again} (judge.north);
  \draw[flow] (judge.east) -- node[midway, above, note] {yes}
    (final.south |- judge.east) -- (final.south);
\end{tikzpicture}
  \caption{Targeted behavior curation pipelines. Iterative rewriting repairs examples of
  a recurring failure mode. After rewriting, quality and patch validity checks ensure that the
  repairs also meet the general training data quality requirements.}
  \label{fig:phase1-targeted-behavior-curation}
\end{figure}
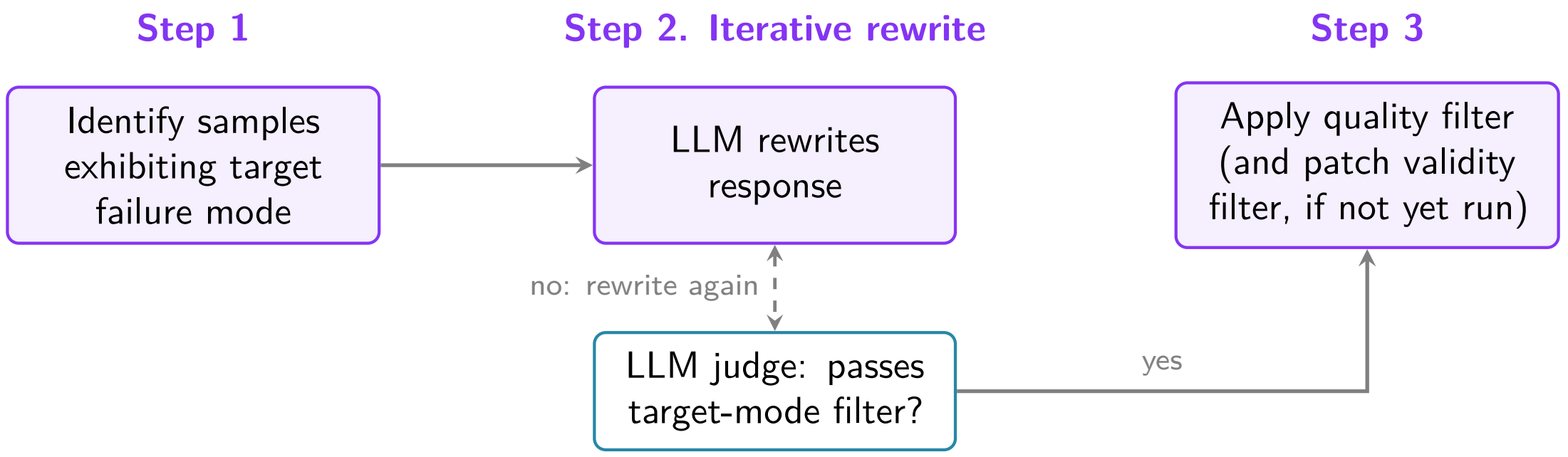

Both curation pipelines follow a similar procedure. First, LLM-based or rules-based
filters identify examples in outputs of previous model iterations that exhibit the target failure mode. An iterative LLM
feedback loop then rewrites the problematic responses.
The LLM judge returns rejected rewrites with a short description of the failure until they pass the target failure mode's filter.
We pass each rewritten example
through the quality filter and, if the example has not already undergone patch validity
filtering, the patch validity filter. Figure~\ref{fig:phase1-targeted-behavior-curation} illustrates this process.

\subsubsection{Data Mixture}
\label{sec:phase1-sft-data}
We construct the 17{,}454-sample SFT dataset from converted NES data,
bootstrapped multi-edit data, and repairs of model outputs that fail
patch validity checks.
No-edit samples account for 45\% of the dataset
(Table~\ref{tab:phase1-sft-data-sources}).
The mixture teaches the model behaviors that reinforcement learning cannot
easily discover from an uninformed initialization: the patch format, anchored
insertions, multi-patch responses, and proximity-first ordering.

\begin{table}[t]
  \caption{SFT dataset composition. Converted NES and bootstrapped data contribute equally, and repaired invalid patch data supplements them.}
  \label{tab:phase1-sft-data-sources}
  \centering
  \small
  \begin{tabular}{lrp{7.4cm}}
    \toprule
    Source & Samples & Description \\
    \midrule
    Converted NES data & 8{,}447 &
      The format converter expresses NES training samples as patches. \\
    Bootstrapped data & 8{,}447 &
      Bootstrapping transfers edits from recorded histories and retains samples with
      at least one jump of 5 or more lines. \\
    Invalid patch repairs & 560 &
      The repair pipeline corrects model outputs that fail a patch validity
      check and uses the corrections as ground truth labels. \\
    \midrule
    Total & 17{,}454 & 55\% edit samples and 45\% no-edit samples. \\
    \bottomrule
  \end{tabular}
\end{table}

We select the patch-count distribution rather than inherit it from the source
datasets.
For the bootstrapped and invalid-patch-repair data, we sweep arithmetic,
exponential, and log-decay profiles over 1 to 9 patches.
Across the resulting full data mixtures, we compare approximate edit to
no-edit ratios of 50/50, 55/45, and 65/35.
Increasing the multi-patch share raises the mean number of predicted patches
but also increases patch validity failures, so the selected distribution
balances these outcomes.
Table~\ref{tab:phase1-sft-patch-count-distribution} reports the resulting distribution among SFT samples that contain at least one edit.

\begin{table}[t]
  \caption{Patch-count distribution among SFT samples that contain at least one edit. Single-patch and multi-patch examples each account for roughly half of samples with edits.}
  \label{tab:phase1-sft-patch-count-distribution}
  \centering
  \small
  \begin{tabular}{rrrrrrrrrr}
    \toprule
    Patches & 1 & 2 & 3 & 4 & 5 & 6 & 7 & 8 & 9 \\
    \midrule
    \% of edit samples &
      48.30 & 29.70 & 12.89 & 5.63 & 2.35 & 1.04 & 0.05 & 0.03 & 0.01 \\
    \bottomrule
  \end{tabular}
\end{table}

\subsubsection{Training Recipe}
\label{sec:phase1-sft-recipe}
The primary goal of supervised fine-tuning is to teach the patch output format
and core edit behaviors.
Reinforcement learning then refines this initialization by reducing residual
failure modes and improving suggestion quality.

We use the same SFT recipe across runs.
Each run trains for 5 epochs with a learning rate of
\(8.1 \times 10^{-5}\) and a batch size of 32.
We evaluate checkpoints every 100 steps.
We linearly warm the learning rate over the first 5\% of training and then
decay it to 0.

\paragraph{Checkpoint Selection.}
We select an SFT checkpoint for reinforcement learning initialization by
applying three criteria in order.
First, the checkpoint must achieve acceptable scores on Simulation Tests,
Augmented HumanEval, and POE.
Second, it must place edits correctly in a set of insertion placement cases that we identify during internal testing in Phase~1; we later formalize these cases by creating the QMH benchmark (Section~\ref{sec:offline-benchmarks}).
Third, it must average approximately 1.3 patches per response and produce
approximately 130 invalid patches on a 15{,}000-sample diagnostic set.
The patch-count target balances follow-up edit frequency with opportunities for caching, while the
invalid-patch target limits the initial patch validity failures that
reinforcement learning must reduce.

\subsubsection{Ablations}
\label{sec:phase1-sft-ablations}

\paragraph{Multi-Patch Behavior and Target.}
Initial SFT models learn the patch format from multi-edit data but still
emit mostly one patch, leaving no follow-up edits for caching.
Adding single-line completion examples does not help, while iterating the
edit-shape distribution establishes multi-patch behavior, indicating that data
distribution rather than format exposure is the limiting factor.

Across 11 SFT checkpoints on the POE benchmark, mean edits range from 1.32 to 2.33 and
the patch validity error rate among editing responses ranges from 2.36\% to
12.23\% (Figure~\ref{fig:phase1-sft-patch-count-validity}).
Mean patch count highlights the tradeoff between opportunities for caching and the frequency of follow-up edits.
Lower mean counts limit opportunities for caching follow-up patches; higher mean counts can offer more caching opportunities, but also risk unnecessary edits.
We also observe higher invalid patch rates as the mean patch count increases; we therefore target approximately 1.3 patches per response to balance caching opportunities against patch validity failures. Both the SFT and RL models achieve this target on POE.

\begin{center}
  \begin{minipage}{\linewidth}
  \centering
  \resizebox{0.88\linewidth}{!}{\begin{tikzpicture}[
  font=\sffamily\small,
  point/.style={circle, fill=gray!70, inner sep=2.4pt},
  highlight/.style={diamond, draw=ghpurple, fill=ghpurple!25, line width=0.8pt, inner sep=3.2pt},
  selected/.style={circle, draw=ghgreen!70!black, fill=ghgreen!25, line width=0.8pt, inner sep=3.2pt}
]
  \foreach \y/\label in {0/2,0.91/4,1.82/6,2.73/8,3.64/10,4.55/12} {
    \draw[gray!18] (0,\y) -- (8.8,\y);
    \draw[gray] (-0.08,\y) -- (0,\y);
    \node[anchor=east, text=gray!80!black] at (-0.14,\y) {\label};
  }
  \foreach \x/\label in {0/1.3,1.6/1.5,3.2/1.7,4.8/1.9,6.4/2.1,8.0/2.3} {
    \draw[gray] (\x,0) -- (\x,-0.08);
    \node[anchor=north, text=gray!80!black] at (\x,-0.14) {\label};
  }

  \draw[->, line width=0.8pt, gray!80!black] (0,0) -- (9.0,0);
  \draw[->, line width=0.8pt, gray!80!black] (0,0) -- (0,5.35);
  \node[anchor=north] at (4.4,-0.72) {Mean patches per response};
  \node[rotate=90, anchor=south] at (-1.0,2.5) {Patch validity error rate (\%)};

  \node[selected] at (9.65,3.45) {};
  \node[anchor=west, font=\sffamily\scriptsize] at (9.90,3.45) {Selected SFT checkpoint};
  \node[point] at (9.65,2.65) {};
  \node[anchor=west, font=\sffamily\scriptsize] at (9.90,2.65) {Other SFT candidates};
  \node[highlight] at (9.65,1.85) {};
  \node[anchor=west, align=left, font=\sffamily\scriptsize] at (9.90,1.85)
    {2-patch-minimum\\ablation};

  \node[point] at (4.56,2.72) {};
  \node[point] at (3.84,1.89) {};
  \node[point] at (3.52,2.24) {};
  \node[point] at (3.84,2.38) {};
  \node[point] at (2.08,1.27) {};
  \node[point] at (1.60,1.23) {};
  \node[point] at (4.16,2.15) {};
  \node[point] at (3.76,2.20) {};
  \node[point] at (3.44,2.07) {};
  \node[highlight] at (8.24,4.65) {};
  \node[selected] at (0.16,0.17) {};
\end{tikzpicture}}
  
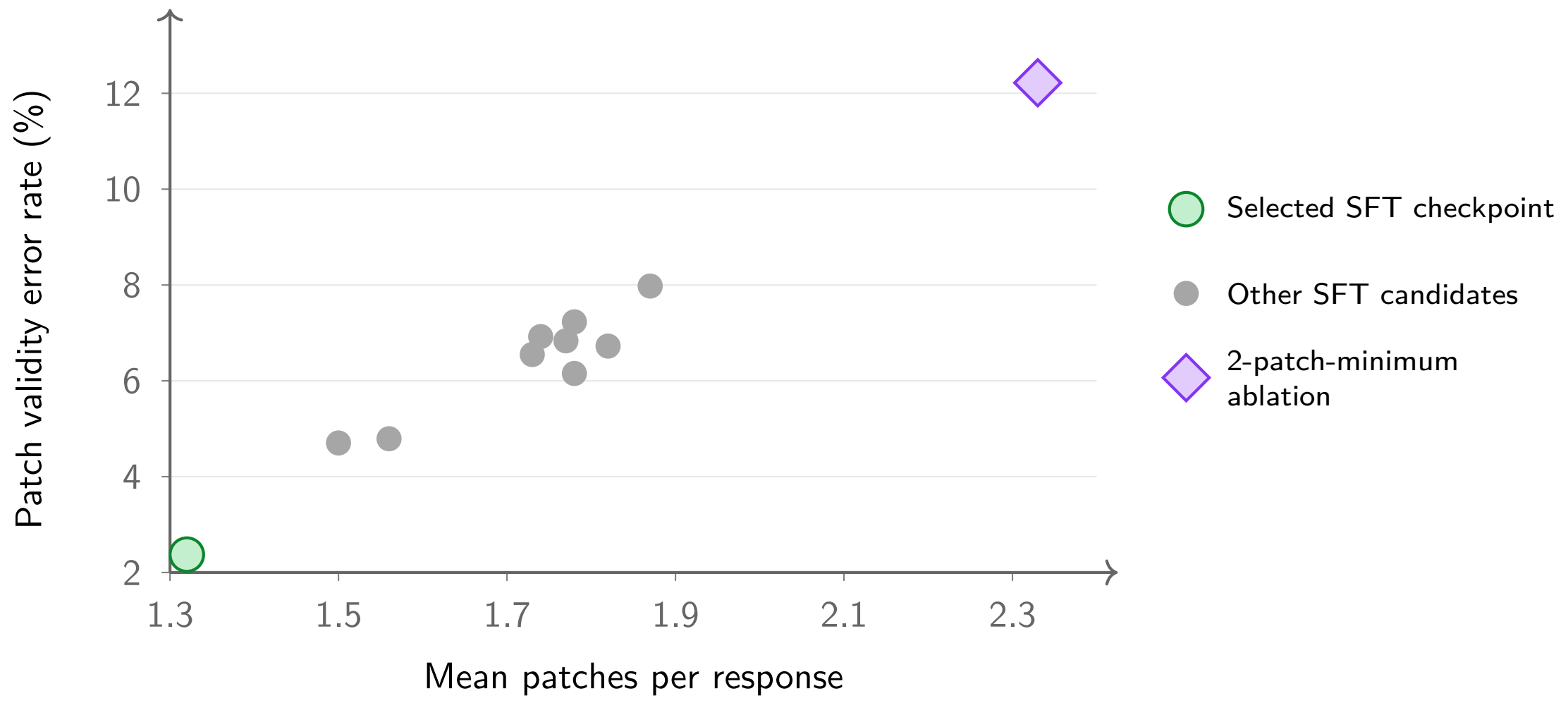
\captionof{figure}{Mean patch counts and patch validity error rates for SFT candidates on the POE benchmark.
  The 2-patch-minimum ablation restricts training to examples whose ground truth labels contain 2 or more patches. Higher mean patch count correlates with higher invalid patch rate.}
  \label{fig:phase1-sft-patch-count-validity}
  \end{minipage}
\end{center}

\paragraph{Rules-Based Data Filtering.}
Rules-based filters alone do not produce a set of workspace examples that
improves training results.
Recorded outcomes provide useful signals, but they do not establish that each
candidate forms a coherent, high-quality training label.
The rules admit false positives that satisfy the heuristics despite weak edit
quality and exclude false negatives that contain useful supervision.
We therefore use an LLM judge to assess each candidate's coherence and quality
before adding it to the training set.

\subsection{Reinforcement Learning}

\subsubsection{Data Mixture}
\label{sec:phase1-rl-data}
We construct a data mixture for the Phase~1 RL stage by
combining examples from multiple sources to balance scenario coverage and quality.
Initial models that use only the original NES data perform on par with the standalone NES models in offline benchmarks, but produce almost exclusively nearby NES edits, with few multi-patch long-distance edits. The same samples sometimes elicit different behavior when we convert them to the diff-patch format, including producing more incremental edits. Below, we discuss the data mixture used alongside the SFT data in Section~\ref{sec:phase1-sft-data} to address the scarcity of multi-patch long-distance edits. In Section~\ref{sec:phase2-accepted-retained-character-data}, we describe the Phase~2 measures that address suggestion length and completeness. We also use early online A/B experiments with these models to improve client integration before evaluating later candidates.

We perform RL in two stages using the on-policy algorithm we describe in Section~\ref{sec:phase1-rl-algorithm}. First, we train on top of the SFT model we discuss in Section~\ref{sec:phase1-sft-recipe}.
We evaluate this model through the fourth stage of the model selection process
illustrated in Figure~\ref{fig:phase1-phase2-model-selection-process}, A/B experiment candidates.
The A/B experiment results show no statistically significant regression in any guardrail metrics,
but continued offline dogfooding reveals the insertion pattern propagation failure mode described in Section~\ref{sec:phase1-targeted-behavior-curation}.
As a result, we add data curated with the pipeline described in the same section to the mixture to target this behavior, and then perform another round of RL on the resulting model.

Tables~\ref{tab:phase1-rl-data-mixture} and~\ref{tab:phase1-rl-data-edit-count-summary} report the dataset-level and
aggregate edit-count distributions. We note a few key characteristics of the data below.

\paragraph{Scenario Diversity.}
As Table~\ref{tab:phase1-rl-data-mixture} shows, we aim to represent a wide variety of
scenarios and edit types across the RL data. We filter all data for quality and
repair it for patch validity. The mixture includes scenarios from the original
NES data and new dedicated datasets that we introduce during 2-in-1 model training,
such as bootstrapped multi-edit data, rewritten invalid patch data, and rewritten insertion pattern propagation data.

\paragraph{Overall Edit Count Distribution.}
Because the SFT model already produces an appropriate edit-count distribution,
the RL mixture need not match the SFT distribution exactly. Instead, the primary goal of the mixture is to
represent the full spectrum of desirable and undesirable edits. For instance, as Table~\ref{tab:phase1-rl-data-mixture} shows, no-edit samples account for
20.9\% of the RL mixture, compared with 45\% of the SFT data.
Single-edit samples account for 48.8\% of the RL mixture, compared with 26.6\% of the SFT data.
This difference reflects the distinct roles of no-edit and single-edit samples in RL and SFT.
In RL, we use separate graders for rollouts containing an edit and those containing no edit.
We find that allowing the grader to drive the decision boundary for when an edit is appropriate,
combined with carefully selected samples that complement this decision boundary, results in
better model learning of user intent and when to suggest an edit or abstain.
For single-edit samples, SFT uses the ground truth response as an explicit label,
while RL uses it as a reference prefix for an acceptable response, as we further discuss in Section~\ref{sec:phase1-grader}. This allows the
model to learn desirable multi-edit behaviors and generate them for single-edit samples without
incurring a penalty. We discuss this further below.

\paragraph{Edit Count Distribution of Newly Added 2-in-1 Data versus NES Data.}
As Table~\ref{tab:phase1-rl-data-edit-count-summary} shows, 50.3\% of the mixture is original NES data and 49.7\% is newly added during 2-in-1 training.
The distribution of edit counts differs between the original NES data and the newly generated data, reflecting
the goal of representing multi-edit scenarios in the new set of data.
The original NES data contributes 34.0\% of the total mixture in single-edit data, compared to a 14.8\% contribution from the new data.
Conversely, the original data contributes 3.4\% of the total mixture in multi-edit data, compared to the 27.0\% contribution from the new data.

\paragraph{Upweighting Rare Cases.}
For some error types, such as patch validity and insertion pattern propagation, negative examples are scarce. After quality filtering, the curated set contains thousands of positive examples, but only a few hundred corrected negative examples that originally exhibit errors (560 and 355 samples, respectively). We handle these two error types differently across training stages. For patch validity, we include the 560 repaired examples in both SFT (Section~\ref{sec:phase1-sft-data}) and RL to establish and reinforce the required diff patch syntax. For insertion pattern propagation, initial experiments show that demonstrating rewritten examples in SFT alone does not noticeably alter model behavior. Because reinforcement learning is particularly effective for unlearning specific problematic behaviors from prior models and the insertion data is particularly scarce, we reserve the 355 insertion pattern propagation examples exclusively for RL. To overcome their scarcity during RL, we introduce two complementary interventions: (1) we upsample the rewritten insertion pattern propagation examples by a factor of 5, from 355 to 1{,}775 samples, as Table~\ref{tab:phase1-rl-data-mixture} shows, and (2) we use RL grader boosting to target this issue with a dedicated auxiliary grader, as we detail in Section~\ref{sec:phase1-grader}.

\begin{table}[!htbp]
  \caption{Phase~1 RL data by scenario and edit count. The mixture combines diverse edit scenarios with targeted repairs of known failure modes.}
  \label{tab:phase1-rl-data-mixture}
  \centering
  \small
  \setlength{\tabcolsep}{3pt}
  \begin{tabular}{@{}p{8cm}rrr|r@{}}
    \toprule
    Category & 0 edits & 1 edit & 2+ edits & Total \\
    \midrule
    NES-converted no-edit & 4{,}832 & 0 & 0 & 4{,}832 \\
    NES-converted single-edit & 0 & 6{,}865 & 0 & 6{,}865 \\
    NES-converted multi-edit & 0 & 0 & 745 & 745 \\
    Small, exact edits & 132 & 1{,}200 & 167 & 1{,}499 \\
    Partial edits & 19 & 2{,}000 & 471 & 2{,}490 \\
    Highly desired edits & 369 & 14{,}485 & 1{,}576 & 16{,}430 \\
    Primarily single-patch ghost text at cursor & 0 & 4{,}698 & 8 & 4{,}706 \\
    Insufficient intent or ambiguous & 2{,}347 & 0 & 0 & 2{,}347 \\
    No edit desired & 5{,}705 & 0 & 0 & 5{,}705 \\
    Bootstrapped multi-edit & 0 & 1{,}227 & 16{,}388 & 17{,}615 \\
    Rewritten invalid patches & 0 & 341 & 219 & 560 \\
    Rewritten incorrect insertion pattern propagation & 345 & 1{,}335* & 440* & 2{,}120 \\
    \midrule
    \textbf{Total} & \textbf{13{,}749} & \textbf{32{,}151} & \textbf{20{,}014} & \textbf{65{,}914} \\
    \multicolumn{5}{@{}p{\linewidth}@{}}{\footnotesize * Counts include fivefold upsampling of insertion pattern propagation negative examples.} \\
    \bottomrule
  \end{tabular}
\end{table}

\begin{table}[!htbp]
  \caption{Phase~1 RL data by source and edit count, with percentages relative to the full mixture. The new data contributes roughly half of the samples but provides the majority of multi-edit examples.}
  \label{tab:phase1-rl-data-edit-count-summary}
  \centering
  \small
  \setlength{\tabcolsep}{2.5pt}
  \begin{tabular}{@{}p{3.1cm}rr|rrrrrr@{}}
    \toprule
    Source & \multicolumn{2}{c|}{All samples} & \multicolumn{2}{c}{0 edits} & \multicolumn{2}{c}{1 edit} & \multicolumn{2}{c}{2+ edits} \\
    \cmidrule(l){2-3}\cmidrule(lr){4-5}\cmidrule(lr){6-7}\cmidrule(lr){8-9}
    & Count & \% & Count & \% & Count & \% & Count & \% \\
    \midrule
    Original NES data & 33{,}177 & 50.3\% & 8{,}572 & 13.0\% & 22{,}383 & 34.0\% & 2{,}222 & 3.4\% \\
    New data & 32{,}737 & 49.7\% & 5{,}177 & 7.9\% & 9{,}768 & 14.8\% & 17{,}792 & 27.0\% \\
    \midrule
    \textbf{Total} & \textbf{65{,}914} & \textbf{100.0\%} & \textbf{13{,}749} & \textbf{20.9\%} & \textbf{32{,}151} & \textbf{48.8\%} & \textbf{20{,}014} & \textbf{30.4\%} \\
    \bottomrule
  \end{tabular}
\end{table}

\subsubsection{RL Algorithm}
\label{sec:phase1-rl-algorithm}
We perform on-policy reinforcement learning for the 2-in-1 model, drawing task examples from the data mixture we describe in Section~\ref{sec:phase1-rl-data}. At each optimization step, we sample a batch of $B$ task instances. For each task $x_i$ in the batch, the policy generates a set of candidate completions (rollouts), $y_{ij} \sim \pi_\theta(\cdot \mid x_i)$, which the grader evaluates.

Each response receives a binary reward $R_{ij} \in \{0, 1\}$ denoting whether it passes the validation checks and criteria. Following prior work \citep{bengio2009curriculum,kumar2010selfpaced,matiisen2017teacherstudent,florensa2017reverse,florensa2018goalgeneration,jiang2021prioritizedlevel,zelikman2022star,lightman2023verify,yu2025dapo,bae2026difficultyfiltering}, we filter tasks by their empirical pass rate to avoid uninformative gradient updates and high variance from tasks that are either intractable or already mastered by the current policy:
\[
\hat{p}_i = \frac{1}{N} \sum_{j=1}^N R_{ij},
\]
where $N$ is the number of evaluated rollouts per instance. We exclude tasks with pass rates outside a configured interval $[\tau_{\min}, \tau_{\max}]$ from the policy update.

For the retained tasks, we follow REINFORCE and use a policy gradient update \citep{williams1992reinforce,sutton1999policygradient} with a task-level baseline to reduce variance:
\[
\nabla_\theta \mathcal{J}(\theta) = \mathbb{E} \left[ \sum_{j} \nabla_\theta \log \pi_\theta(y_{ij} \mid x_i) \left( R_{ij} - b_i \right) \right],
\]
where $b_i$ denotes a baseline computed across the rollouts for task $x_i$. This use of same-task rollouts to construct a baseline is shared with REINFORCE leave-one-out (RLOO) \citep{ahmadian2024backtobasics}, which specifically averages the rewards of the other rollouts, excluding the rollout being scored. The gradients are aggregated across all active instances in the batch, and model parameters are updated using the learning rate $\eta$.

In our 2-in-1 RL experiments, we train with a batch size of $B=64$, evaluate $N=8$ rollouts per task, and use a learning rate of $\eta=1\times10^{-5}$. We set the pass-rate filtering thresholds separately for each dataset category, with lower bounds falling within $0.20 \le \tau_{\min} \le 0.50$ and upper bounds falling within $0.90 \le \tau_{\max} \le 0.95$.

\subsubsection{Grader}
\label{sec:phase1-grader}
For the $j$-th rollout response $y_{ij}$ that the policy generates for task instance $i$,
the grader returns a binary outcome $R_{ij}\in\{0,1\}$. Here, $R_{ij}=1$ denotes a passing response 
and $R_{ij}=0$ a failing response. We write $y_{ij}\sim\pi_\theta(\cdot\mid x_i)$, where $x_i$ 
is the task input for instance $i$ and $\pi_\theta$ is the current policy.

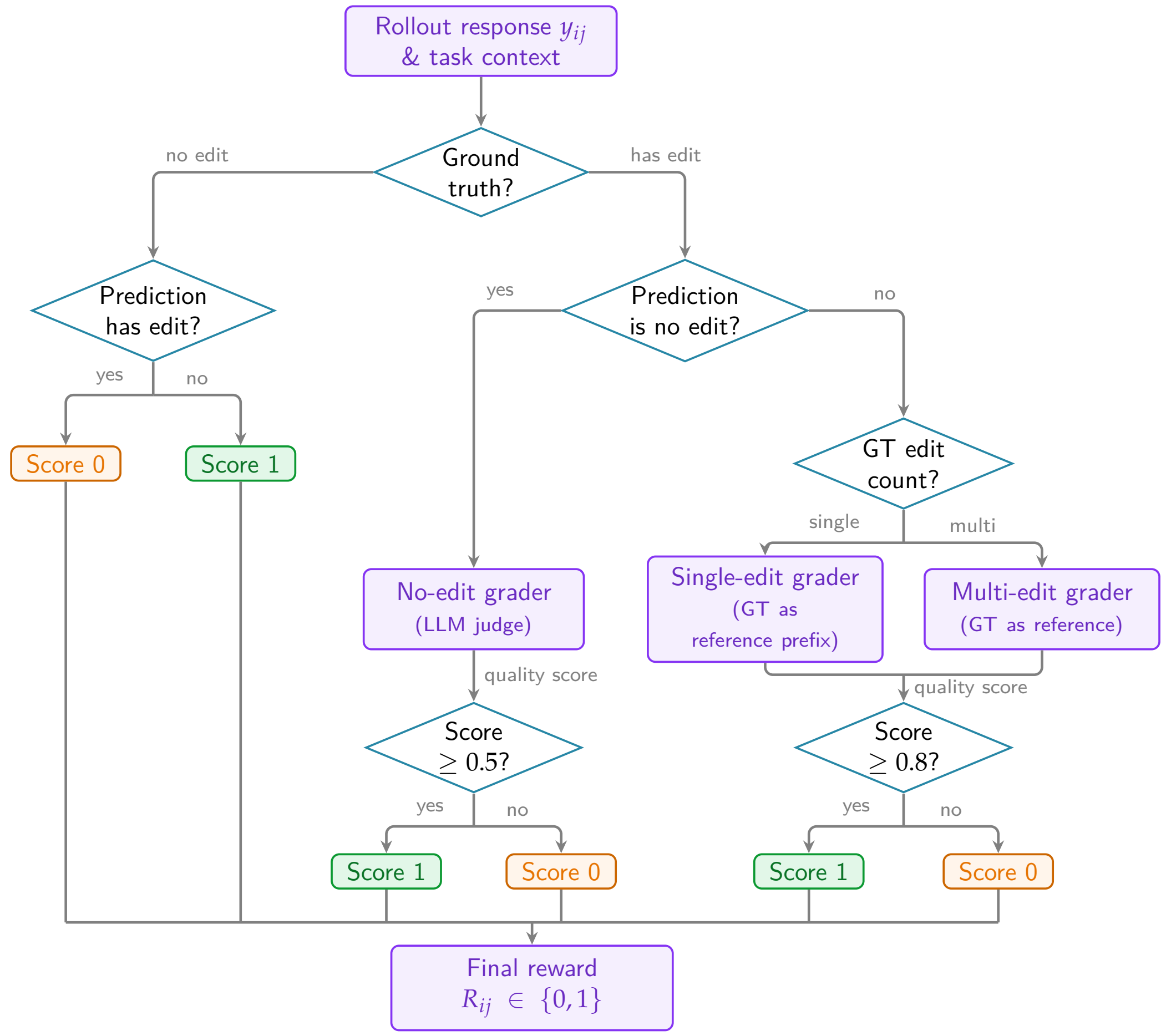
\begin{figure}[!ht]
  \centering
  \resizebox{0.82\linewidth}{!}{\begin{tikzpicture}[
  font={\sffamily\small\hyphenpenalty=10000\exhyphenpenalty=10000\relax},
  box/.style={
    draw=ghpurple,
    fill=ghpurple!8,
    text=ghpurple!80!black,
    rounded corners=3pt,
    line width=0.8pt,
    align=center,
    inner xsep=6pt,
    inner ysep=4pt
  },
  decision/.style={
    diamond,
    aspect=2.4,
    draw=cyan!60!black,
    fill=white,
    text=black,
    line width=0.8pt,
    align=center,
    inner sep=1pt,
    minimum width=2.4cm,
    minimum height=0.9cm
  },
  leafzero/.style={
    draw=orange!80!black,
    fill=orange!8,
    text=orange!90!black,
    rounded corners=3pt,
    line width=0.8pt,
    align=center,
    minimum width=1.5cm,
    inner xsep=4pt,
    inner ysep=3pt
  },
  leafone/.style={
    draw=ghgreen!80!black,
    fill=ghgreen!12,
    text=ghgreen!60!black,
    rounded corners=3pt,
    line width=0.8pt,
    align=center,
    minimum width=1.5cm,
    inner xsep=4pt,
    inner ysep=3pt
  },
  reward/.style={
    draw=ghpurple,
    fill=ghpurple!8,
    text=ghpurple!80!black,
    rounded corners=3pt,
    line width=0.8pt,
    align=center,
    inner xsep=8pt,
    inner ysep=5pt
  },
  flow/.style={->, draw=gray, line width=1pt, rounded corners=3pt, >={Stealth[length=5pt, width=4.5pt]}},
  trunk/.style={draw=gray, line width=1pt, rounded corners=3pt},
  note/.style={font=\sffamily\scriptsize, text=gray}
]

  \node[box, text width=3.3cm] (input) at (6.5, 12.0)
    {Rollout response $y_{ij}$\\\& task context};

  \node[decision] (gt-check) at (6.5, 10.2)
    {Ground\\truth?};

  \node[decision] (pred-edit-check) at (2.0, 8.3)
    {Prediction\\has edit?};

  \node[decision] (pred-noedit-check) at (9.3, 8.3)
    {Prediction\\is no edit?};

  \node[leafzero] (leaf-0-gtno) at (0.8, 6.2) {Score 0};
  \node[leafone] (leaf-1-gtno) at (3.2, 6.2) {Score 1};

  \node[decision] (gt-count) at (12.3, 6.2)
    {GT edit\\count?};

  \node[box, text width=2.6cm, minimum height=1.1cm] (noedit-grader) at (6.4, 4.2)
    {No-edit grader\\{\scriptsize(LLM judge)}};

  \node[box, text width=2.8cm, minimum height=1.1cm] (single-grader) at (10.4, 4.2)
    {Single-edit grader\\{\scriptsize(GT as reference prefix)}};
  \node[box, text width=2.8cm, minimum height=1.1cm] (multi-grader) at (14.2, 4.2)
    {Multi-edit grader\\{\scriptsize(GT as reference)}};

  \node[decision] (threshold-noedit) at (6.4, 2.3)
    {Score\\$\ge 0.5$?};

  \node[decision] (threshold) at (12.3, 2.3)
    {Score\\$\ge 0.8$?};

  \node[leafone] (leaf-1-noedit) at (5.2, 0.6) {Score 1};
  \node[leafzero] (leaf-0-noedit) at (7.6, 0.6) {Score 0};

  \node[leafone] (leaf-1-thresh) at (11.0, 0.6) {Score 1};
  \node[leafzero] (leaf-0-thresh) at (13.6, 0.6) {Score 0};

  \node[reward, text width=3.3cm] (reward) at (7.2, -1.0)
    {Final reward $R_{ij} \in \{0, 1\}$};

  \draw[flow] (input.south) -- (gt-check.north);

  \draw[flow] (gt-check.west) -| node[pos=0.4, above, note] {no edit} (pred-edit-check.north);
  \draw[flow] (gt-check.east) -| node[pos=0.4, above, note] {has edit} (pred-noedit-check.north);

  \path (pred-edit-check.south) -- ++(0,-0.45) coordinate (fork-pred-edit);
  \draw[trunk] (pred-edit-check.south) -- (fork-pred-edit);
  \draw[flow] (fork-pred-edit) -- node[above, note] {yes} (leaf-0-gtno.north |- fork-pred-edit) -- (leaf-0-gtno.north);
  \draw[flow] (fork-pred-edit) -- node[above, note] {no} (leaf-1-gtno.north |- fork-pred-edit) -- (leaf-1-gtno.north);

  \draw[flow] (pred-noedit-check.west) -| node[pos=0.35, above, note] {yes} (noedit-grader.north);
  \draw[flow] (pred-noedit-check.east) -| node[pos=0.4, above, note] {no} (gt-count.north);

  \draw[flow] (noedit-grader.south) -- node[midway, right, note] {quality score} (threshold-noedit.north);

  \path (threshold-noedit.south) -- ++(0,-0.45) coordinate (fork-thresh-noedit);
  \draw[trunk] (threshold-noedit.south) -- (fork-thresh-noedit);
  \draw[flow] (fork-thresh-noedit) -- node[above, note] {yes} (leaf-1-noedit.north |- fork-thresh-noedit) -- (leaf-1-noedit.north);
  \draw[flow] (fork-thresh-noedit) -- node[above, note] {no} (leaf-0-noedit.north |- fork-thresh-noedit) -- (leaf-0-noedit.north);

  \path (gt-count.south) -- ++(0,-0.45) coordinate (fork-gt-count);
  \draw[trunk] (gt-count.south) -- (fork-gt-count);
  \draw[flow] (fork-gt-count) -- node[above, note] {single} (single-grader.north |- fork-gt-count) -- (single-grader.north);
  \draw[flow] (fork-gt-count) -- node[above, note] {multi} (multi-grader.north |- fork-gt-count) -- (multi-grader.north);

  \draw[trunk] (single-grader.south) -- (10.4, 3.3) -- (12.3, 3.3);
  \draw[trunk] (multi-grader.south) -- (14.2, 3.3) -- (12.3, 3.3);
  \draw[flow] (12.3, 3.3) -- node[midway, right, note] {quality score} (threshold.north);

  \path (threshold.south) -- ++(0,-0.45) coordinate (fork-thresh);
  \draw[trunk] (threshold.south) -- (fork-thresh);
  \draw[flow] (fork-thresh) -- node[above, note] {yes} (leaf-1-thresh.north |- fork-thresh) -- (leaf-1-thresh.north);
  \draw[flow] (fork-thresh) -- node[above, note] {no} (leaf-0-thresh.north |- fork-thresh) -- (leaf-0-thresh.north);

  \draw[trunk] (leaf-0-gtno.south) -- (0.8, -0.1);
  \draw[trunk] (leaf-1-gtno.south) -- (3.2, -0.1);
  \draw[trunk] (leaf-1-noedit.south) -- (5.2, -0.1);
  \draw[trunk] (leaf-0-noedit.south) -- (7.6, -0.1);
  \draw[trunk] (leaf-1-thresh.south) -- (11.0, -0.1);
  \draw[trunk] (leaf-0-thresh.south) -- (13.6, -0.1);

  \draw[trunk] (0.8, -0.1) -- (13.6, -0.1);

  \draw[flow] (7.2, -0.1) -- (reward.north);

\end{tikzpicture}}
  \caption{Core logic behind the reinforcement learning custom grader, expressed as a decision tree focused on edit type. Given a rollout response $y_{ij}$ and task ground truth, the grader branches on whether the ground truth requires no edit or contains edits, routing to rules-based checks, the no-edit LLM judge (thresholded at 0.5), or single- and multi-edit LLM judges (thresholded at 0.8) to determine the final binary reward $R_{ij}$. We omit additional granular rules-based checks such as patch validity mentioned in Section~\ref{sec:phase1-grader} for brevity.}
  \label{fig:phase1-grader-decision-tree}
\end{figure}

We develop a custom grader that combines rules-based patch validity checks with an LLM judge. We also use both components in the data quality filters detailed in Section~\ref{sec:phase1-data-quality-filtering}.
Suggestion quality covers criteria for desirable edit behavior along with patch amounts and ordering.
Figure~\ref{fig:phase1-grader-decision-tree} illustrates the decision tree logic of this custom grader.
The evaluation first branches on whether the ground truth requires no edit or contains edits.
If the ground truth requires no edit, the grader checks the response for edits: predicting an edit receives a score of 0, whereas predicting no edit receives a score of 1.
If the ground truth contains an edit, the evaluation branches on whether the rollout prediction abstains or suggests an edit.
When the prediction is no edit, our no-edit grader variant evaluates whether abstention satisfies the decision boundary articulated in Section~\ref{sec:phase1-data-quality-filtering}.
The judge assigns an LLM quality score between 0 and 1, and an inclusive threshold of 0.5 assigns a passing score of 1, while lower scores receive 0.

When the prediction contains an edit, the evaluation routes by ground truth edit count to either our single-edit or multi-edit grader variant.
For instances where the ground truth has a single edit, the single-edit grader provides the ground truth as an example of a \textit{prefix} of a desirable response rather than enforcing an exact match or identical patch count in order to encourage generalizable taste in quality and patch count.
For instances where the ground truth has multiple edits, the multi-edit grader provides the ground truth as an example of a desirable response, rather than enforcing an exact match, for the same reason.
The single- and multi-edit graders assign responses an LLM quality score between 0 and 1 inclusive in increments of 0.1, and an inclusive threshold of 0.8 assigns a passing score of 1, while lower scores receive 0.
All scoring leaves across these branches define the final binary reward $R_{ij} \in \{0, 1\}$.

We primarily use rules-based checks to assign a score of 0, while the LLM judge determines whether a response merits a score of 1.
Thus, we also apply additional rules-based checks to assign a score of 0, such as patch validity, code compilation (requiring that the resulting code after patch application compiles successfully if the code compiles successfully prior to the edit), patch length constraints, and scenario-specific requirements.
We also experiment with raising the LLM judge score to the power of 4, which sharpens the distinction between higher- and lower-quality edits; in these cases, the inclusive threshold becomes 0.35. Our final grader does not use exponentiation.

One challenge in developing a centralized LLM judge for the custom grader is balancing multiple objectives.
When we add objectives, the grader may not satisfy all of them consistently, which can create blind spots in its judgments.
We address this challenge with \textit{grader boosting}: auxiliary graders strengthen the residual signal while the main grader focuses on criteria for the most common scenarios.

\paragraph{Grader Boosting.}
We name grader boosting after the statistical concept of gradient boosting \citep{friedman2001greedy}.
We generalize it as a directed acyclic graph of main and auxiliary graders. The graph's topological execution order dictates grader orchestration, including each grader's invocation conditions and potential inputs. Each grader can take input from any node that precedes it topologically.
Main graders typically provide canonical or general assessments, while auxiliary graders focus on specific aspects of response quality that the main grader may underrepresent or overlook. The final score assembly combines these auxiliary assessments with the main grader's assessment.

In the case of the 2-in-1 RL grader, we observe upon qualitative inspection of rollouts that the main grader does not reliably flag insertion pattern propagation errors despite the targeted data. We therefore develop an auxiliary grader specifically for this behavior.
We hypothesize that adding insertion pattern guidance to the main grader, which already covers 14 quality criteria, may reduce its judgment quality on other aspects of suggestion quality.
Because this pattern has a narrow scope, we gate the auxiliary grader on relevant instances to focus its feedback on these cases.
The reported results do not isolate the auxiliary grader's contribution from that of upsampling the insertion pattern propagation examples.

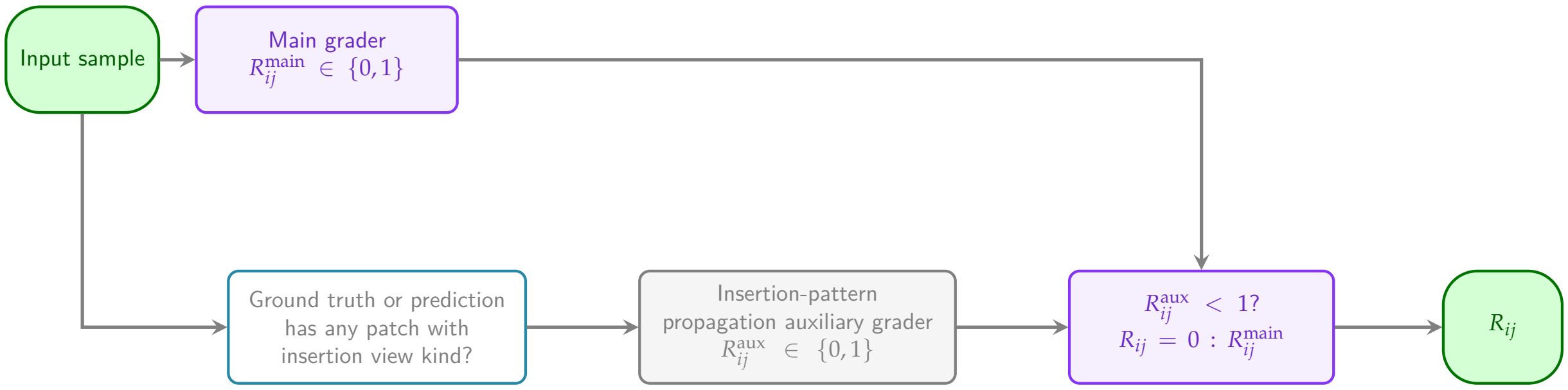
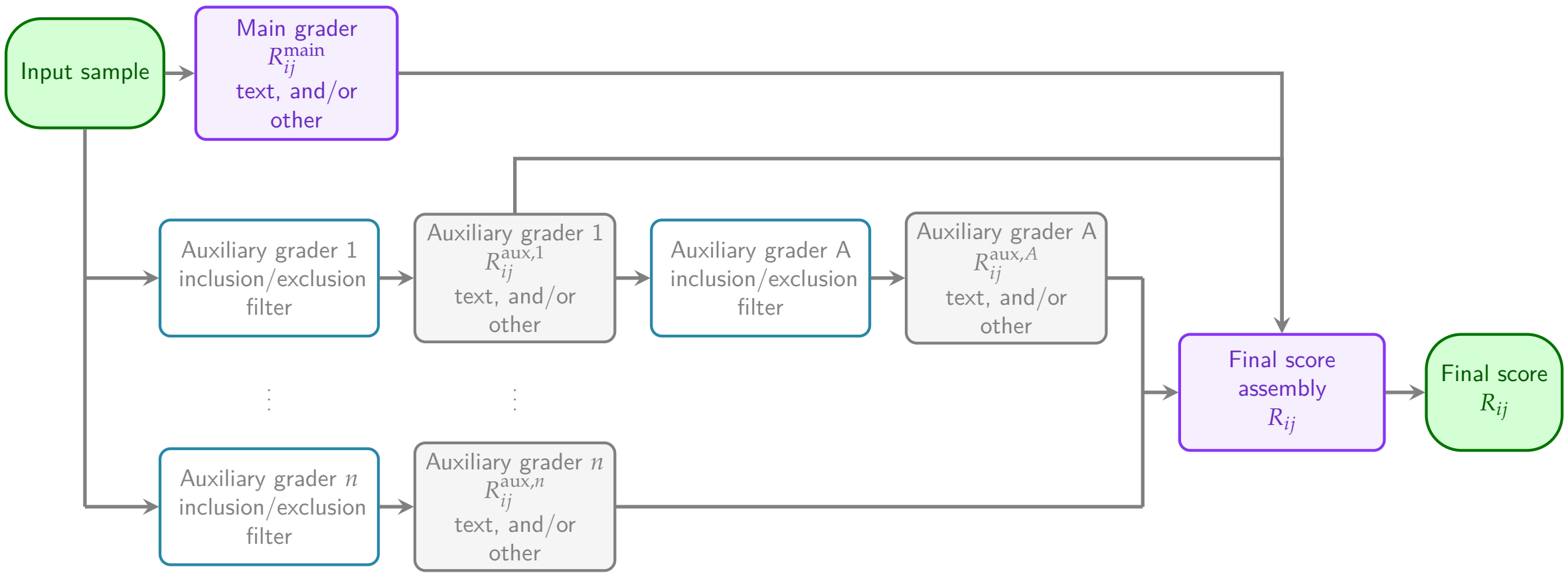
\begin{figure}[!ht]
  \centering
  \captionsetup[subfigure]{font=small,justification=centering}
  \begin{subfigure}[t]{\linewidth}
  \centering
  \begin{tikzpicture}[
    scale=0.68, transform shape,
    font={\sffamily\small\hyphenpenalty=10000\exhyphenpenalty=10000\relax},
    sample/.style={draw=green!45!black, fill=green!17, text=green!35!black, rounded corners=10pt, line width=0.8pt, align=center, inner xsep=6pt, inner ysep=5pt, minimum width=1.8cm, minimum height=0.85cm},
    box/.style={draw=ghpurple, fill=ghpurple!8, text=ghpurple!80!black, rounded corners=3pt, line width=0.8pt, align=center, inner xsep=5pt, inner ysep=5pt},
    filter/.style={draw=cyan!60!black, fill=white, text=gray, rounded corners=3pt, line width=0.8pt, align=center, inner sep=4pt, text width=4.2cm, minimum width=4.5cm, minimum height=1.7cm},
    auxiliary/.style={draw=gray, fill=gray!8, text=gray, rounded corners=3pt, line width=0.8pt, align=center, inner sep=4pt},
    assembly/.style={draw=ghpurple, fill=ghpurple!8, text=ghpurple!80!black, rounded corners=3pt, line width=0.8pt, align=center, inner sep=4pt, text width=3.4cm, minimum width=4cm, minimum height=1.7cm},
    flow/.style={->, draw=gray, line width=1pt, >=stealth}
  ]
    \node[sample, minimum height=1.6cm] (sample) at (1.3,2.7) {Input sample};
    \node[box, text width=3.6cm, minimum width=3.8cm, minimum height=1.6cm] (main) at (5,2.7)
      {Main grader\\$R_{ij}^{\mathrm{main}}\in\{0,1\}$};
    \node[filter] (gate) at (5.76,-1.35)
      {Ground truth or prediction has any patch with insertion view kind?};
    \node[auxiliary, text width=4.5cm, minimum height=1.7cm] (insertion) at (12.125,-1.35)
      {Insertion-pattern\\propagation auxiliary grader\\$R_{ij}^{\mathrm{aux}}\in\{0,1\}$};
    \node[assembly] (assembly) at (18.24,-1.35)
      {$R_{ij}^{\mathrm{aux}}<1$?\\$R_{ij}=0:R_{ij}^{\mathrm{main}}$};
    \node[sample, minimum width=1.8cm, minimum height=1.7cm] (output) at (22.8,-1.35) {$R_{ij}$};

    \draw[flow] (sample.east) -- (main.west);
    \draw[flow] (sample.south) -- (1.3,-1.35) -- (gate.west);
    \draw[flow] (gate.east) -- (insertion.west);
    \draw[flow] (main.east) -| (assembly.north);
    \draw[flow] (insertion.east) -- (assembly.west);
    \draw[flow] (assembly.east) -- (output.west);
  \end{tikzpicture}
  \caption{Insertion-pattern propagation example.}
  \label{fig:phase1-grader-boosting-insertion}
\end{subfigure}

\medskip

\begin{subfigure}[t]{\linewidth}
  \centering
  \begin{tikzpicture}[
    scale=0.68, transform shape,
    font={\sffamily\small\hyphenpenalty=10000\exhyphenpenalty=10000\relax},
    sample/.style={draw=green!45!black, fill=green!17, text=green!35!black, rounded corners=10pt, line width=0.8pt, align=center, inner xsep=6pt, inner ysep=5pt, minimum width=1.8cm, minimum height=0.85cm},
    box/.style={draw=ghpurple, fill=ghpurple!8, text=ghpurple!80!black, rounded corners=3pt, line width=0.8pt, align=center, inner xsep=5pt, inner ysep=5pt},
    filter/.style={draw=cyan!60!black, fill=white, text=gray, rounded corners=3pt, line width=0.8pt, align=center, inner sep=4pt, text width=2.65cm, minimum width=3.2cm, minimum height=1.7cm},
    auxiliary/.style={draw=gray, fill=gray!8, text=gray, rounded corners=3pt, line width=0.8pt, align=center, inner sep=4pt},
    assembly/.style={draw=ghpurple, fill=ghpurple!8, text=ghpurple!80!black, rounded corners=3pt, line width=0.8pt, align=center, inner sep=4pt, text width=2cm, minimum width=3cm, minimum height=1.7cm},
    flow/.style={->, draw=gray, line width=1pt, >=stealth},
    trunk/.style={draw=gray, line width=1pt},
    note/.style={font=\sffamily\scriptsize, text=gray}
  ]
    \node[sample, minimum height=1.6cm] (sample) at (0.9,2.25) {Input sample};
    \node[box, text width=2.6cm, minimum height=1.6cm] (main) at (4,2.25)
      {Main grader\\$R_{ij}^{\mathrm{main}}$\\text, and/or other};
    \node[filter] (gate-a) at (10.8,-0.75)
      {Auxiliary grader A\\inclusion/exclusion filter};
    \node[filter] (gate-one) at (3.6,-0.75)
      {Auxiliary grader 1\\inclusion/exclusion filter};
    \node[filter] (gate-n) at (3.6,-4.1)
      {Auxiliary grader $n$\\inclusion/exclusion filter};
    \node[auxiliary, text width=2.65cm, minimum height=1.7cm] (aux-a) at (14.4,-0.75)
      {Auxiliary grader A\\$R_{ij}^{\mathrm{aux},A}$\\text, and/or other};
    \node[auxiliary, text width=2.65cm, minimum height=1.7cm] (aux-one) at (7.2,-0.75)
      {Auxiliary grader 1\\$R_{ij}^{\mathrm{aux},1}$\\text, and/or other};
    \node[auxiliary, text width=2.65cm, minimum height=1.7cm] (aux-n) at (7.2,-4.1)
      {Auxiliary grader $n$\\$R_{ij}^{\mathrm{aux},n}$\\text, and/or other};
    \node[assembly] (assembly) at (18.44,-2.425) {Final score assembly\\$R_{ij}$};
    \node[sample, minimum width=1.8cm, minimum height=1.7cm] (output) at (21.55,-2.425) {Final score\\$R_{ij}$};

    \draw[flow] (sample.east) -- (main.west);
    \draw[trunk] (sample.south) -- (0.9,-4.1);
    \draw[flow] (0.9,-0.75) -- (gate-one.west);
    \draw[flow] (0.9,-4.1) -- (gate-n.west);
    \node[note] at (3.6,-2.425) {$\vdots$};
    \node[note] at (7.2,-2.425) {$\vdots$};

    \draw[flow] (gate-one.east) -- (aux-one.west);
    \draw[flow] (gate-n.east) -- (aux-n.west);
    \draw[flow] (aux-one.east) -- (gate-a.west);
    \draw[flow] (gate-a.east) -- (aux-a.west);
    \draw[trunk] (aux-a.east) -- (16.4,-0.75);
    \draw[trunk] (aux-n.east) -- (16.4,-4.1);
    \draw[trunk] (16.4,-0.75) -- (16.4,-4.1);
    \draw[flow] (16.4,-2.425) -- (assembly.west);
    \draw[flow] (aux-one.north) -- (7.2,1) -| (assembly.north);
    \draw[flow] (main.east) -| (assembly.north);
    \draw[flow] (assembly.east) -- (output.west);
  \end{tikzpicture}
  \caption{General grader-boosting graph with input-gated auxiliary graders and final score assembly.}
  \label{fig:phase1-grader-boosting-general}
\end{subfigure}
  \caption{Grader boosting technique, illustrated by the insertion pattern propagation example (a) and the generalized RL grader-boosting mechanism (b).}
  \label{fig:phase1-grader-boosting}
\end{figure}

As shown in Figure~\ref{fig:phase1-grader-boosting-insertion}, we invoke the insertion pattern propagation auxiliary grader whenever a well-formed diff patch contains an insertion view kind in any ground truth or predicted patch.
The final score assembly takes on the following logic:
\[
R_{ij} =
\begin{cases}
0 & \text{if } R_{ij}^{\text{aux}} < 1 \\
R_{ij}^{\text{main}} & \text{otherwise}
\end{cases}
\]

Figure~\ref{fig:phase1-grader-boosting-general} shows the general form of grader boosting as a directed acyclic graph.

\paragraph{Grader Calibration.}
\label{sec:phase1-grader-calibration}
There exist many ways to calibrate graders against human judgment. In this work, we align our graders with human feedback, or quality signals for particular editing contexts (e.g., acceptance and dismissal).
For the main grader, we employ a self-refinement loop against the data in our POE offline benchmark (Section~\ref{sec:offline-benchmarks}) using the GitHub Copilot agent harness and an agent skill for grader optimization.
First, we obtain the grader's judgment for each POE sample. We then construct the confusion matrix of false positives, false negatives, true positives, and true negatives.
Next, we randomly sample examples from each confusion matrix category to understand patterns and failure modes in the grader's judgments.
We then invoke the agent to adapt the current grader's instructions to incorporate feedback from the confusion matrix analysis.
We repeat this process until the changes in the true positive and false positive rates between consecutive rounds fall below a target threshold.

Algorithm~\ref{alg:phase1-grader-calibration} formalizes this procedure. Let $z_i$ denote the binary quality label for POE example $i$, and let $\hat z_i^{(t)}=1$ when the main grader passes that example under instructions $I_t$.
\begin{algorithm}[t]
  \caption{Quality signal-based main grader calibration}
  \label{alg:phase1-grader-calibration}
  \begin{algorithmic}[1]
    \Require POE examples $\mathcal D=\{(x_i,z_i)\}_{i=1}^{N}$, initial instructions $I_0$, tolerance $\epsilon>0$
    \Ensure Calibrated instructions $I_t$
    \State $t\gets 0$
    \While{true}
      \State Evaluate the main grader on all $x_i\in\mathcal D$ using $I_t$ to obtain $\hat z_i^{(t)}$
      \State Compute $\mathrm{TP}_t=\sum_i\mathds{1}[z_i=1,\hat z_i^{(t)}=1]$ and $\mathrm{FP}_t=\sum_i\mathds{1}[z_i=0,\hat z_i^{(t)}=1]$
      \State Compute $\mathrm{FN}_t=\sum_i\mathds{1}[z_i=1,\hat z_i^{(t)}=0]$ and $\mathrm{TN}_t=\sum_i\mathds{1}[z_i=0,\hat z_i^{(t)}=0]$
      \State Compute $\mathrm{TPR}_t=\frac{\mathrm{TP}_t}{\mathrm{TP}_t+\mathrm{FN}_t}$ and $\mathrm{FPR}_t=\frac{\mathrm{FP}_t}{\mathrm{FP}_t+\mathrm{TN}_t}$
      \If{$t>0$ and $\max\{|\mathrm{TPR}_t-\mathrm{TPR}_{t-1}|,|\mathrm{FPR}_t-\mathrm{FPR}_{t-1}|\}\leq\epsilon$}
        \State \Return $I_t$
      \EndIf
      \For{each nonempty confusion-matrix category $c\in\{\mathrm{TP},\mathrm{FP},\mathrm{FN},\mathrm{TN}\}$}
        \State Uniformly sample a subset $\mathcal S_t^c$ of examples in category $c$ for LLM-based qualitative analysis for common failure modes
      \EndFor
      \State $I_{t+1}\gets\textsc{Refine}(I_t,\{\mathcal S_t^c\}_c)$ using the agent and the qualitative analysis
      \State $t\gets t+1$
    \EndWhile
  \end{algorithmic}
\end{algorithm}

We run this calibration process for 24 rounds before convergence. The resulting main grader reduces false negatives by 8.4\% and false positives by 1.1\%.
We note that quality signals can be noisy, which means that one set of data often does not delineate a perfect decision boundary.
However, these results demonstrate that this iterative calibration process improves grader agreement with the quality signals on the calibration samples.

\subsubsection{Ablations}
\label{sec:phase1-rl-ablations}

\paragraph{Patch Count Bonuses.}
\label{sec:phase1-patch-count-bonuses}
In addition to adding more multi-edit data to the SFT and RL recipes, we also test whether adding a conditional programmatic bonus to the RL reward encourages the model to produce more high-quality multi-edit responses.

The programmatic adjustment rewards multi-patch edit responses only when their format score is nonzero and their LLM quality score is at least 0.7; single-patch edit responses with a nonzero format score receive a penalty regardless of that quality threshold.
We test adjustments both before and after a score exponentiation, which sharpens the reward's preference for high-quality responses by raising edit scores to the power of 4 and no-edit scores to the power of 3.
Before exponentiation, the adjustment adds 0.1 for 2 patches or 0.2 for 3 or more patches, and subtracts 0.1 for a single patch.
After exponentiation, the corresponding bonuses are 0.05 and 0.1, and the single-patch penalty is 0.03.
We cap bonuses at 1 and floor penalties at 0. Qualifying multi-patch edits whose first edit is a ghost text view kind edit may also receive an additional bonus.
We also test continuous reward configurations with and without a translated Heaviside function gate that sets edit rewards to 0 below a configurable quality floor, evaluating the gate either on the adjusted pre-exponentiation score or on the post-exponentiation score.

\begin{figure}[!ht]
  \centering
  \includegraphics[width=\linewidth]{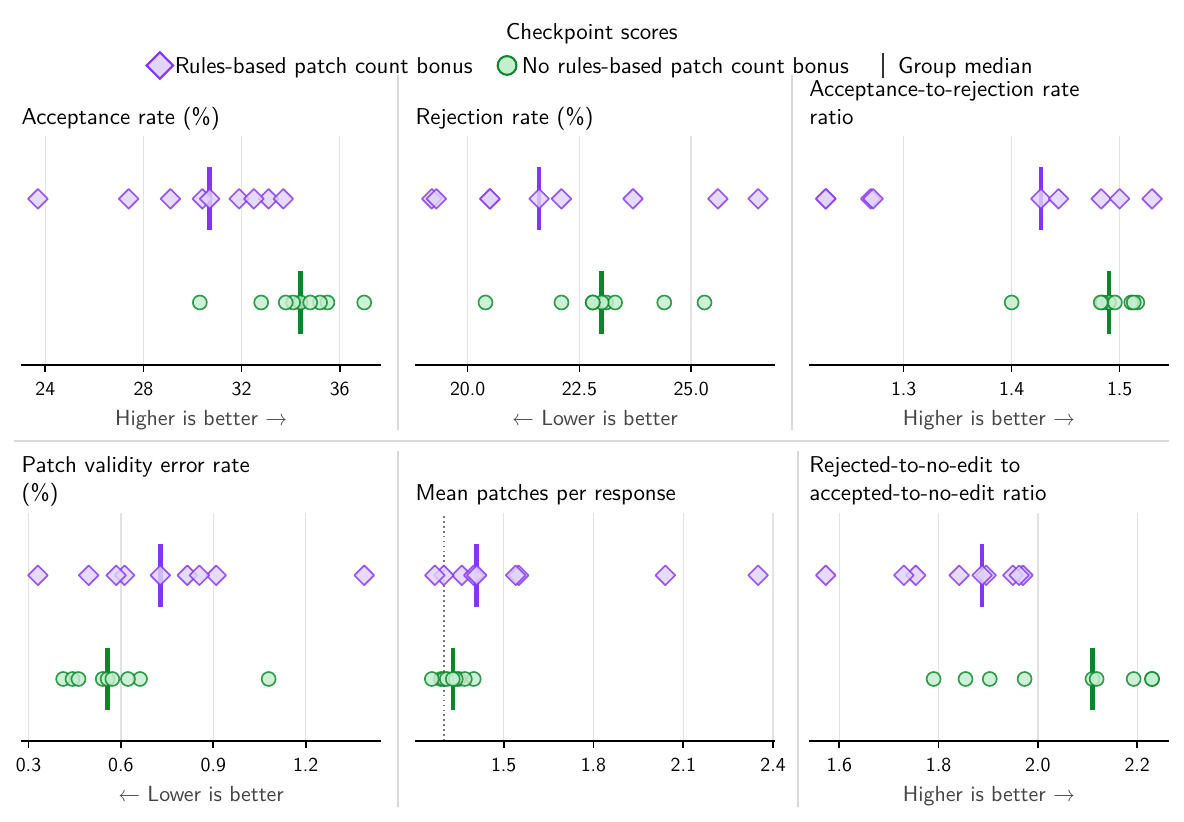}
  \caption{Comparison of patch count reward configurations on key POE metrics after RL training. Candidates trained with the rules-based patch count bonus have a higher median patch count but lower median acceptance rate, patch validity, and acceptance-to-rejection ratio. Though these descriptive comparisons do not isolate causal effects, they motivate removing the bonus from the grader.}
  \label{fig:phase1-patch-count-ablation}
\end{figure}

We compare nine 2-in-1 offline evaluation candidates trained with different variations of the patch count bonus and nine candidates trained without the patch count bonus. As shown in Figure~\ref{fig:phase1-patch-count-ablation}, we observe that while patch counts trend higher for the former group (a median of 1.41 edits compared to 1.33), quality metrics suffer. The latter group has a lower median patch validity error rate (0.56\% versus 0.73\%), higher median POE acceptance rate (34.4\% versus 30.7\%), higher acceptance-to-rejection rate ratio (1.49 vs 1.43), and higher median ratio of rejected-to-no-edit to accepted-to-no-edit counts (2.11 versus 1.89).
Rejection rate is an exception to this pattern: the no-bonus group has a higher median (23.0\% versus 21.6\%), while the bonus group spans a wider range (19.2\% to 26.5\% versus 20.4\% to 25.3\%). The spread and variance are arguably greatest for this metric.
We qualitatively attribute the negative impact of this rules-based reward to reward hacking, in which optimization exploits a reward specification without satisfying its intended objective \citep{amodei2016concrete}. Models learn undesired behaviors, such as adding meaningless patches like trivial comments, to increase the reward at the cost of response quality.

These trends support removing rules-based reward elements for patch counts. However, these comparisons include few candidates and describe groups rather than a matched, single-factor experiment, so we make no statistical claims. We use the results to inform subsequent grader designs: we encode nuanced or conditional behavior, especially tradeoffs, in the LLM judge rather than in rules, which may be brittle and more prone to reward hacking.

\paragraph{Critical and Suboptimal Patch Penalties.}
We also ablate whether the RL grader should penalize only critical patch validity issues or also suboptimal ones, as discussed in Section~\ref{sec:phase1-data-quality-filtering}. We hypothesize that penalizing only critical issues provides a better middle ground than penalizing all suboptimally expressed patches. Allowing less efficient patches that do not affect end-user experience may give the model more flexibility to learn more important characteristics of high-quality responses. In some cases, these patches may even benefit the model; for instance, no-op patches in which deletions match additions may serve as a chain-of-thought for considering viable, even if ultimately non-optimal, edit locations.

\begin{figure}[htbp]
  \centering
  \begin{subfigure}[t]{0.96\linewidth}
    \centering
    \includegraphics[width=\linewidth]{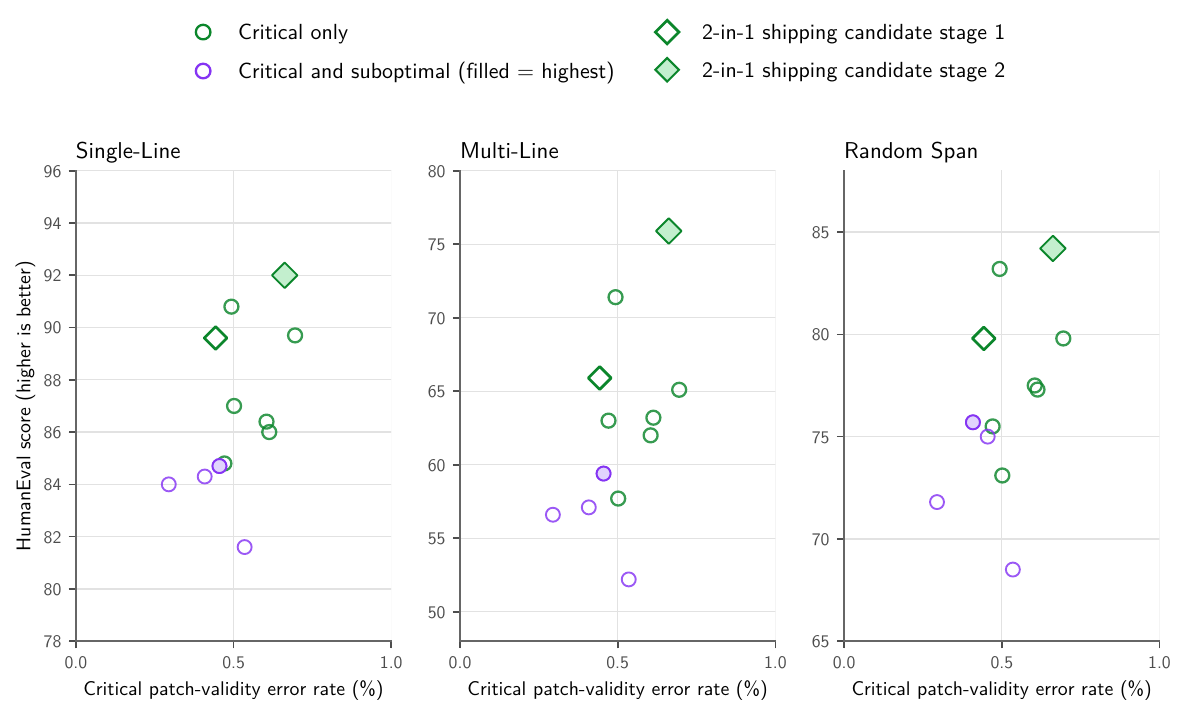}
    \caption{Critical patch validity error rate.}
  \end{subfigure}
  \vspace{3pt}
  \begin{subfigure}[t]{0.96\linewidth}
    \centering
    \includegraphics[width=\linewidth]{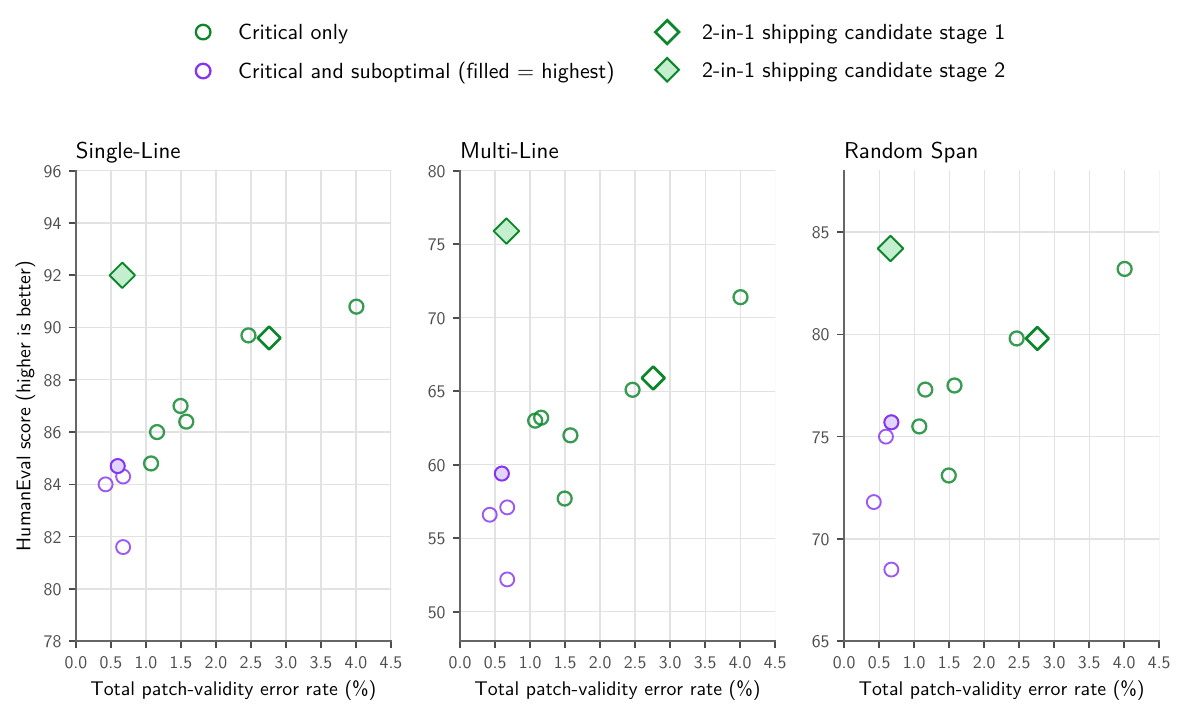}
    \caption{Total patch validity error rate, including critical and suboptimal errors.}
  \end{subfigure}
  \caption{Augmented HumanEval subdataset scores against (a) critical and (b) total patch validity error rates. These results motivate our choice of critical-only penalties. Model checkpoints trained with them generally achieve higher completion scores with comparable critical error rates, though they exhibit more suboptimal patch errors. Notably, the shipped 2-in-1 checkpoint still maintains a low total error rate despite using critical-only penalties.}
  \label{fig:phase1-patch-validity-ablation}
\end{figure}

Figure~\ref{fig:phase1-patch-validity-ablation} compares twelve offline evaluation candidate checkpoints across the three Augmented HumanEval subdatasets. We use this benchmark because it measures the model's ability to provide complete and working suggestions, which we qualitatively find declines when we penalize correct but suboptimally efficient patches. Overall, critical-only checkpoints tend to achieve higher Augmented HumanEval accuracy than checkpoints that receive both critical and suboptimal patch validity constraints during training. Critical patch validity error rates overlap across the two training conditions, while critical-only models generally achieve higher Augmented HumanEval scores. Their total error rates are higher, consistent with a greater contribution from suboptimal patch validity errors, which the critical-only condition does not penalize. The final 2-in-1 shipping candidate (stage 2) uses only critical patch validation and scores 92.0, 75.9, and 84.2 on the three subdatasets, the highest score in each subset, while maintaining a relatively low total error rate. The contrast between stages 1 and 2 illustrates the tradeoff between suboptimal patch validity errors and Augmented HumanEval accuracy across the three subdatasets.
These selected-checkpoint comparisons are descriptive and do not isolate the effect of penalty strictness.

\subsection{Results}

\subsubsection{Offline Benchmarks}
Table~\ref{tab:phase1-offline-results} compares the SFT checkpoint with checkpoints from
two sequential RL stages for the 2-in-1 model.
Across both stages, reinforcement learning teaches the model to avoid poor-quality
suggestions (STests MustAvoid), improve suggestion quality (STests Total, Augmented HumanEval), improve abstention discernment (POE rejection-to-no-edit to acceptance-to-no-edit ratio, R/A $\rightarrow$ NE), and maintain the SFT multi-patch operating point (POE mean patches).
Stage~2 further improves completion performance (Augmented HumanEval) compared to Stage~1.
We show QMH for reference only. Because we use QMH to select RL checkpoints (Section~\ref{sec:model-selection-process}), the RL checkpoints' scores of 18/19 and 19/19 partly reflect that selection and are not independent evidence of quality.
We formalize the 19-case QMH set late in RL stage~1, so we score the SFT checkpoint on it retrospectively.

\begin{table}[t]
  \caption{Phase~1 offline results for SFT and two RL stages. RL improves STests, Augmented HumanEval, and POE over SFT while maintaining a similar mean patch count; we use QMH for checkpoint selection and show it for reference.}
  \label{tab:phase1-offline-results}
  \centering
  \footnotesize
  \setlength{\tabcolsep}{2.6pt}
  \begin{tabular}{lrrrrrrrrrr}
    \toprule
    & \multicolumn{2}{c}{STests v1} & \multicolumn{2}{c}{STests v2}
    & \multicolumn{3}{c}{Augmented HumanEval} & \multicolumn{2}{c}{POE}
    & \multicolumn{1}{c}{QMH} \\
    \cmidrule(lr){2-3}\cmidrule(lr){4-5}\cmidrule(lr){6-8}
    \cmidrule(lr){9-10}\cmidrule(lr){11-11}
    Model & MustAvoid & Total & MustAvoid & Total
          & Single & Multi & Span & Mean patches & R/A $\rightarrow$ NE & Passes \\
    \midrule
    SFT &
      89.6\% & 48.0/77 & 86.7\% & 36.8/79 &
      79.6 & 53.9 & 72.5 & 1.32 & 1.59 & 13/19 \\
    RL stage~1 &
      91.2\% & 51.6/77 & 91.4\% & 45.4/79 &
      89.6 & 65.9 & 79.8 & 1.29 & 2.11 & 18/19 \\
    RL stage~2 &
      91.2\% & 51.7/77 & 89.1\% & 45.6/79 &
      92.0 & 75.9 & 84.2 & 1.30 & 2.23 & 19/19 \\
    \bottomrule
  \end{tabular}
\end{table}

\subsubsection{Online Evaluation}
Over the course of Phase~1, we evaluate 14 model candidates across four rounds of online A/B experiments.
These experiments validate the output format and client integration, screen model
candidates against guardrail metrics, and refine the recipe to target and correct insertion placement errors we discover during internal testing.

\paragraph{Benchmark Refinement through Qualitative Testing.}
The second round of experiments demonstrates that existing offline benchmark parity and a clean online scorecard do not alone establish deployment readiness. During this round, we have not yet introduced the QMH benchmark and rely on the other four quantitative benchmarks and online A/B experimentation as the primary means of evaluation.
Although two candidates meet the quantitative criteria in this round, further testing
identifies overly frequent suggestions, incorrect insertion placement, and code
duplication.
This discrepancy motivates the development of the QMH benchmark as an initial quality filter, the POE benchmark as the
primary offline benchmark, and Simulation Tests, Augmented HumanEval, and
Output View Kind as guardrail benchmarks.
These results also underscore the importance of treating benchmark coverage as iterative: qualitative findings from internal testing can expose behaviors that existing measures miss and guide the addition or refinement of targeted evaluation cases.

\paragraph{2-in-1 Model Results.}
Table~\ref{tab:phase1-online-results-2-in-1} reports results for the final 2-in-1 model candidate paired with the separate completion model against the
three-model baseline in an A/B experiment.
The experiment randomly assigns a fraction of VS Code stable users to each condition.
The experiment finds no statistically significant regression in shown, accepted, or dismissed suggestions, recurring engagement, or accepted and retained characters.
Mean time to display and mean time to produce
decrease by 6.77\% and 5.95\%, while output and prompt tokens decrease by
58.93\% and 4.83\%, respectively.
These reductions coincide with a more compact output format, a smaller input
context, and fewer sequential model calls.
Cached suggestion share increases by 8.86\%.

\begin{table}[t]
  \caption{Phase~1 online A/B experiment results for the 2-in-1 model paired with the completion model relative to the three-model baseline. Latency and token counts decrease, with no statistically significant regression in other quality and engagement metrics.}
  \label{tab:phase1-online-results-2-in-1}
  \centering
  \small
  \begin{tabular}{lrr}
    \toprule
    Metric & Rel.\ change (\%) & $p$-value \\
    \midrule
    Shown suggestions                  & $+2.25 \pm 8.58$  & 0.6069 \\
    Accepted suggestions               & $-2.63 \pm 9.35$  & 0.5819 \\
    Dismissed suggestions              & $+16.00 \pm 19.10$ & 0.1006 \\
    Ignored suggestions                & $+3.40 \pm 8.50$  & 0.4327 \\
    ARC at 30 seconds & $+2.94 \pm 9.08$ & 0.5257 \\
    ARC at 5 minutes & $+1.26 \pm 9.56$ & 0.7966 \\
    Recurring engagement   & $-2.50 \pm 4.31$  & 0.2558 \\
    Time to display & $\mathbf{-6.77 \pm 1.98}$  & $\approx 0$ \\
    Time to produce & $\mathbf{-5.95 \pm 1.75}$  & $\approx 0$ \\
    Output tokens          & $\mathbf{-58.93 \pm 1.54}$ & $\approx 0$ \\
    Prompt tokens          & $\mathbf{-4.83 \pm 0.78}$ & $\approx 0$ \\
    Cached suggestion share & $\mathbf{+8.86 \pm 1.16}$ & $\approx 0$ \\
    \bottomrule
  \end{tabular}
\end{table}

\section{Phase 2: Unifying the 2-in-1 Model with Code Completion}
\label{sec:phase2-unifying-2-in-1-with-code-completion}

The diff patch format can express completion suggestions as well as nearby and long-distance edits.
We therefore first test whether the 2-in-1 model can handle all three tasks by removing the separate completion model from the client.
An online experiment against the 2-in-1 model paired with the completion model shows a 22.45\% decrease in accepted suggestions per user and an 18.99\% decrease in accepted and retained characters (ARC) at 30 seconds, both statistically significant.
The experiment randomly assigns a fraction of VS Code stable users to each condition.
Table~\ref{tab:phase2-completion-removal-results} reports these effects with their confidence interval half-widths.
These regressions show that the completion model still plays an important role in suggestion acceptance and retained characters.
Supporting completion in the output format is not sufficient to preserve the quality of the paired system.
Phase~2 therefore aims to improve completion behavior and address these regressions through completion-focused reinforcement learning and client adaptation, producing the 3-in-1 model we present here.

\begin{table}[htbp]
  \caption{Initial Phase~2 online experiment removing the separate completion model. Relative changes compare the 2-in-1 model alone with the 2-in-1 model paired with the completion model. Using the 2-in-1 model for completion without additional training results in statistically significant regressions.}
  \label{tab:phase2-completion-removal-results}
  \centering
  \small
  \begin{tabular}{lr}
    \toprule
    Metric & Rel.\ change (\%) \\
    \midrule
    Accepted suggestions & $\mathbf{-22.45 \pm 11.24}$ \\
    ARC at 30 seconds & $\mathbf{-18.99 \pm 11.73}$ \\
    \bottomrule
  \end{tabular}
\end{table}

We develop the completion-focused data mixture, completeness grader, and client adaptation together, iterating on data sources, sample weights, grading criteria, and client settings.
We describe these components separately below, although their development is interleaved.

\subsection{Reinforcement Learning}

\subsubsection{Data Mixture}
\label{sec:phase2-data-mixture}
We construct the Phase~2 RL mixture by extending the Phase~1 mixture with completion examples.
The added data targets ghost text at the cursor while retaining coverage of nearby and long-distance edits.
Table~\ref{tab:phase2-data-mixture} lists the five datasets we add in this phase.
Sample counts are before upsampling.
We describe their roles and the changes to the inherited mixture below.

\begin{table}[htbp]
  \caption{Datasets added to the Phase~2 RL mixture. Added examples target ghost text at the cursor and completion length. Completion length examples receive tenfold upsampling to increase their representation.}
  \label{tab:phase2-data-mixture}
  \centering
  \small
  \begin{tabular}{@{}lrr@{}}
    \toprule
    Source & Samples & Upsampling \\
    \midrule
    Converted completion & 40{,}000 & 1x \\
    Converted NES & 33{,}000 & 1x \\
    Converted joint model & 11{,}069 & 1x \\
    Multi-edit with first edit ghost text at cursor & 2{,}653 & 1x \\
    Completion length & 908 & 10x \\
    \bottomrule
  \end{tabular}
\end{table}

\begin{table}[htbp]
  \caption{Upsampling of inherited patch validity and insertion pattern propagation data. Phase~2 increases both factors to maintain coverage of known failure modes as the mixture expands.}
  \label{tab:phase2-data-upsampling}
  \centering
  \small
  \begin{tabular}{@{}lrrr@{}}
    \toprule
    Source & Samples & Phase~1 Upsampling & Phase~2 Upsampling \\
    \midrule
    Rewritten invalid patches & 560 & 1x & 5x \\
    Rewritten incorrect insertion pattern propagation & 355 & 5x & 50x \\
    \bottomrule
  \end{tabular}
\end{table}

In addition to these data, we upsample examples that address two Phase~1 failure categories: patch validity and insertion pattern propagation. Through experimentation, we find that this additional upsampling adequately maintains the model's exposure to these cases to counterbalance the volume of examples from the additional sources in Table~\ref{tab:phase2-data-mixture}. Table~\ref{tab:phase2-data-upsampling} summarizes the upsampling we apply to these datasets in the Phase~1 and Phase~2 RL recipes.
Factors in both phases are total multipliers relative to the original sample counts shown.

In Phase~1, the completion model produces only ghost text, while the 2-in-1 model supports all seven view kinds.
The 3-in-1 model replaces both models and therefore must produce ghost text more often than the 2-in-1 model alone to preserve completion behavior alongside next edit suggestions.
One purpose of the added data is to increase the ghost text share in the training mixture, as defined by the ghost text ratio discussed in Section~\ref{sec:output-view-kind}.
Table~\ref{tab:phase2-ghost-text-share} summarizes sample counts and ghost text shares after upsampling for the inherited data, the added data, and the combined mixture.

\begin{table}[htbp]
  \caption{Phase~2 RL sample counts and ghost text shares after upsampling. We use the same view-kind classifier as Output View Kind v1 (Section~\ref{sec:output-view-kind}). We compute the ghost text share by dividing the ghost text count by the total sample response count excluding no edit, whitespace-only, and unknown responses. The added data raises the mixture's ghost text share from approximately 32\% to 66\%.}
  \label{tab:phase2-ghost-text-share}
  \centering
  \small
  \begin{tabular}{@{}lrr@{}}
    \toprule
    Source & Samples & Ghost text share \\
    \midrule
    Inherited 2-in-1 RL data & 84{,}129 & 32.09\% \\
    Added Phase~2 data & 95{,}802 & 92.38\% \\
    \midrule
    \textbf{Total} & \textbf{179{,}931} & \textbf{65.52\%} \\
    \bottomrule
  \end{tabular}
\end{table}

\paragraph{Converted Ghost Text Data.}
We use the first pipeline described in Section~\ref{sec:phase1-data-format-conversion} to convert existing prompt-response samples into diff patches.
The source data comes from three types of models: a completion model, an NES model, and a joint completion-and-NES model.
We filter suggestions from both the NES model and the joint model to the ghost text view kind.
The joint model supports both code completion and nearby next edit suggestion using the NES window-rewrite prompt and output format, without predicting long-distance edits.
We explore this approach as an intermediate step to generate more ghost text in unified models before adopting the diff patch format. It retains the rewrite-window representation and adapts training to increase the model's tendency to produce ghost text.
Distilling knowledge from this intermediate model into the diff patch 3-in-1 model helps improve ghost text generation by using edits sampled from the distribution of a unified model that combines both completion and nearby next edit capabilities.

\paragraph{Cursor-First Multi-Edit Examples.}
We add examples whose recorded editing action begins with a ghost text completion at the cursor and spans multiple patches.
This subset supports completion at the cursor followed by related edits elsewhere, rather than encouraging the model to move away from the cursor before completing the local code.

\paragraph{Completion Length Examples.}
\label{sec:phase2-accepted-retained-character-data}
We investigate suggestion completeness as one of several ways to improve completion behavior and address ARC regressions.
The ARC regression in Table~\ref{tab:phase2-completion-removal-results}, along with observed lower predicted character counts in ghost text suggestions, motivates the hypothesis that some suggestions are too short or incomplete; developers may accept a promising start to a suggestion, but later rewrite or remove it when the rest of the code is missing.
To target this behavior, an LLM identifies short or incomplete ghost text suggestions and rewrites them into longer, complete suggestions.
We convert the rewritten examples into diff patches and upsample this dataset by a factor of 10.
We introduce the data together with the ghost text completeness auxiliary grader described below.

\subsubsection{Training Recipe}
\label{sec:phase2-training-recipe}

\paragraph{Ghost Text Completeness Auxiliary Grader.}
To target the incomplete suggestions described in Section~\ref{sec:phase2-accepted-retained-character-data}, we extend the Phase~1 grader with an auxiliary LLM judge focused on ghost text completeness.
The main judge continues to assess general suggestion quality, while the auxiliary judge evaluates whether a completion provides a complete, usable unit of code for the apparent editing intent.
We retain the insertion pattern propagation auxiliary grader and use the same grader boosting approach described in Section~\ref{sec:phase1-grader}.

We gate the completeness judge on examples designated as requiring ghost text, and it runs only when both the reference and the candidate response are ghost text edits.
Given the editing context, reference edit, and candidate suggestion, it identifies the intended code unit and checks whether the candidate is at least as complete as the reference.
For example, it rejects a function signature without the body supplied by the reference, a test without its assertion, or a partially completed argument list.
The candidate need not match the reference text and may be shorter or use a different structure if it provides a complete solution to the same intent.

The auxiliary judge returns a binary score: 1 for a sufficiently complete suggestion and 0 for an incomplete one.
As in Phase~1, an auxiliary score below 1 sets the final reward to 0; otherwise, we retain the main judge's score.
We use the same completeness check in both the single-edit and multi-edit grader variants.

To track suggestion size while addressing the ARC regression, we also introduce Output View Kind v2 (Section~\ref{sec:output-view-kind}).
Its average inserted characters and lines per patch and average patches per response describe the amount of code suggested, which is relevant to ARC.

\paragraph{Preceding-Line Duplication Check.}
When we increase the share of completion examples in the training mixture, the
model sometimes repeats the line immediately before the insertion point or a
larger block of preceding context.
The model should return no edit rather than introduce this redundant code.
We track this failure mode in QMH and add a preceding-line duplication rule to
the patch validity checker.
The rule marks these patches as invalid, so the training signal penalizes
duplication and candidates perform better on the corresponding QMH cases.

\subsection{Client Adaptation}

Inline suggestions depend on both the model and the client that presents its output.
The client constructs prompts, schedules model requests, caches responses, and renders the suggestion user interface, determining when suggestions appear and how it displays them to the developer.
Replacing the completion and 2-in-1 models with a single model therefore requires adapting client behavior to the combined task.
We describe five candidate changes to the VS Code client below and evaluate them through online experiments.
The final model-plus-client configuration combines four adaptations: the 3-in-1 speculative decoding draft, progressive ghost text reveal, diff-based view kind rendering, and cross-mode suppression of ignored suggestions.

\paragraph{Speculative Decoding Draft Text.}
Speculative decoding accelerates autoregressive generation by proposing draft tokens and verifying them in parallel with the target model, allowing multiple tokens to be accepted in a single model pass \citep{stern2018blockwise,xia2023speculative,leviathan2023speculative,chen2023speculative}.
For inline suggestions, the client constructs the speculative decoding draft by predicting a likely prefix of the diff patch response from the current file content.
The 2-in-1 client uses a file path header as the draft.
For the 3-in-1 model, we use a draft that anticipates a completion at the cursor by supplying the prefix of a diff patch anchored to the current line.
The draft spans three lines: the patch header, the anchoring deletion, and the beginning of the addition line.
Figure~\ref{fig:phase2-speculative-decoding-drafts} compares the two drafts for a function body completion.
We select the 3-in-1 draft format based on a load test comparing five draft formats (Appendix~\ref{sec:speculative-decoding-draft-comparison}).
Although the 3-in-1 draft matches a longer response prefix in Figure~\ref{fig:phase2-speculative-decoding-drafts}, it can mismatch the response when the suggested edit is not an insertion at the current line, wasting computation and reducing the speedup from speculative decoding.
However, the higher ghost text ratio of the 3-in-1 model relative to the 2-in-1 model in Table~\ref{tab:phase2-offline-results} and its completion-focused training motivate our expectation of more frequent matches to the cursor-anchored draft.
The 3-in-1 speculative decoding draft significantly reduces mean time to display and mean time to produce relative to the 2-in-1 draft, as shown by the rendering and speculative decoding ablations in Table~\ref{tab:phase2-rendering-speculative-decoding-ablation}.

\begin{figure}[tp]
  \centering
  \footnotesize
  \setlength{\fboxsep}{0pt}
  \begin{minipage}{\linewidth}
    \textbf{File content:} \texttt{/path/to/util.py}
    \par\smallskip
    {\ttfamily
    def count\_words(text: str) -> int:\\
    \colorbox{ghgreen!18}{\strut\ \ \ \ return len(text.split())}
    }
    \par\medskip
    \hrule
    \medskip
    \begin{minipage}[t]{0.49\linewidth}
      \textbf{Speculative decoding draft (2-in-1)}
      \par\smallskip
      \begin{tabular}{@{}l@{}}
        \texttt{\colorbox{ghpurple!15}{\strut/path/to/util.py:}0} \\
        \texttt{-def count\_words(text: str) -> int:} \\
        \texttt{+def count\_words(text: str) -> int:} \\
        \texttt{+\ \ \ \ return len(text.split())}
      \end{tabular}
    \end{minipage}\hfill
    \begin{minipage}[t]{0.49\linewidth}
      \textbf{Speculative decoding draft (3-in-1)}
      \par\smallskip
      \begin{tabular}{@{}l@{}}
        \texttt{\colorbox{ghpurple!15}{\strut/path/to/util.py:0}} \\
        \texttt{\colorbox{ghpurple!15}{\strut-def count\_words(text: str) -> int:}} \\
        \texttt{\colorbox{ghpurple!15}{\strut+def count\_words(text: str) -> int:}} \\
        \texttt{+\ \ \ \ return len(text.split())}
      \end{tabular}
    \end{minipage}
    \par\medskip
    \textbf{Legend:}\quad
    \colorbox{ghgreen!18}{\strut Inserted text}
    \qquad
    \colorbox{ghpurple!15}{\strut Speculative decoding draft}
  \end{minipage}
  \caption{Speculative decoding drafts for a function body completion.
  The function signature is present when the client triggers the request; the body is the suggested insertion.
  Both lower panels show the same diff patch response and highlight the respective draft prefix.}
  \label{fig:phase2-speculative-decoding-drafts}
\end{figure}

\paragraph{Progressive Ghost Text Reveal.}
We extend the completion client's progressive reveal behavior to the unified model.
The client streams the response from the model, displays the immediately relevant portion as soon as it arrives, and progressively reveals the remainder.
In addition to reducing the time to initial display, progressive reveal allows the developer to inspect the suggestion incrementally rather than evaluate the entire block at once.

\paragraph{Diff-Based View Kind Rendering.}
The raw structure of a patch does not always imply the most suitable presentation.
For example, a replacement patch may reproduce existing text while adding new content, even though the effective change consists only of insertions.
The client therefore computes the diff between the original and patched code for each patch and derives the view kind from that diff.
This preserves the proposed code change and allows the client to display it as ghost text or another insertion view instead of a replacement.
Figure~\ref{fig:phase2-diff-based-rendering} illustrates how the client can present a replacement-shaped patch as two insertions.

\begin{figure}[tp]
  \centering
  \footnotesize
  \setlength{\tabcolsep}{5pt}
  \begin{tabular}{@{}p{0.46\linewidth}p{0.50\linewidth}@{}}
    \toprule
    \textbf{Model response} & \textbf{Diff-based rendering} \\
    \midrule
    \texttt{/path/to/aquarium.py:2}
    & Insertion before \texttt{cuttlefish}: \\
    \texttt{-cuttlefish}
    & \texttt{+axolotl} \\
    \texttt{+axolotl} & \texttt{+blobfish} \\
    \texttt{+blobfish} & \\[0.25em]
    \texttt{+cuttlefish} & Insertion after \texttt{cuttlefish}: \\
    \texttt{+dolphin} & \texttt{+dolphin} \\
    \bottomrule
  \end{tabular}
  \caption{Diff-based view kind rendering of a replacement-shaped patch.
  The raw patch replaces \texttt{cuttlefish} with four lines, which the client can display as a side-by-side replacement.
  Computing the diff identifies \texttt{cuttlefish} as unchanged and separates the effective change into two insertions: \texttt{axolotl} and \texttt{blobfish} before it, and \texttt{dolphin} after it.
  The first insertion is at the cursor, so the client can render it as ghost text.
  Both presentations produce the same final code; the right column shows the inserted text, not additional model-generated patches.}
  \label{fig:phase2-diff-based-rendering}
\end{figure}

\paragraph{Cross-Mode Suppression of Ignored Suggestions.}
The completion client behavior removes ghost text when the cursor moves, whereas the NES client persists it because NES suggestions can arrive asynchronously during cursor movement.
The 3-in-1 model retains the NES client behavior, so an ignored ghost text suggestion persists after cursor movement. In certain cases, it can reappear as a next edit suggestion.
We change the client to suppress this reappearance, preventing it from presenting the same ignored suggestion again in a different view kind.

\paragraph{Cache Delay.}
The client caches streamed edits locally and retrieves them when the editor requests a suggestion.
Caching is particularly useful for multi-patch responses: the client shows the first patch and keeps subsequent patches available to display after the developer accepts the first.
Because displaying a cached suggestion requires no new model request, the next suggestion can appear with little delay.
However, suggestions that appear too quickly can interrupt the developer's typing.
The existing client applies a 200-millisecond delay before displaying cached suggestions.
We hypothesize that removing this delay can offset the additional time needed to emit patch structure before completion text, compared with the standalone completion model's suffix-only output.
We test a zero-delay configuration against the existing 200-millisecond configuration (Table~\ref{tab:phase2-cache-delay-ablation}).

\subsubsection{Ablations}
\label{sec:phase2-client-ablations}

We evaluate the client changes through three controlled online A/B experiments using the VS Code client.
The experiments follow the randomized assignment, metric aggregation, and significance criteria described in Section~\ref{sec:online-experiment-design}, and report the same metrics as the Phase~1 online experiment in Table~\ref{tab:phase1-online-results-2-in-1}.
Within each client ablation, we hold the model checkpoint and other client settings fixed to isolate the change under evaluation.

\paragraph{Rendering and Speculative Decoding.}
This experiment evaluates three features: the speculative decoding draft text designed for the 3-in-1 model, progressive ghost text reveal, and diff-based view kind rendering.
It randomly assigns a fraction of VS Code stable users to each condition.
It compares a configuration with all three features enabled against three variants, each removing one feature.
The speculative decoding variant uses the earlier 2-in-1 draft text; the other variants disable progressive reveal or diff-based rendering, respectively.
These comparisons measure each feature's contribution when the other two remain enabled.
Table~\ref{tab:phase2-rendering-speculative-decoding-ablation} reports each variant relative to the full configuration.

\begin{table}[htbp]
  \caption{Rendering and speculative decoding ablations relative to the full configuration. The 3-in-1 draft and progressive reveal each increase accepted suggestions, while the draft also reduces latency. Diff-based rendering shows no statistically significant effect on measured outcomes.}
  \label{tab:phase2-rendering-speculative-decoding-ablation}
  \centering
  \footnotesize
  \setlength{\tabcolsep}{2pt}
  \begin{tabular}{lrrrrrr}
    \toprule
    & \multicolumn{6}{c}{Change from full configuration} \\
    \cmidrule(lr){2-7}
    & \multicolumn{2}{c}{Use 2-in-1 draft}
    & \multicolumn{2}{c}{Disable progressive reveal}
    & \multicolumn{2}{c}{Disable diff-based rendering} \\
    \cmidrule(lr){2-3}\cmidrule(lr){4-5}\cmidrule(lr){6-7}
    Metric & Rel.\ change (\%) & $p$-value & Rel.\ change (\%) & $p$-value & Rel.\ change (\%) & $p$-value \\
    \midrule
    Shown suggestions & $\mathbf{-8.40 \pm 5.03}$ & 0.0011 & $\mathbf{-6.51 \pm 5.03}$ & 0.0113 & $+0.16 \pm 5.48$ & 0.9534 \\
    Accepted suggestions & $\mathbf{-7.87 \pm 5.50}$ & 0.0051 & $\mathbf{-10.22 \pm 5.44}$ & 0.0002 & $-1.28 \pm 6.12$ & 0.6830 \\
    Dismissed suggestions & $-4.47 \pm 8.93$ & 0.3268 & $+1.92 \pm 9.15$ & 0.6807 & $+2.67 \pm 9.29$ & 0.5727 \\
    Ignored suggestions & $\mathbf{-8.80 \pm 4.98}$ & 0.0005 & $\mathbf{-5.57 \pm 5.00}$ & 0.0290 & $+0.57 \pm 5.45$ & 0.8385 \\
    ARC at 30 seconds & $-2.72 \pm 5.50$ & 0.3321 & $-1.29 \pm 5.48$ & 0.6441 & $-0.44 \pm 5.49$ & 0.8757 \\
    ARC at 5 minutes & $-2.98 \pm 5.75$ & 0.3109 & $-2.39 \pm 5.71$ & 0.4123 & $-0.97 \pm 5.77$ & 0.7427 \\
    Recurring engagement & $\mathbf{-3.09 \pm 2.36}$ & 0.0103 & $-1.19 \pm 2.37$ & 0.3239 & $+0.37 \pm 2.39$ & 0.7583 \\
    Time to display & $\mathbf{+11.44 \pm 1.16}$ & $\approx 0$ & $+1.00 \pm 1.12$ & 0.0779 & $+0.03 \pm 1.12$ & 0.9524 \\
    Time to produce & $\mathbf{+11.19 \pm 1.07}$ & $\approx 0$ & $+0.98 \pm 1.03$ & 0.0620 & $+0.12 \pm 1.03$ & 0.8206 \\
    Output tokens & $\mathbf{-12.67 \pm 1.24}$ & $\approx 0$ & $+0.16 \pm 1.33$ & 0.8135 & $+0.25 \pm 1.29$ & 0.7036 \\
    Prompt tokens & $+0.32 \pm 0.66$ & 0.3387 & $-0.29 \pm 0.66$ & 0.3904 & $+0.65 \pm 0.66$ & 0.0525 \\
    Cached suggestion share & $\mathbf{+0.60 \pm 0.57}$ & 0.0402 & $\mathbf{-1.82 \pm 0.54}$ & $\approx 0$ & $-0.28 \pm 0.61$ & 0.3713 \\
    \bottomrule
  \end{tabular}
\end{table}

Using the earlier 2-in-1 speculative decoding draft instead of the 3-in-1 draft increases mean time to display by 11.44\% and mean time to produce by 11.19\%.
It also reduces shown suggestions by 8.40\%, accepted suggestions by 7.87\%, ignored suggestions by 8.80\%, and recurring engagement by 3.09\%.
All of these changes are statistically significant.
The acceptance and engagement results support the use of the 3-in-1 draft.

Disabling progressive ghost text reveal reduces accepted suggestions by 10.22\%, shown suggestions by 6.51\%, and ignored suggestions by 5.57\%, all statistically significant.
The acceptance result supports displaying the immediately relevant portion of a response as it arrives rather than waiting for the full response.
Neither latency metric changes significantly in this comparison.
Disabling diff-based view kind rendering produces no statistically significant changes in the measured outcomes.

\paragraph{Cross-Mode Suppression of Ignored Suggestions.}
We evaluate cross-mode suppression by comparing a suppression-enabled treatment group with a suppression-disabled control group.
Both groups use the same final 3-in-1 checkpoint and the other three client adaptations: the 3-in-1 speculative decoding draft, progressive ghost text reveal, and diff-based view kind rendering.
Table~\ref{tab:phase2-suppression-ablation} reports the treatment relative to the control, isolating the effect of enabling suppression.

\begin{table}[htbp]
  \caption{Cross-mode suppression ablation with the final 3-in-1 checkpoint held fixed and the other three client adaptations enabled in both groups. Relative changes compare the suppression-enabled treatment with the suppression-disabled control. Suppression reduces dismissed and ignored suggestions but increases latency.}
  \label{tab:phase2-suppression-ablation}
  \centering
  \small
  \begin{tabular}{lrr}
    \toprule
    Metric & Rel.\ change (\%) & $p$-value \\
    \midrule
    Shown suggestions & $-5.31 \pm 5.35$ & 0.0517 \\
    Accepted suggestions & $-1.53 \pm 5.91$ & 0.6106 \\
    Dismissed suggestions & $\mathbf{-22.40 \pm 8.87}$ & $\approx 0$ \\
    Ignored suggestions & $\mathbf{-5.77 \pm 5.30}$ & 0.0328 \\
    ARC at 30 seconds & $+1.41 \pm 5.77$ & 0.6325 \\
    ARC at 5 minutes & $+2.03 \pm 6.07$ & 0.5119 \\
    Recurring engagement & $-0.56 \pm 2.43$ & 0.6511 \\
    Time to display & $\mathbf{+4.27 \pm 1.15}$ & $\approx 0$ \\
    Time to produce & $\mathbf{+4.47 \pm 1.05}$ & $\approx 0$ \\
    Output tokens & $+0.38 \pm 1.16$ & 0.5246 \\
    Prompt tokens & $+0.06 \pm 0.63$ & 0.8514 \\
    Cached suggestion share & $\mathbf{-3.90 \pm 0.73}$ & $\approx 0$ \\
    \bottomrule
  \end{tabular}
\end{table}

Enabling suppression reduces dismissed suggestions per user by 22.40\% and ignored suggestions by 5.77\%, both statistically significant.
Accepted suggestions, ARC at both retention windows, and recurring engagement do not change significantly.
Mean time to display increases by 4.27\% and mean time to produce by 4.47\%, both statistically significant, indicating a latency tradeoff.
The direct comparison isolates the contribution of the client change to reducing dismissals.

\paragraph{Cache Delay.}
This experiment tests removing the existing 200-millisecond cache delay using an earlier 3-in-1 checkpoint rather than the final model.
It randomly assigns a fraction of VS Code stable users to each condition.
The hypothesis is that removing this delay improves responsiveness and suggestion outcomes by offsetting the latency of generating patch structure before completion text.
Table~\ref{tab:phase2-cache-delay-ablation} reports the zero-delay treatment relative to the existing 200-millisecond control.

\begin{table}[htbp]
  \caption{Cache delay ablation using an earlier 3-in-1 checkpoint. Relative changes compare the zero-delay treatment with the existing 200-millisecond control. Removing the delay reduces time to display and time to produce, without statistically significant differences in suggestion quality or recurring engagement.}
  \label{tab:phase2-cache-delay-ablation}
  \centering
  \small
  \begin{tabular}{lrr}
    \toprule
    Metric & Rel.\ change (\%) & $p$-value \\
    \midrule
    Shown suggestions & $+2.09 \pm 11.55$ & 0.7226 \\
    Accepted suggestions & $-4.66 \pm 11.87$ & 0.4410 \\
    Dismissed suggestions & $+14.41 \pm 25.86$ & 0.2748 \\
    Ignored suggestions & $+3.76 \pm 11.53$ & 0.5231 \\
    ARC at 30 seconds & $-2.79 \pm 11.44$ & 0.6321 \\
    ARC at 5 minutes & $-3.12 \pm 11.80$ & 0.6043 \\
    Recurring engagement & $+0.75 \pm 6.00$ & 0.8068 \\
    Time to display & $\mathbf{-8.90 \pm 2.66}$ & $\approx 0$ \\
    Time to produce & $\mathbf{-9.85 \pm 2.42}$ & $\approx 0$ \\
    Output tokens & $+0.67 \pm 3.94$ & 0.7392 \\
    Prompt tokens & $+0.17 \pm 1.12$ & 0.7659 \\
    Cached suggestion share & $\mathbf{+4.39 \pm 1.35}$ & $\approx 0$ \\
    \bottomrule
  \end{tabular}
\end{table}

Removing the existing 200-millisecond cache delay configuration reduces mean time to display by 8.90\% and mean time to produce by 9.85\%. Both reductions are statistically significant.
The reduction in time to display is consistent with removing the delay before presenting cached suggestions.
The relative change in dismissed suggestions is $+14.41\% \pm 25.86\%$ and is not statistically significant ($p=0.2748$).
Removing the delay reduces latency but does not produce statistically significant improvements in suggestion quality or recurring engagement.
We retain the existing 200-millisecond delay in the final configuration.

\subsection{Results}

\subsubsection{Training Dynamics}

\begin{figure}[!b]
  \centering
  \includegraphics[width=\textwidth]{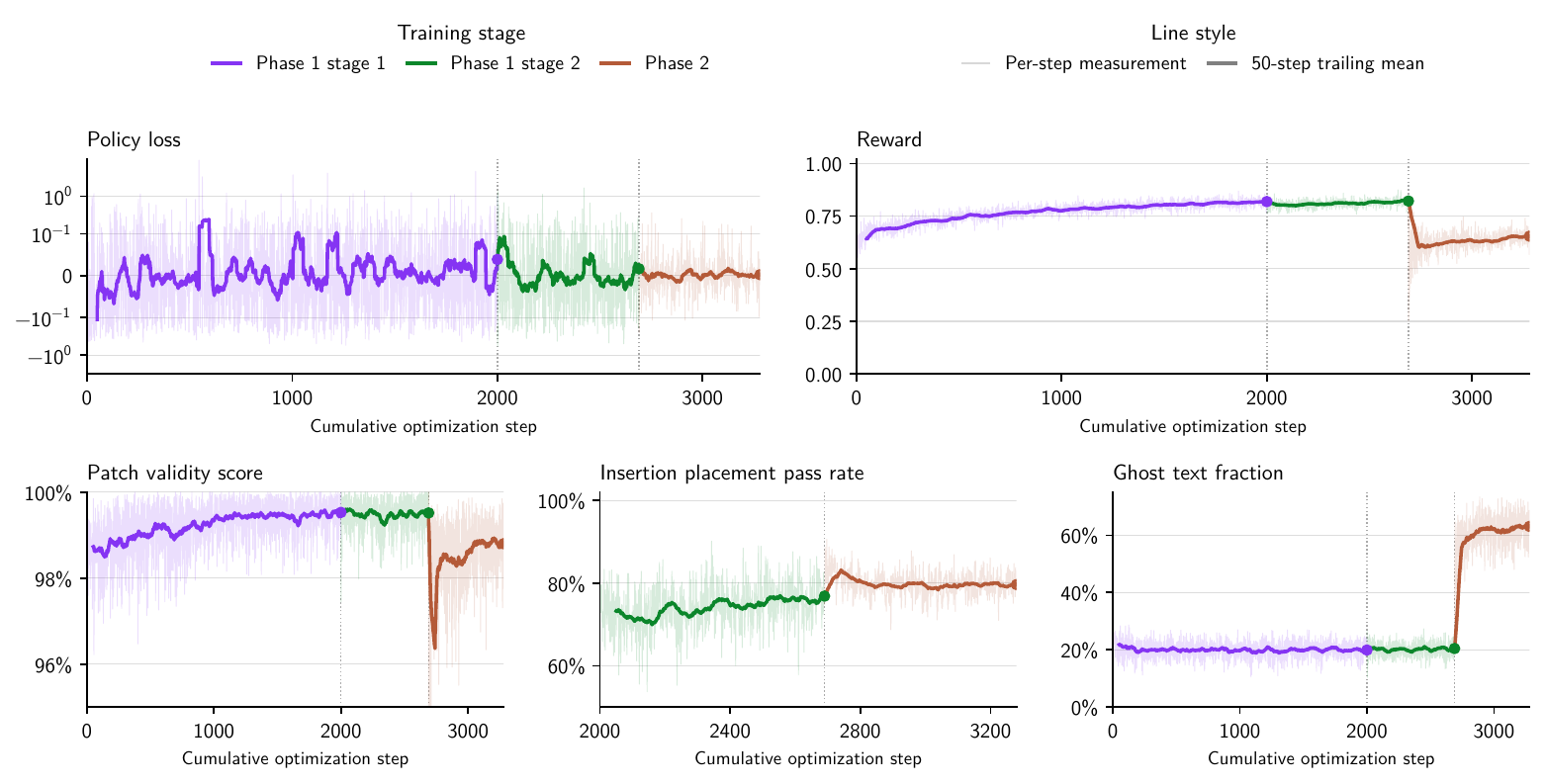}
  \caption{Learning curves of training-time metrics on sampled training data across three successive training stages. Phase~1 stage~2
  improves insertion placement, while Phase~2 increases ghost text generation.
  Phase~2 changes the task mixture and reward criterion, so its lower starting
  reward is not directly comparable with Phase~1.}
  \label{fig:phase1-phase2-training-curves}
\end{figure}

Figure~\ref{fig:phase1-phase2-training-curves} separates gains in overall reward from changes in specific editing behaviors across all three training stages: Phase~1 stage~1, Phase~1 stage~2, and Phase~2, all measured during training. Note that training experiences high variance from step to step and we take these measurements on training data rather than held-out evaluation data. We show these results to illustrate metric trends in training that corroborate evaluation results and the purpose of each training stage.
Each stage starts from the preceding stage's selected checkpoint, giving boundaries at 2{,}000 and 2{,}690 and a final checkpoint at 3{,}280 combined steps.

Phase~1 stage~1 accounts for the largest within-stage reward increase, from 0.642 to 0.819.
Stage~2 adds a smaller overall reward gain, from 0.803 to 0.822, while the insertion placement pass rate (Section~\ref{sec:phase1-grader}) improves from 73.2\% to 76.9\% and average patch validity remains near 99.5\%.

Phase~2 changes both the training task mixture (Section~\ref{sec:phase2-data-mixture}) and the reward criterion (Section~\ref{sec:phase2-training-recipe}) by emphasizing ghost text and adding the ghost-text completeness requirement.
The lower reward at the stage boundary therefore reflects these changes to the training setting, rather than a like-for-like comparison of model quality.
Within Phase~2, reward increases from 0.602 to 0.655 and patch validity recovers from 96.4\% to 98.8\%.
The ghost text fraction rises to 63.1\%, compared with 20.4\% at the end of Phase~1 stage~2, showing a shift toward ghost text suggestions.

\subsubsection{Offline Benchmarks}
Table~\ref{tab:phase2-offline-results} compares the 3-in-1 release candidate with the 2-in-1 RL stage~2 checkpoint using the same benchmark versions and the metrics in Table~\ref{tab:phase1-offline-results}.
The comparison measures completion capability on the three Augmented HumanEval subsets, scenario coverage and avoidance of undesirable suggestions on Simulation Tests, mean patches per response on POE, recurring failure modes on QMH, and no-edit and ghost text ratios on Output View Kind v1.
The baseline is the 2-in-1 model alone, not the 2-in-1 model paired with the completion model.
The online experiments below evaluate replacement of that paired configuration.

\begin{table}[htbp]
  \caption{Phase~2 offline comparison against the standalone 2-in-1 RL stage~2 checkpoint. The 3-in-1 model improves completion accuracy on all three Augmented HumanEval subsets and achieves the desired higher ghost text ratio, while matching or improving Simulation Tests scores. QMH passes decrease from 19/19 to 18/19.}
  \label{tab:phase2-offline-results}
  \centering
  \footnotesize
  \setlength{\tabcolsep}{2.6pt}
  \resizebox{\textwidth}{!}{%
  \begin{tabular}{lrrrrrrrrrrr}
    \toprule
    & \multicolumn{2}{c}{STests v1} & \multicolumn{2}{c}{STests v2}
    & \multicolumn{3}{c}{Augmented HumanEval} & \multicolumn{1}{c}{POE}
    & \multicolumn{1}{c}{QMH} & \multicolumn{2}{c}{Output View Kind v1} \\
    \cmidrule(lr){2-3}\cmidrule(lr){4-5}\cmidrule(lr){6-8}
    \cmidrule(lr){9-9}\cmidrule(lr){10-10}\cmidrule(lr){11-12}
    Model & MustAvoid & Total & MustAvoid & Total
          & Single & Multi & Span & Mean patches & Passes & No-edit ratio & Ghost text ratio \\
    \midrule
    2-in-1 &
      91.2\% & 51.7/77 & 89.1\% & 45.6/79 &
      92.0 & 75.9 & 84.2 & 1.30 & 19/19 & 55.46\% & 47.47\% \\
    3-in-1 &
      92.0\% & 52.8/77 & 89.1\% & 48.5/79 &
      93.7 & 82.3 & 88.0 & 1.40 & 18/19 & 47.10\% & 72.51\% \\
    \bottomrule
  \end{tabular}%
  }
\end{table}

Relative to the 2-in-1 model alone, the release candidate achieves higher accuracy across all three Augmented HumanEval subsets.
Single-Line accuracy increases from 92.0\% to 93.7\%, Multi-Line accuracy from 75.9\% to 82.3\%, and Random Span accuracy from 84.2\% to 88.0\%.
The ghost text ratio also increases from 47.47\% to 72.51\%, while the no-edit ratio decreases from 55.46\% to 47.10\%.
However, the 3-in-1 model passes 18/19 QMH cases compared with 19/19 for the 2-in-1 model due to a code duplication failure case that persists across all evaluated 3-in-1 checkpoints.
We use QMH to screen and select checkpoints, so its scores are not independent evidence of overall quality.
We nevertheless review this failure case and judge its impact limited based on subsequent qualitative evaluations and internal testing, accepting this tradeoff for improved completion accuracy.
These results show improved completion accuracy on this benchmark and a shift toward completion-shaped suggestions, not a quality improvement over the paired completion-and-2-in-1 system.

\subsubsection{Online Evaluation}
\label{sec:phase2-online-evaluation}

We evaluate the final 3-in-1 checkpoint in two online A/B experiments: a model-only comparison and a model-plus-client comparison with all four client adaptations applied.
The final 3-in-1 model with all four client adaptations is the deployed system.
Both experiments use the 2-in-1 model paired with the completion model as the control and follow the randomized user assignment and analysis procedure described in Section~\ref{sec:online-experiment-design}.
Neither experiment compares the deployed system with the original three-model baseline, so these results do not measure the cumulative effect of both phases.
The model-plus-client comparison uses the same suppression-enabled treatment group and exposure period as the cross-mode suppression ablation (Table~\ref{tab:phase2-suppression-ablation}), but compares that group with the paired-model control rather than the suppression-disabled configuration.
Table~\ref{tab:phase2-online-results-3-in-1} reports each experiment's relative changes with 95\% confidence intervals and $p$-values against its control.
As in Phase~1, we assess suggestion quality through shown, accepted, and dismissed suggestions, recurring engagement, and ARC.

\begin{table}[htbp]
  \caption{Phase~2 online results from two experiments, each relative to the 2-in-1 model paired with the completion model. The model-plus-client experiment applies all four client adaptations.}
  \label{tab:phase2-online-results-3-in-1}
  \centering
  \small
  \begin{tabular}{lrrrr}
    \toprule
    & \multicolumn{2}{c}{Model only}
    & \multicolumn{2}{c}{Model + client} \\
    \cmidrule(lr){2-3}\cmidrule(lr){4-5}
    Metric & Rel.\ change (\%) & $p$-value & Rel.\ change (\%) & $p$-value \\
    \midrule
    Shown suggestions & $\mathbf{+7.73 \pm 6.85}$ & 0.0270 & $+4.78 \pm 6.22$ & 0.1326 \\
    Accepted suggestions & $+2.92 \pm 7.50$ & 0.4447 & $+4.43 \pm 7.41$ & 0.2408 \\
    Dismissed suggestions & $\mathbf{+12.24 \pm 12.20}$ & 0.0492 & $\mathbf{-10.06 \pm 9.85}$ & 0.0453 \\
    Ignored suggestions & $\mathbf{+9.31 \pm 6.91}$ & 0.0082 & $+5.64 \pm 6.05$ & 0.0675 \\
    ARC at 30 seconds & $\mathbf{-13.27 \pm 7.45}$ & 0.0005 & $-3.25 \pm 5.67$ & 0.2613 \\
    ARC at 5 minutes & $\mathbf{-11.56 \pm 7.83}$ & 0.0038 & $-2.42 \pm 6.00$ & 0.4288 \\
    Recurring engagement & $+2.04 \pm 3.96$ & 0.3134 & $\mathbf{+2.93 \pm 2.50}$ & 0.0217 \\
    Time to display & $\mathbf{-10.82 \pm 1.59}$ & $\approx 0$ & $\mathbf{-11.00 \pm 1.12}$ & $\approx 0$ \\
    Time to produce & $\mathbf{-5.63 \pm 1.48}$ & $\approx 0$ & $\mathbf{-5.18 \pm 1.04}$ & $\approx 0$ \\
    Output tokens & $\mathbf{+19.62 \pm 1.77}$ & $\approx 0$ & $\mathbf{+20.98 \pm 1.43}$ & $\approx 0$ \\
    Prompt tokens & $\mathbf{+1.79 \pm 0.79}$ & $\approx 0$ & $+0.43 \pm 0.65$ & 0.1942 \\
    Cached suggestion share & $\mathbf{+13.60 \pm 0.91}$ & $\approx 0$ & $\mathbf{+9.38 \pm 0.87}$ & $\approx 0$ \\
    \bottomrule
  \end{tabular}
\end{table}

The model-only experiment improves latency: mean time to display and mean time to produce decrease by 10.82\% and 5.63\%, respectively, relative to the paired control.
However, dismissed suggestions increase by 12.24\%, and ARC decreases by 13.27\% at 30 seconds and 11.56\% at 5 minutes.
All of these changes are statistically significant.
With all four client adaptations also applied, dismissed suggestions decrease by 10.06\% (unadjusted $p = 0.0453$), while mean time to display and mean time to produce decrease by 11.00\% and 5.18\%, respectively.
The latency reductions remain statistically significant after Bonferroni correction across the 12 reported metrics, but the dismissal reduction does not (see Section~\ref{sec:limitations}).

In the model-plus-client experiment, ARC changes by $-3.25\%$ at 30 seconds and $-2.42\%$ at 5 minutes; neither change is statistically significant (Table~\ref{tab:phase2-online-results-3-in-1}).
ARC is a character count, so longer suggestions can increase ARC even if the fraction of accepted characters retained stays the same.
The number of accepted suggestions also does not differ significantly between the treatment and control; these results therefore do not establish an overall improvement in response quality.

Output tokens per response increase by 19.62\% in the model-only experiment and 20.98\% in the model-plus-client experiment relative to the 2-in-1 model alone; both increases are statistically significant.
As discussed in Section~\ref{sec:online-metrics}, this metric averages tokens per response within each user, then across users, so the treatment model handling more requests does not by itself increase it.
This increase is consistent with our training emphasis on longer, more complete ghost text suggestions (Section~\ref{sec:phase2-accepted-retained-character-data}), but increases decoding work per model call.

As noted in Section~\ref{sec:online-experiment-design}, we do not directly multiply relative changes across the two phases to estimate a cumulative effect, since the experiments are not directly comparable.
Together with the 58.93\% output-token reduction in Phase~1, these results suggest that the 3-in-1 model emits fewer tokens per response than the original NES model, though we do not directly measure this comparison.

Cached suggestion share increases by 13.60\% in the model-only experiment and 9.38\% in the model-plus-client experiment.
Displaying cached suggestions requires no additional model call, but a higher cache share does not necessarily indicate better suggestion quality.
For example, cross-mode suppression of ignored suggestions reduces both cached suggestion share and dismissals (Table~\ref{tab:phase2-suppression-ablation}).

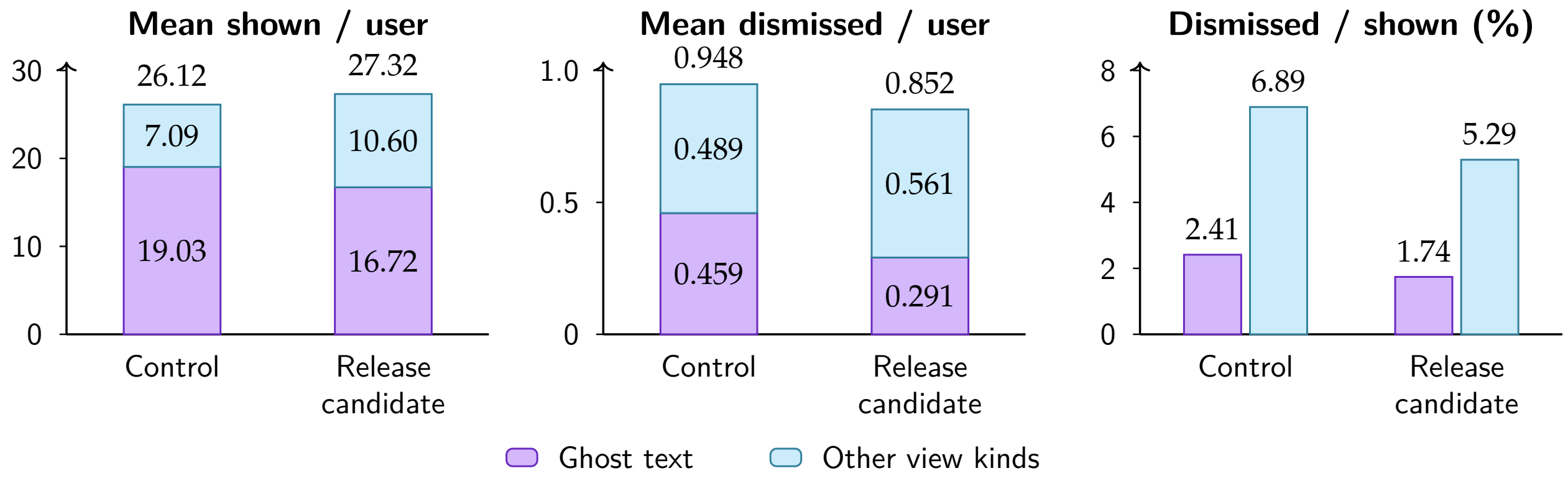
\begin{figure}[htbp]
  \centering
  \resizebox{\linewidth}{!}{\begin{tikzpicture}[
  font=\sffamily\small,
  text=black,
  ghost/.style={fill=ghpurple!35,draw=ghpurple!80!black,line width=0.6pt},
  other/.style={fill=cyan!18,draw=cyan!55!black,line width=0.6pt},
  axis/.style={draw=black,line width=0.7pt},
  tick/.style={draw=black,line width=0.5pt}
]
  \begin{scope}[x=1cm,y=0.1cm]
    \node[font=\sffamily\bfseries] at (2.4,35) {Mean shown / user};
    \draw[axis,->] (0,0) -- (0,31);
    \draw[axis] (0,0) -- (4.8,0);
    \foreach \v in {0,10,20,30} {
      \draw[tick] (-0.08,\v) -- (0,\v);
      \node[anchor=east] at (-0.14,\v) {\v};
    }
    \foreach \x/\ghost/\other/\total/\name in {
      1.2/19.0319/7.0878/26.1197/Control,
      3.6/16.7167/10.6044/27.3211/Release candidate} {
      \draw[ghost] (\x-0.55,0) rectangle (\x+0.55,\ghost);
      \draw[other] (\x-0.55,\ghost) rectangle (\x+0.55,\total);
      \node at (\x,\ghost/2) {\pgfmathprintnumber[fixed,fixed zerofill,precision=2]{\ghost}};
      \node at (\x,{\ghost+\other/2}) {\pgfmathprintnumber[fixed,fixed zerofill,precision=2]{\other}};
      \node[anchor=south] at (\x,\total+0.7) {\pgfmathprintnumber[fixed,fixed zerofill,precision=2]{\total}};
      \node[anchor=north,align=center,text width=2.4cm] at (\x,-1) {\name};
    }
  \end{scope}
  \begin{scope}[shift={(6.1cm,0)},x=1cm,y=3cm]
    \node[font=\sffamily\bfseries] at (2.4,1.1667) {Mean dismissed / user};
    \draw[axis,->] (0,0) -- (0,1.0333);
    \draw[axis] (0,0) -- (4.8,0);
    \foreach \v in {0,0.5,1.0} {
      \draw[tick] (-0.08,\v) -- (0,\v);
      \node[anchor=east] at (-0.14,\v) {\v};
    }
    \foreach \x/\ghost/\other/\total/\name in {
      1.2/0.4594/0.4885/0.9479/Control,
      3.6/0.2910/0.5614/0.8524/Release candidate} {
      \draw[ghost] (\x-0.55,0) rectangle (\x+0.55,\ghost);
      \draw[other] (\x-0.55,\ghost) rectangle (\x+0.55,\total);
      \node at (\x,\ghost/2) {\pgfmathprintnumber[fixed,precision=3]{\ghost}};
      \node at (\x,{\ghost+\other/2}) {\pgfmathprintnumber[fixed,precision=3]{\other}};
      \node[anchor=south] at (\x,\total+0.015) {\pgfmathprintnumber[fixed,precision=3]{\total}};
      \node[anchor=north,align=center,text width=2.4cm] at (\x,-0.0333) {\name};
    }
  \end{scope}
  \begin{scope}[shift={(12.2cm,0)},x=1cm,y=0.375cm]
    \node[font=\sffamily\bfseries] at (2.4,9.3333) {Dismissed / shown (\%)};
    \draw[axis,->] (0,0) -- (0,8.2667);
    \draw[axis] (0,0) -- (4.8,0);
    \foreach \v in {0,2,4,6,8} {
      \draw[tick] (-0.08,\v) -- (0,\v);
      \node[anchor=east] at (-0.14,\v) {\v};
    }
    \foreach \x/\ghost/\other/\name in {
      1.2/{100*0.4594/19.0319}/{100*0.4885/7.0878}/Control,
      3.6/{100*0.2910/16.7167}/{100*0.5614/10.6044}/Release candidate} {
      \pgfmathsetmacro{\g}{\ghost}
      \pgfmathsetmacro{\o}{\other}
      \draw[ghost] (\x-0.7,0) rectangle (\x-0.05,\g);
      \draw[other] (\x+0.05,0) rectangle (\x+0.7,\o);
      \node[anchor=south] at (\x-0.375,\g+0.1) {\pgfmathprintnumber[fixed,fixed zerofill,precision=2]{\g}};
      \node[anchor=south] at (\x+0.375,\o+0.1) {\pgfmathprintnumber[fixed,fixed zerofill,precision=2]{\o}};
      \node[anchor=north,align=center,text width=2.4cm] at (\x,-0.2667) {\name};
    }
  \end{scope}
  \draw[ghost,rounded corners=2pt] (5,-1.5) rectangle (5.35,-1.3);
  \node[anchor=west] at (5.45,-1.4) {Ghost text};
  \draw[other,rounded corners=2pt] (8,-1.5) rectangle (8.35,-1.3);
  \node[anchor=west] at (8.45,-1.4) {Other view kinds};
\end{tikzpicture}}
  \caption{Suggestions by view kind for the 2-in-1-plus-completion control and
  the 3-in-1 release candidate with all four client adaptations applied, with counts reported as per-user means.
  The release candidate and client configuration together show fewer ghost text suggestions per user and lower dismissal rates in both view kind groups.}
  \label{fig:view-kind-dismissals}
\end{figure}

Figure~\ref{fig:view-kind-dismissals} shows suggestion counts per user and dismissal rates grouped by view kind for the model-plus-client experiment.
Dismissal rate is the fraction of shown suggestions that developers dismiss.
Ghost text includes suggestions at and away from the cursor, while other view kinds are non-ghost-text NES suggestions.
These UI-specific metrics count events rather than the unique opportunities counted in Table~\ref{tab:phase2-online-results-3-in-1}.
Ghost text has a lower dismissal rate than other view kinds in both the control (2.41\% versus 6.89\%) and the release candidate (1.74\% versus 5.29\%).
This pattern is consistent with the less intrusive ghost text UI, although suggestion content also differs.

The 3-in-1 model replaces both models in the control and must therefore provide the ghost text suggestions that the separate completion model supplies in the control.
Compared with the control, the release candidate shows 12.16\% fewer ghost text suggestions per user and reduces overall dismissals per user by 10.06\% (Table~\ref{tab:phase2-online-results-3-in-1}).
Although ghost text has a lower dismissal rate than other view kinds in the UI event counts (Figure~\ref{fig:view-kind-dismissals}), the reduction in dismissal events comes from fewer ghost text dismissals, not from showing more ghost text.
The decrease in ghost text dismissals more than offsets the increase in dismissals of other suggestions.
For other view kinds, the release candidate shows more suggestions, so dismissals per user rise even though their dismissal rate falls.
These measurements describe dismissal behavior for the combined model-and-client configuration; they do not isolate an improvement in the predicted edits themselves.

Note that the offline evaluation compares the 3-in-1 model with the 2-in-1 model alone (Table~\ref{tab:phase2-offline-results}).
The online experiment instead compares it with the 2-in-1 model paired with the completion model; the latter produces only ghost text suggestions.
The offline increase in ghost text ratio and online decrease in ghost text suggestions per user are therefore measured against different baselines.

\subsubsection{Qualitative Perception Survey Results}
\label{sec:phase2-perception-survey}

\begin{figure}[t]
  \centering
  \begin{subfigure}[t]{0.49\textwidth}
    \centering
    \resizebox{\linewidth}{!}{\begin{tikzpicture}[
  font=\sffamily\small,
  legend swatch/.style={rounded corners=2pt, minimum width=0.28cm, minimum height=0.2cm, inner sep=0pt, line width=0.6pt},
  legend text/.style={anchor=west, text=black},
  legend value/.style={anchor=east, text=black}
]
  \path[fill=ghpurple!35] (0,2.7) rectangle (0.28,3.25);
  \path[fill=ghpurple!10] (0.28,2.7) rectangle (0.84,3.25);
  \path[fill=black!8] (0.84,2.7) rectangle (7,3.25);
  \draw[black, line width=0.7pt] (0,2.7) rectangle (7,3.25);
  \draw[black, line width=0.5pt] (0.28,2.7) -- (0.28,3.25);
  \draw[black, line width=0.5pt] (0.84,2.7) -- (0.84,3.25);
  \node[text=black] at (3.92,2.975) {88\%};

  \node[legend swatch, draw=ghpurple!80!black, fill=ghpurple!35] at (0.14,2.15) {};
  \node[legend text] at (0.42,2.15) {Much better};
  \node[legend value] at (7,2.15) {1 (4\%)};

  \node[legend swatch, draw=ghpurple, fill=ghpurple!10] at (0.14,1.68) {};
  \node[legend text] at (0.42,1.68) {Slightly better};
  \node[legend value] at (7,1.68) {2 (8\%)};

  \node[legend swatch, draw=black, fill=black!8] at (0.14,1.21) {};
  \node[legend text, align=left] at (0.42,1.21) {Same, or did not notice a\\difference};
  \node[legend value] at (7,1.21) {22 (88\%)};

  \node[legend swatch, draw=cyan!55!black, fill=cyan!18] at (0.14,0.58) {};
  \node[legend text] at (0.42,0.58) {Slightly worse};
  \node[legend value] at (7,0.58) {0 (0\%)};

  \node[legend swatch, draw=cyan!75!black, fill=cyan!35] at (0.14,0.12) {};
  \node[legend text] at (0.42,0.12) {Much worse};
  \node[legend value] at (7,0.12) {0 (0\%)};
\end{tikzpicture}}
    \caption{Perceived quality (25 responses).}
  \end{subfigure}\hfill
  \begin{subfigure}[t]{0.49\textwidth}
    \centering
    \resizebox{\linewidth}{!}{\begin{tikzpicture}[
  font=\sffamily\small,
  axis/.style={draw=black, line width=0.6pt},
  count/.style={text=black, anchor=south, align=center},
  category/.style={text=black, align=center, anchor=north}
]
  \draw[axis] (0.9,0.8) -- (7,0.8);
  \draw[axis] (0.9,0.8) -- (0.9,3.1);
  \draw[axis] (0.84,0.8) -- (0.9,0.8) node[left=3pt, text=black] {0};
  \draw[axis] (0.84,1.8) -- (0.9,1.8) node[left=3pt, text=black] {5};
  \draw[axis] (0.84,2.8) -- (0.9,2.8) node[left=3pt, text=black] {10};

  \filldraw[draw=black, fill=black!8] (1.2,0.8) rectangle (2.0,1.4);
  \filldraw[draw=black, fill=black!8] (2.85,0.8) rectangle (3.65,1.6);
  \filldraw[draw=black, fill=black!8] (4.5,0.8) rectangle (5.3,2.8);
  \filldraw[draw=black, fill=black!8] (6.15,0.8) rectangle (6.95,1.2);

  \node[count] at (1.6,1.4) {3\\(15.8\%)};
  \node[count] at (3.25,1.6) {4\\(21.1\%)};
  \node[count] at (4.9,2.8) {10\\(52.6\%)};
  \node[count] at (6.55,1.2) {2\\(10.5\%)};

  \node[category] at (1.6,0.68) {Less than\\1 day};
  \node[category] at (3.25,0.68) {1 or 2\\days};
  \node[category] at (4.9,0.68) {3 or 4\\days};
  \node[category] at (6.55,0.68) {5 days\\or more};
\end{tikzpicture}}
    \caption{Days of use per week (19 responses).}
  \end{subfigure}
  \caption{Perceived quality and days of use among survey
  respondents in the 3-in-1 treatment group. Respondents report either no change or an improvement
  relative to their recollection of the previous 2-in-1-plus-completion setup.}
  \label{fig:phase2-perception-survey-results}
\end{figure}
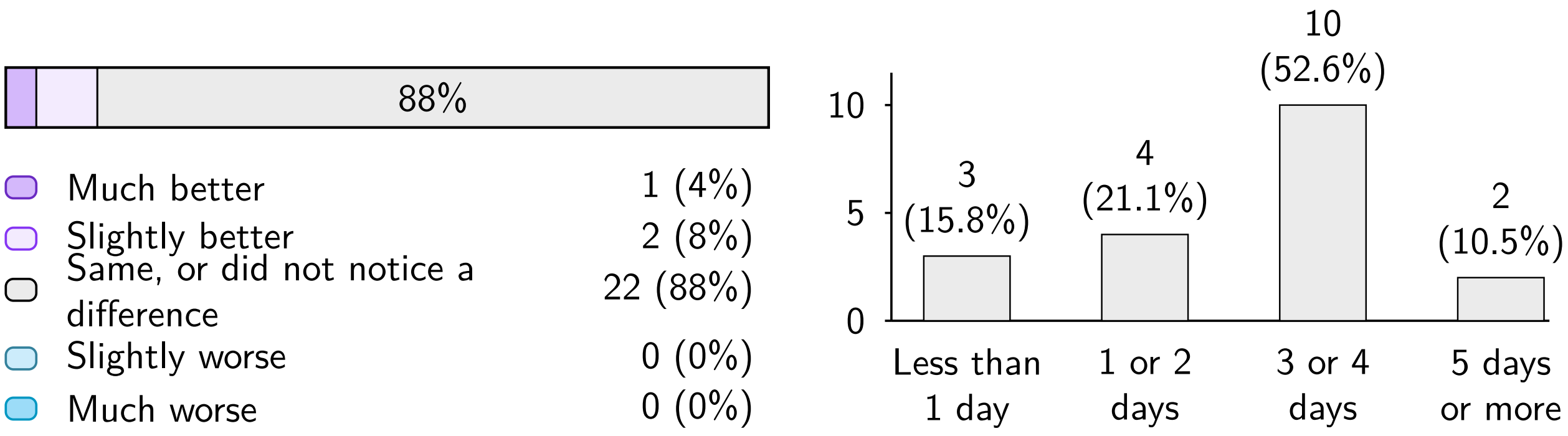

We also conduct a retrospective perception survey during the internal A/B experiment of the 3-in-1 model candidate. We send the survey only to users assigned to the treatment configuration. The survey targets approximately 200 of the most active users of GitHub Copilot Inline Suggestions over one week, from July 29, 2026, through August 4, 2026. We ask respondents how the quality of inline suggestions during that week compares with the weeks before July 29, 2026. The questionnaire also asks about satisfaction, desired improvements and problematic suggestions, typical days of use, and active hours of use. Table~\ref{tab:phase2-perception-survey-questions} lists the five questions in more detail. Figure~\ref{fig:phase2-perception-survey-results} shows the perceived quality responses and days of use per week that respondents report ($n=25$ in total and $n=19$ for the usage question).

\begin{table}[t]
  \caption{Perception survey questionnaire for the 3-in-1 treatment group.}
  \label{tab:phase2-perception-survey-questions}
  \centering
  \small
  \begin{tabular}{@{}r>{\raggedright\arraybackslash}p{0.54\linewidth}>{\raggedright\arraybackslash}p{0.32\linewidth}@{}}
    \toprule
    & Question & Response format \\
    \midrule
    1 & How satisfied are you with the quality of inline suggestions in the past week (07/29 - 08/04, last Wednesday to this Tuesday)?
      & Numerical rating, 1 to 5 \\
    \addlinespace
    2 & How does the quality of inline suggestions in the past week (07/29 - 08/04) compare with previous weeks (before 07/29)?
      & Much better; slightly better; same or no noticed difference; slightly worse; much worse \\
    \addlinespace
    3 & What improvements would you like to see for inline suggestions? (e.g., what are some common bad suggestions that you've encountered lately?)
      & Free text \\
    \addlinespace
    4 & In a typical week, on how many days do you use inline suggestions while coding?
      & Less than 1 day; 1 to 2 days; 3 to 4 days; 5 days or more \\
    \addlinespace
    5 & On days when you use inline suggestions, approximately how many hours do you actively use them?
      & Less than 1 hour; 1 to 2 hours; 2 to 4 hours; 4 to 6 hours; more than 6 hours \\
    \bottomrule
  \end{tabular}
\end{table}

The comparison item asks respondents to compare the A/B experiment week with their recollection of previous weeks, not with a blind side-by-side alternative or a task-based evaluation. Respondents know they are participating in an internal online A/B experiment. The 25 responses represent approximately 12.5\% of the targeted users, and respondents self-select; thus, recall, awareness of participation in the experiment, and nonresponse may affect the results.

Among the 25 responses, 12\% (3 respondents) report a perceived improvement in quality, and the remaining 88\% (22 respondents) report no perceived difference in quality. No respondents report a perceived decline in quality. Under an independent, common-probability binomial model, the two-sided exact Clopper-Pearson 95\% confidence interval for the improvement share is 2.5\% to 31.2\%. With zero reports of decline, the one-sided exact 95\% upper confidence bound on the decline share is 11.3\%. Among the 19 respondents who report their days of use per week, 63.2\% (12 respondents) indicate using inline suggestions at least 3 days per week, while the remaining 36.8\% (7 respondents) report using it fewer than 3 days per week. Most respondents perceive no quality difference. These self-selected responses do not establish population-level equivalence or improvement in response quality.

Sixteen of the twenty-five respondents answer the free-text question about desired improvements and problematic suggestions. Because the question solicits problems, the comments may overrepresent negative experiences. In the following analysis, one author assigns one or more themes to each response without a second rater, so theme counts are indicative. Six respondents describe interaction friction, including suggestions intercepting Tab or Enter, appearing during typing, or resisting dismissal; these comments concern editor integration. Three describe Markdown or prose suggestions, and three request editor features such as per-language controls and MCP support. Both groups of comments concern areas outside the model behavior evaluated here.

Two respondents report inconsistent coverage, including missing suggestions for apparent edits or when repeating an edit while moving upward through a file. Two describe repetitive suggestions, and one describes suggestions that restore text after they delete it. These comments point to opportunities for further study, but the questionnaire does not measure quality, latency, and rendering separately, so we cannot attribute perceived changes to a single intervention.

Three respondents praise suggestion quality. The respondent who selects ``Much better'' describes improved SQL dialect correction during the A/B experiment week and says the behavior later reverts.

\section{Iteration Flywheel}
\subsection{Iteration Velocity}
One critical result of the experimentation we conduct in developing the unified inline suggestions model is the compounding benefit of learnings that decrease iteration time in each phase.
The 2-in-1 model requires 15 SFT training runs, 173 RL training runs, and several times as many checkpoint-selection evaluation runs. We evaluate 14 candidates in A/B experiments before selecting the final shipping candidate.
In contrast, the 3-in-1 model requires only RL on top of the 2-in-1 model, with 35 RL training runs and 10 A/B experiment candidates.
Over the same period, we also merge 113 PRs to the VS Code client to improve and support the inline suggestions experience.
Each modeling phase benefits from and builds on our earlier learnings, which accelerates the iteration process.
Table~\ref{tab:phase1-phase2-development-iteration-counts} summarizes the training runs, A/B experiment candidates, and client-side PRs by phase.

\begin{table}[htbp]
  \caption{Development iterations by phase. Building on the 2-in-1 model, Phase~2 requires no SFT runs, about one-fifth as many RL runs, and fewer A/B experiment candidates and client-side PRs.}
  \label{tab:phase1-phase2-development-iteration-counts}
  \centering
  \begin{tabular}{lrrrr}
    \toprule
    Phase & SFT runs & RL runs & A/B experiment candidates & Client-side PRs \\
    \midrule
    Phase~1 (2-in-1) & 15 & 173 & 14 & 96 \\
    Phase~2 (3-in-1) & 0 & 35 & 10 & 17 \\
    \bottomrule
  \end{tabular}
\end{table}

Here, we share additional analyses of tools that we introduce in this work to illustrate the importance of building strong evaluation components that accelerate this iteration flywheel.

\subsection{A/B Experiment Metrics over Model Iterations}
\begin{figure}[!ht]
  \centering
  \resizebox{\linewidth}{!}{\input{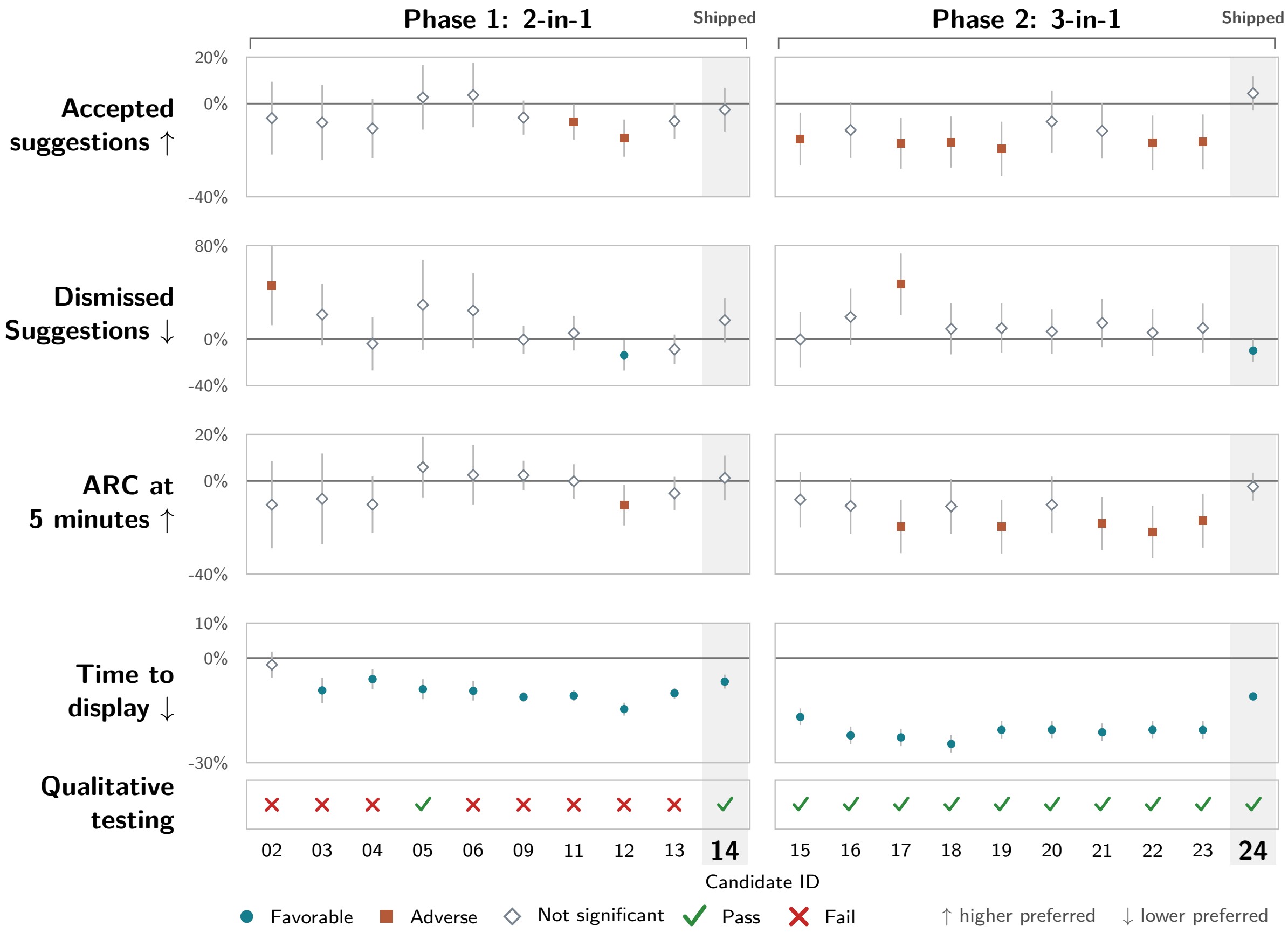}}
  \caption{Key online metric relative changes compared to the control (expressed in \% with 95\% confidence intervals) and qualitative assessments across model candidates. ARC uses the 5-minute retention window. Controls are the three standalone models in Phase~1 and the 2-in-1-plus-completion model in Phase~2.
  Multiple rounds of iteration on data, model training, offline and online evaluation methods, client implementation, and other factors in each phase produce the final shipped configurations that meet both quantitative and qualitative quality standards.}
  \label{fig:phase1-phase2-candidate-online-metrics}
\end{figure}

Figure~\ref{fig:phase1-phase2-candidate-online-metrics} shows four online metrics and qualitative regression assessments for candidates whose A/B experiments run to completion, ordered chronologically within each phase.
Phase~1 uses the three-model baseline as its control, while Phase~2 uses the 2-in-1 model (candidate 14 in Phase~1 of Figure~\ref{fig:phase1-phase2-candidate-online-metrics}) paired with completion as its control.
We omit candidates 1, 7, 8, and 10 because we stop their experiments early, in most cases after interim results show clear regressions.
Section~\ref{sec:limitations} discusses how selecting candidates by their A/B results affects the reported estimates.
For the purpose of this analysis, we consider changes to be statistically significant if $p < 0.05$.
We share this visual of our model development process to transparently communicate the learning process and dynamic nature of scientific inquiry alongside the end results and successful recipes.

For model unification, our primary goal is to maintain performance relative to the control rather than to outperform it.
As described in Section~\ref{sec:online-metrics}, a non-significant result alone does not establish that a candidate is acceptable.
We also defer deployment when confidence intervals or qualitative testing suggest potential regressions.
Most Phase~1 candidates that do not meet the shipping bar exhibit qualitative regressions, including malformed patches, over-eager edits, code duplication, or misplaced insertions. Most Phase~2 candidates that do not meet the shipping bar exhibit statistically significant decreases in accepted suggestions or accepted and retained characters (ARC).
These complementary signals guide five iterations of model development across the two unification phases \citep{gong2026inlineone,gong2026inlinetwo}. While we do not elaborate on many of the 113 client-side PRs that complement the model improvements in this work to improve the user experience, it is important to recognize that these changes are foundational to the improvements we make during the model iterations.

We note that successive candidates test different, sometimes independent, changes rather than a single cumulative sequence.
Each selected configuration combines changes we retain from earlier experiments, so the figure does not isolate the effect of any individual training change.
Comparisons across experiments are descriptive because their exposure periods and audiences differ.

\paragraph{Phase 1 Iteration.}
The first model iteration (candidates 02 to 03 in Figure~\ref{fig:phase1-phase2-candidate-online-metrics}) establishes the baseline performance of using only single-edit data from the NES recipe, as we discuss in Section~\ref{sec:phase1-rl-data}, and identifies initial failure modes such as the code duplication we describe in Section~\ref{sec:phase1-data-quality-filtering}.
A subsequent iteration (candidates 04 to 06) introduces multi-edit supervision through the pseudolabeling and edit history playback pipelines in Section~\ref{sec:phase1-data-bootstrapping}, along with the rules-based multi-edit RL grader bonus we discuss in Section~\ref{sec:phase1-patch-count-bonuses}.
Online results improve, but elevated dismissals and qualitative failures motivate the POE offline benchmark we introduce in Section~\ref{sec:offline-benchmarks}, grader calibration as we describe in Section~\ref{sec:phase1-grader-calibration}, and targeted patch-validity repairs using the filtering and repair process in Section~\ref{sec:phase1-data-quality-filtering}.
The next iteration of model refinement (candidates 09 to 14), which produces the final Phase~1 candidate, addresses the remaining insertion placement failures through the targeted-example pipeline in Section~\ref{sec:phase1-targeted-behavior-curation}, hard-example upweighting in Section~\ref{sec:phase1-rl-data}, and RL grader boosting in Section~\ref{sec:phase1-grader}.
These failures also motivate the QMH canary benchmark we introduce in Section~\ref{sec:offline-benchmarks}, which supports the checkpoint selection process in Section~\ref{sec:model-selection-process} by screening for behaviors that aggregate online A/B experiment metrics miss.

\paragraph{Phase 2 Iteration.}
In Phase~2, where the focus shifts to integrating completion data and refining client-side behavior, candidates 15 to 24 use the completion data we describe in Section~\ref{sec:phase2-data-mixture}.
All of these candidates also use the upweighted longer-output examples in Section~\ref{sec:phase2-accepted-retained-character-data} and the ghost text completeness auxiliary grader in Section~\ref{sec:phase2-training-recipe} to target incomplete ghost text.
After we add the completion length examples and the completeness grader, the mean number of inserted characters per patch increases from 38.04 to 58.56 on Output View Kind v2 (Section~\ref{sec:output-view-kind}).
An online experiment with candidate 15 against the 2-in-1 model paired with the completion model reports a non-significant relative change in 30-second ARC of $-7.94\%$.
Figure~\ref{fig:phase1-phase2-candidate-online-metrics} shows the same candidate's 5-minute ARC result.
From candidate 16 onward, we also increase upsampling of the patch validity and insertion pattern propagation repair examples to maintain coverage of these failure modes as the mixture expands (Table~\ref{tab:phase2-data-upsampling}).
Candidates 17, 19, and 21 to 23 still show statistically significant regressions in 5-minute ARC (Figure~\ref{fig:phase1-phase2-candidate-online-metrics}) despite the data and grader interventions, suggesting that these interventions do not consistently eliminate ARC regressions.
Candidate 24 also adds the cursor-first multi-edit examples we describe in Section~\ref{sec:phase2-data-mixture} to encourage completion at the cursor before related follow-up edits.
In the model-only experiment, candidate 24 increases dismissed suggestions per user by 12.24\% and reduces ARC relative to control.
With all four client adaptations, candidate 24 instead reduces dismissed suggestions per user by 10.06\%.
Though these candidate comparisons do not isolate the contributions of individual training changes to ARC, the final configuration combines these data, grader, and client changes and shows no statistically significant ARC regression relative to the paired control (Table~\ref{tab:phase2-online-results-3-in-1}).
\FloatBarrier

\subsection{RL Checkpoint Stability}
Figure~\ref{fig:phase1-phase2-qmh-checkpoint-stability} shows QMH pass counts across checkpoints for the two RL runs that produce the shipped 2-in-1 and 3-in-1 models. Scores in these runs remain relatively stable compared with other experimental runs, which show greater variability and lower averages. Although all scores are at least $14/19$, they still span a wide range. Each run contains a contiguous sequence of checkpoints with scores of at least $18/19$. Both shipped models attain the maximum score in their respective runs. Our checkpoint expansion technique evaluates these models at a higher granularity in order to capture the best checkpoints in each run, as described in Section~\ref{sec:model-selection-process}. This checkpoint-level variation underscores the need for a lightweight canary benchmark that efficiently identifies local performance optima during checkpoint expansion.

\begin{center}
  \centering
  \begin{tikzpicture}[x=0.0105cm,y=0.38cm,font=\small]
  \foreach \score in {14,15,16,17,18,19} {
    \draw[gray!25,line width=0.4pt] (100,\score) -- (1300,\score);
    \draw[gray!70,line width=0.5pt] (92,\score) -- (100,\score);
    \node[anchor=east,text=gray!70] at (80,\score) {\score};
  }

  \draw[gray!75,line width=0.6pt] (100,14) -- (1300,14);
  \draw[gray!75,line width=0.6pt] (100,14) -- (100,19);
  \foreach \step in {100,300,...,1300} {
    \draw[gray!70,line width=0.5pt] (\step,14) -- (\step,13.82);
    \node[anchor=north,text=gray!70,font=\scriptsize] at (\step,13.72) {\step};
  }
  \node[anchor=north,font=\small] at (700,12.35) {Training step};
  \node[rotate=90,font=\small] at (-115,14) {QMH passes (out of 19)};

  \draw[ghpurple,line width=1.15pt] plot coordinates {
    (100,17) (200,16) (300,17) (400,17) (500,18) (530,18) (600,15)
    (650,17) (660,17) (670,16) (680,18) (690,19) (700,18) (710,16)
    (720,18) (730,16) (740,18) (750,17) (800,16) (900,15) (1000,17)
    (1100,17) (1200,19) (1300,19)
  };
  \foreach \step/\score in {
    100/17,200/16,300/17,400/17,500/18,530/18,600/15,650/17,660/17,
    670/16,680/18,690/19,700/18,710/16,720/18,730/16,740/18,750/17,
    800/16,900/15,1000/17,1100/17,1200/19,1300/19
  } {
    \fill[white] (\step,\score) circle (1.45pt);
    \draw[ghpurple,line width=0.8pt] (\step,\score) circle (1.45pt);
  }

  \draw[ghgreen!80!black,line width=1.15pt] plot coordinates {
    (100,16) (200,16) (300,16) (350,17) (360,17) (370,17) (380,16)
    (390,17) (400,17) (410,16) (420,17) (430,15) (440,15) (450,17)
    (500,16) (550,16) (560,17) (570,17) (580,17) (590,18) (600,18)
    (610,17) (620,18) (630,18) (640,18) (650,18) (660,18) (670,16)
    (680,18) (690,17) (700,17) (710,17) (720,18) (730,17) (740,18)
    (750,17) (760,17) (770,18) (780,17) (790,16) (800,18) (810,16)
    (820,17) (830,15) (840,14) (850,16) (860,14) (870,15) (880,18)
    (890,17) (900,17) (910,18) (920,16) (930,17) (940,16) (950,16)
    (1000,16) (1100,16) (1200,16) (1250,17) (1260,15) (1270,16)
    (1280,17) (1290,17) (1300,17)
  };
  \foreach \step/\score in {
    100/16,200/16,300/16,350/17,360/17,370/17,380/16,390/17,400/17,
    410/16,420/17,430/15,440/15,450/17,500/16,550/16,560/17,570/17,
    580/17,590/18,600/18,610/17,620/18,630/18,640/18,650/18,660/18,
    670/16,680/18,690/17,700/17,710/17,720/18,730/17,740/18,750/17,
    760/17,770/18,780/17,790/16,800/18,810/16,820/17,830/15,840/14,
    850/16,860/14,870/15,880/18,890/17,900/17,910/18,920/16,930/17,
    940/16,950/16,1000/16,1100/16,1200/16,1250/17,1260/15,1270/16,
    1280/17,1290/17,1300/17
  } {
    \fill[white] (\step,\score) circle (1.45pt);
    \draw[ghgreen!80!black,line width=0.8pt] (\step,\score) circle (1.45pt);
  }

  \node[star,star points=5,draw=ghpurple,fill=white,line width=0.8pt,
    minimum size=7pt,inner sep=0pt] at (690,19) {};
  \node[star,star points=5,draw=ghgreen!80!black,fill=white,line width=0.8pt,
    minimum size=7pt,inner sep=0pt] at (590,18) {};

  \draw[ghpurple,line width=1.15pt] (150,20.4) -- (200,20.4);
  \node[anchor=west] at (220,20.4) {2-in-1 model};
  \draw[ghgreen!80!black,line width=1.15pt] (470,20.4) -- (520,20.4);
  \node[anchor=west] at (540,20.4) {3-in-1 model};
  \node[star,star points=5,draw=gray!75,fill=white,line width=0.8pt,
    minimum size=7pt,inner sep=0pt] at (850,20.4) {};
  \node[anchor=west] at (880,20.4) {Shipped};
\end{tikzpicture}
  
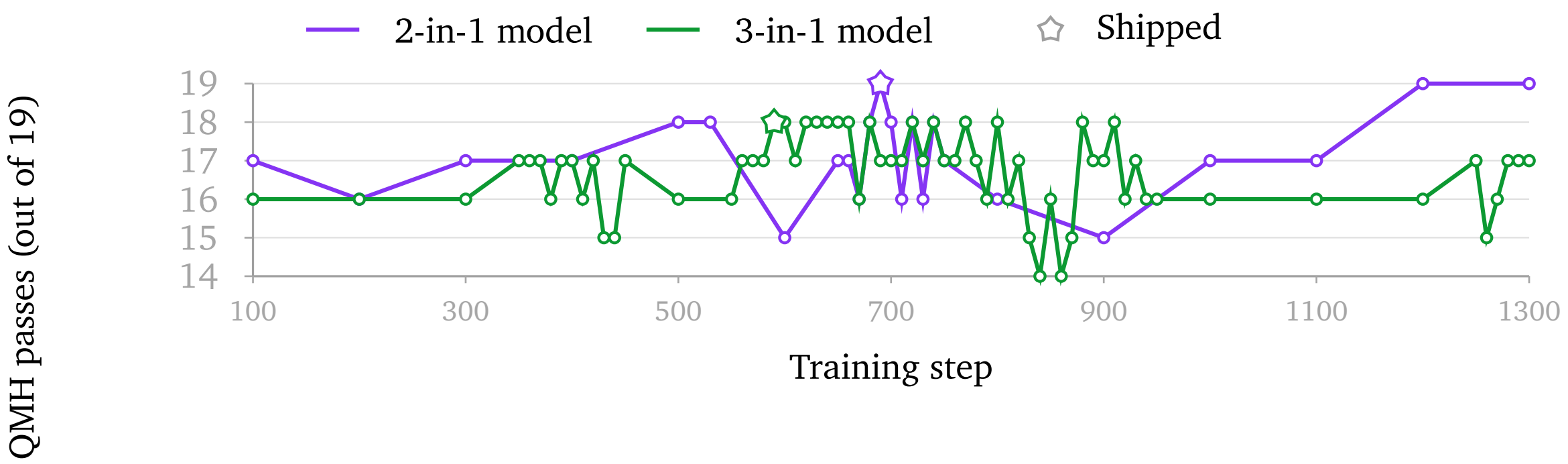
\captionof{figure}{QMH pass counts across training checkpoints for the 2-in-1
  and 3-in-1 models. Stars mark the shipped checkpoints at steps 690 and 590,
  respectively. Scores fluctuate between nearby checkpoints, motivating the use of a lightweight canary benchmark (QMH) for fast, finer-grained checkpoint evaluation.}
  \label{fig:phase1-phase2-qmh-checkpoint-stability}
\end{center}

\subsection{Offline-Online Evaluation Metric Correlation}
\label{sec:phase1-phase2-offline-online-correlation}
Figure~\ref{fig:phase1-phase2-offline-online-correlation-selected-pairs} compares representative metrics from each of the five offline benchmarks with online metrics (quantified as relative changes from each A/B experiment's control), using all 20 candidates across Phase~1 and Phase~2 for which we perform a full A/B experiment.
Each scatterplot reports Pearson's linear correlation $r$ and Spearman's rank correlation $\rho$. We compute both separately for the 2-in-1 and 3-in-1 candidates and plot a phase-specific least-squares line.
Figure~\ref{fig:phase1-phase2-offline-online-correlation-selected-pairs} shows one representative pairing per benchmark. Appendix~\ref{app:phase1-phase2-offline-online-correlation-details} contains 18 selected offline-online metric pairings, shown separately by phase.
Note that these are descriptive associations among selected A/B experiments, not estimates of the effect of changing an offline score or evidence of prospective predictiveness.

\begin{figure}[!t]
  \centering
  \includegraphics[width=\linewidth]{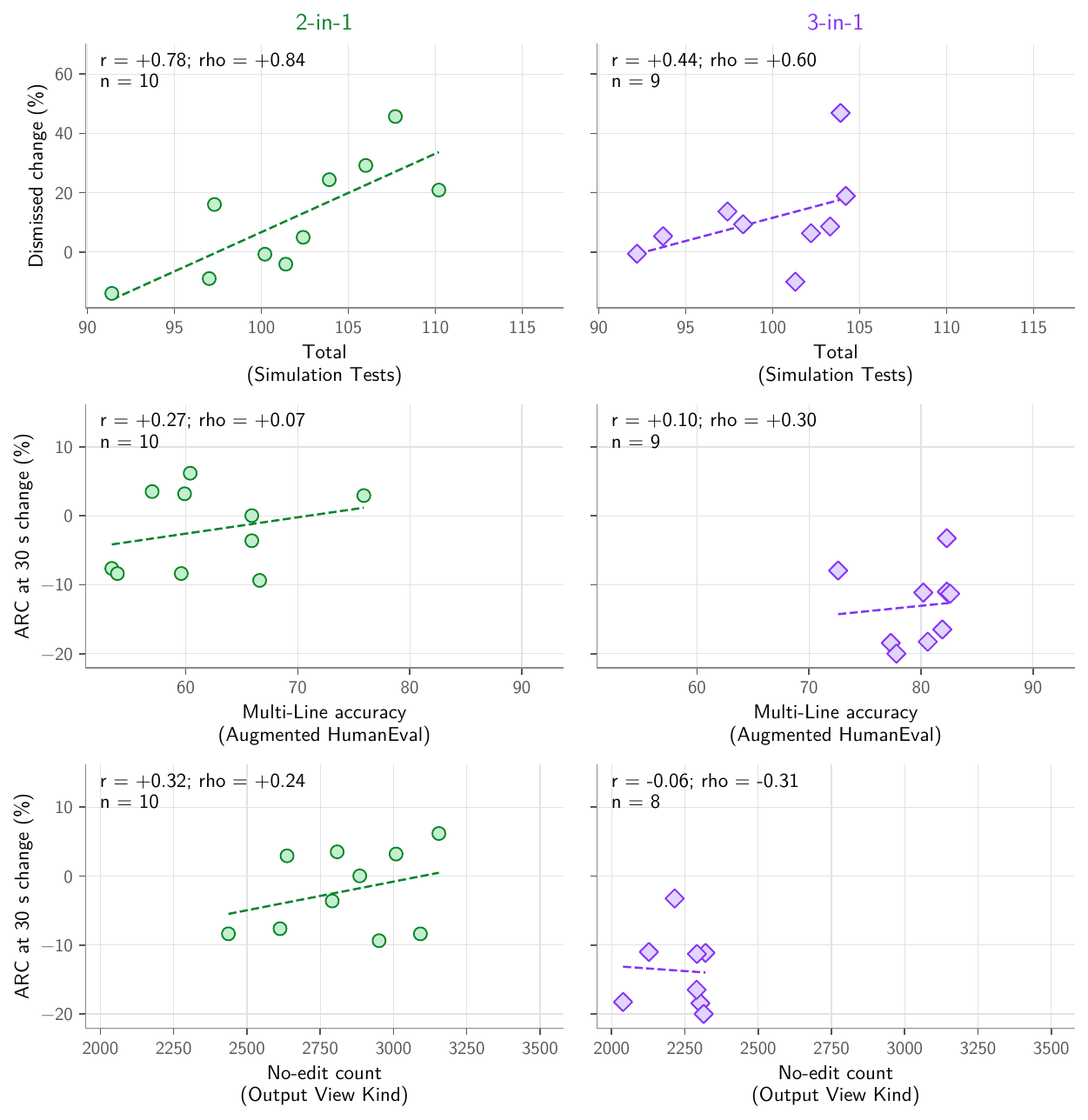}
  \caption{Representative offline-online associations for Simulation Tests,
  Augmented HumanEval, and Output View Kind, separated by phase.
  Online changes are relative to each experiment's control.
  Associations vary by benchmark and phase; none meets the adjusted significance
  threshold. Appendix~\ref{app:phase1-phase2-offline-online-correlation-details}
  shows all pairings.}
  \label{fig:phase1-phase2-offline-online-correlation-selected-pairs}
\end{figure}

\begin{figure}[!t]
  \ContinuedFloat
  \centering
  \includegraphics[width=\linewidth]{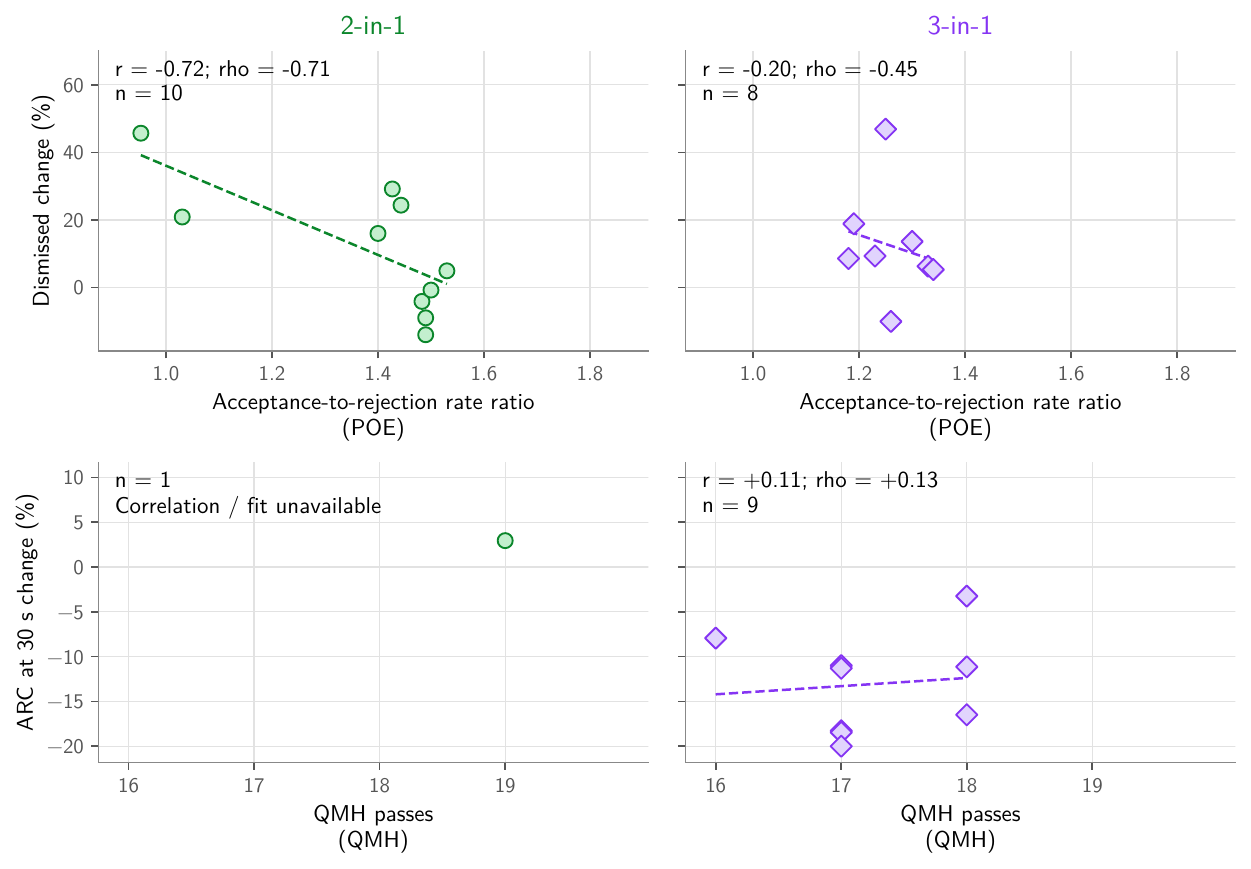}
  \caption[]{Representative offline-online associations for POE and QMH
  (continued), with online changes relative to each experiment's control.
  Higher POE acceptance-to-rejection ratios accompany fewer dismissals in both
  phases, while QMH has little association with ARC in Phase~2.
  The single Phase~1 QMH observation does not support a correlation estimate;
  none of the displayed associations meets the adjusted significance threshold.}
\end{figure}

\paragraph{Note on Scope of Inference.}
Each observation is an A/B experiment where the treatment is the candidate model setup and the control is the corresponding baseline model (the then-production setup). We use pairwise complete cases: the displayed correlations use 8 to 10 experiments per phase, apart from the 2-in-1 QMH cases with only one observation.
We retain all reported treatment-control changes regardless of their statistical significance since filtering would select observations using the outcomes being correlated.
Separating phases avoids conflating between-phase shifts with within-phase associations, but does not control for differences in client settings, controls, or checkpoint ancestry within a phase.
We adjust two-sided Spearman permutation-test $p$-values jointly across all tested offline-online metric pairs in both phases using the Benjamini-Hochberg procedure.
None of the displayed phase-level associations passes the adjusted 0.05 threshold; agreement in direction is descriptive rather than evidence of a statistically established or causal effect.
Paired percentile bootstrap intervals from 5{,}000 resamples also show substantial uncertainty: for instance, the association between POE A/R and dismissed suggestions has a 95\% rank-correlation interval of approximately $[-0.95,-0.09]$ for 2-in-1 and $[-0.92,+0.34]$ for 3-in-1.
We report pointwise intervals without selection or multiplicity adjustment, and permutation and bootstrap inference assume exchangeable, independent A/B experiments.
The fitted lines do not represent a held-out candidate test, and we do not formally test whether the phase-specific coefficients differ.
The results support outcome-specific hypotheses rather than a validated offline ranking of online performance.

\paragraph{Simulation Tests.}
The Total score correlates positively with shown and accepted suggestions in the 2-in-1 group ($\rho=0.52$ and $0.25$) and the 3-in-1 group ($\rho=0.65$ and $0.05$), but also with dismissed and ignored suggestions (2-in-1: $\rho=0.84$ and $0.45$; 3-in-1: $\rho=0.60$ and $0.60$).
The positive shown and dismissed suggestion associations also occur within each model phase.
This pattern is consistent with the benchmark capturing variation in suggestion frequency, but does not establish how well the benchmark distinguishes wanted from unwanted suggestions.

\paragraph{Augmented HumanEval.}
Multi-Line accuracy has weak positive correlations with accepted and retained characters (ARC) at 30 seconds and 5 minutes in the 2-in-1 group ($\rho=0.07$ and $0.22$) and the 3-in-1 group ($\rho=0.30$ and $0.36$).
These weak associations do not establish a predictive relationship.

\paragraph{Output View Kind.}
The number of no-edit responses has weak positive correlations with both ARC outcomes in the 2-in-1 group ($\rho=0.24$ and $0.28$), but negative correlations in the 3-in-1 group ($\rho=-0.31$ and $-0.19$).
The association changes sign between phases, providing no consistent within-phase relationship between no-edit counts and ARC.
ARC measures retained-character volume rather than a conditional persistence rate.

\paragraph{Pseudo-Online Evaluation.}
POE rejection correlates positively with online A/B experiment dismissals in the 2-in-1 and 3-in-1 groups ($\rho=0.38$ and $0.62$), and its acceptance-to-rejection rate ratio correlates negatively with online A/B experiment dismissals ($\rho=-0.71$ and $-0.45$).
These associations suggest POE's assessment of undesirable suggestions corresponds to the changes in dismissed suggestions in online A/B experiments.

The acceptance-to-rejection rate ratio has little rank association with the two ARC outcomes in the 2-in-1 group ($\rho=0.06$ and $0.10$), and negative associations in the 3-in-1 group ($\rho=-0.55$ and $-0.24$).
The counts of previously accepted suggestions converted to no edit correlate positively with ARC in the 2-in-1 group ($\rho=0.35$ and $0.41$), but negatively in the 3-in-1 group ($\rho=-0.26$ and $-0.10$).
Together with the Output View Kind analysis, these results support examining restraint alongside suggestion quality, but do not establish a consistent restraint-to-ARC relationship across phases.

\paragraph{Qualitative Must-Haves.}
Only ten of the twenty A/B experiments have QMH scores available for this analysis, and only one comes from a Phase~1 model. We introduce the 19-case QMH set late in Phase~1 RL stage~1 (Section~\ref{sec:model-selection-process}). The single Phase~1 observation does not support a correlation estimate, so we omit it from the analysis but show the data point for reference.
Within the 3-in-1 group, QMH scores have weak associations with ARC ($r=0.11$ and $0.12$, with $\rho=0.13$ and $0.06$), and with shown and accepted suggestions ($\rho=0.11$ and $0.07$).
These correlations do not establish QMH as a strong predictor of aggregate online outcomes; we find it useful for screening candidates for known failure modes.

\paragraph{Discussion.}
We observe exploratory offline-online associations in this analysis, but do not establish prospective predictive validity.
POE associations have consistent directions across phases: higher rejection rates and lower acceptance-to-rejection ratios accompany more dismissed suggestions.
These patterns suggest that POE captures aspects of suggestion undesirability, although we need further validation to establish its predictive value.

These results also motivate further refining offline evaluation around the online outcome each benchmark is intended to anticipate.
For instance, we may want to improve Simulation Tests to better capture the decision boundary between desirable and undesirable edits, and investigate whether POE's assessment of desirable edits can better anticipate acceptance and retained character outcomes.
We aim to continuously refine benchmarks and evaluation metrics to lessen the gap between offline assessment and the online developer experience.
\FloatBarrier

\section{Conclusion, Limitations, and Future Directions}
In this work, we introduce the techniques and approaches we use to train the unified inline suggestions model for GitHub Copilot, including data pipelines, model training, offline evaluation, and online A/B experimentation. We share lessons from creating the model development flywheel and plan to apply them to future extensions of our models.
The final Phase~1 model-client configuration reduces time to display by 6.77\% and output tokens by 58.93\%, with no statistically significant changes in quality and engagement metrics relative to the three-model control.
The Phase~2 configuration with all four client adaptations also reduces dismissals by 10.06\% relative to the 2-in-1 model paired with the completion model, although this result is not statistically significant under a Bonferroni correction.
These results show that the model-client configuration in each phase of unification achieves its intended effect on online metrics relative to its corresponding control.

\subsection{Limitations}
\label{sec:limitations}

\paragraph{Online Effect Estimates.}
Each A/B experiment directly compares one candidate with the then-production configuration over the same period. Because experiments for different candidates run at different times, their results do not directly compare candidates with each other.
The results we report for the shipped models (Tables~\ref{tab:phase1-online-results-2-in-1} and~\ref{tab:phase2-online-results-3-in-1}) come from the same experiments we use to select them, and we do not adjust $p$-values for the 12 metrics reported per experiment or the 24 candidates across both phases (Table~\ref{tab:phase1-phase2-development-iteration-counts}).
This candidate selection process can inflate reported effect sizes, and unadjusted testing increases the risk of false positives across comparisons. Under Bonferroni correction across the 12 reported metrics ($p < 0.05/12$), the decreases in time to display and time to produce remain statistically significant, but the Phase~2 decrease in dismissed suggestions ($p = 0.045$) and increase in recurring engagement ($p = 0.022$) do not. In addition, this correction does not account for the comparisons across different candidates and phases.

\paragraph{Sources of Supervision.}
We use a general-purpose frontier reasoning model as the LLM judge. The judge contributes to binary rewards in RL (Section~\ref{sec:phase1-grader}) and provides pass or fail decisions in both data filtering (Section~\ref{sec:phase1-data-quality-filtering}) and edit history playback (Section~\ref{sec:phase1-data-bootstrapping}).
Because we use this frontier reasoning model as the LLM judge, part of the improvement could reflect the judge model's capabilities rather than the training recipe.
In addition, we use the same POE samples for grader calibration and the reported offline evaluations, so POE evaluation results may not be independent of LLM judge development.

Training labels come from existing NES and completion training data that we convert to the patch format, recorded edit histories, and our own models, including pseudolabels from our strongest model checkpoint (Section~\ref{sec:phase1-data-bootstrapping}).
A general-purpose frontier reasoning model writes ground truth labels only in three targeted behavior curation datasets: rewritten invalid patches, rewritten insertion pattern propagation examples, and completion length examples.
Rewritten invalid patches make up about 3\% of the SFT data, and together with rewritten insertion pattern propagation examples, they make up about 4\% of the Phase~1 RL mixture after upsampling. Completion length examples make up about 5\% of the Phase~2 RL mixture after upsampling.
We do not ablate judge strength, so these results do not separate the judge's contribution from that of the training recipe.
Several evaluation measures also do not use the judge to score predictions: for instance, Augmented HumanEval is used for evaluation, not SFT or RL training, and scores predictions using unit tests rather than the LLM judge. Its scores improve from SFT to RL (Table~\ref{tab:phase1-offline-results}). Likewise, the online metrics measure developer behavior directly without relying on an LLM judge. When we hold the checkpoint fixed, a client change alone shifts dismissed suggestions (Section~\ref{sec:phase2-online-evaluation}).

\subsection{Future Directions}

One direction for future work is personalized eagerness. Our Phase~1 and Phase~2 ablations and qualitative testing, in addition to developer feedback, identify tradeoffs between follow-up edit frequency, suggestion quality, and developer preferences, motivating the ability to personalize and adapt model behavior (Section~\ref{sec:phase1-nes-long-distance-unification}).
As a starting point, we are adding the capability to condition the unified model on developer preferences for factors such as suggestion frequency and edit size.

Looking forward, this work establishes a foundation for a unified model flywheel that we plan to continue to improve.
We will make continual updates to use developer feedback and build on components we introduce in this work, from the evaluation methods and checkpoint selection process (Sections~\ref{sec:evaluation-methodology} and~\ref{sec:model-selection-process}) to targeted data curation and RL grader boosting (Sections~\ref{sec:phase1-targeted-behavior-curation} and~\ref{sec:phase1-grader}).
Our current prompt's recent edit history can be viewed as a short editing trajectory; we plan to extend training to longer trajectories to improve how the inline suggestions model works together with developers across successive interactions.
We plan to evaluate these advances against the existing quality guardrails, retaining useful suggestions while reducing unwanted interruptions.

We see this model unification as just one step in our continued journey toward more intelligent and context-aware AI-assisted coding, and we look forward to sharing our learnings as we make further progress toward our vision of meeting every builder in their workflow with the right assistance at the right time.

\phantomsection
\section*{Contributors and Acknowledgments}
\label{sec:contributors}
\addcontentsline{toc}{section}{Contributors and Acknowledgments}

The unified inline suggestions model for GitHub Copilot is the result of a collaboration among many people at Microsoft. Below, we list core contributors and contributors, followed by acknowledgments.

\subsection*{Core Contributors}
\subsubsection*{Model Training, Data Curation, and Evaluation}
\namelist{Julia Gong, Benjamin Liggett, Shengjie Ma, Yu Hu, Elsie Nallipogu}

\subsubsection*{Client Engineering}
\namelist{Ulugbek Abdullaev, Alexandru Dima}

\subsubsection*{Science Product Management}
\namelist{Zhehui Liao, Luciana Abud, Nathan Halstead}

\smallskip
\subsection*{Contributors}
\namelist{Benjamin Steenhoek, Gaurav Mittal, Fuhui Fang, Satarupa Guha, Hang Zhao, Anand Mooga, Cheng Long, Benjamin Simmonds, Soojin Choi}

\bigskip
\subsection*{Acknowledgments}
We thank the following collaborators whose insights and support made this work possible.

\subsubsection*{Science Tooling and Data Pipelines}
\namelist{Kevin Merchant, Sarah Widder, Abhijith Chatra, Suman Bhagavathula, Maoliang Huang, KwangJe Cho, Carl Brochu, Baoping Zhang, Shubhangi Thakur, Martin Valle, Farzad Eskafi}

\subsubsection*{Experimentation, Product, and Review}
\namelist{Cindy Chiu, Amy Hartwig, Nick Trogh, Brigit Murtaugh, John Mogensen, Rahul Pandita, Mario Rodriguez, Kai Maetzel}

\subsubsection*{Platform and Infrastructure}
\namelist{Jesten Herrild, Joel Dunham, Joseph Leinsdorf, Alejandro Oviedo Cruz, Boshi Lian, Shuhan Yan, Shiven Raina, Vidur Vatsalya, Vijetha Vijayendran, Meng Lan, Eddie Ailijiang, Ming Liu, Dharmesh Guna, Mostafa Elzoghbi}

\bigskip
We thank the VS Code and GitHub Copilot developer community for the ongoing feedback that informs and improves our work.

\clearpage
\bibliography{main}
\bibliographystyle{plainnat}

\clearpage
\appendix

\section{Citation Reference}
\label{app:citation-reference}

Please use this \hologo{BibTeX} entry when citing this report:
\begin{lstlisting}[
  style=promptstyle,
  backgroundcolor=\color{ghpurple!5!white},
  rulecolor=\color{ghpurple!20!white}
]
@techreport{unified_inline_code_editing_model,
    title={Building a Unified Inline Code Editing Model for {GitHub Copilot}},
    author={{Microsoft CodeAI Team} and {Microsoft VS Code Team}},
    institution={Microsoft},
    year={2026},
}
\end{lstlisting}

\section{Speculative Decoding Draft Comparison}
\label{sec:speculative-decoding-draft-comparison}

We compare five speculative decoding draft formats while holding the 2-in-1 model fixed.
This comparison evaluates how the draft prefix affects request latency, rather than comparing the 2-in-1 and 3-in-1 models.
Figure~\ref{fig:speculative-decoding-draft-variants} illustrates all five formats using the same function body completion as Figure~\ref{fig:phase2-speculative-decoding-drafts}.
The no-draft baseline supplies no speculative decoding draft.
The file path draft, used by the 2-in-1 client, contains the file path and a trailing colon.
The current line draft adds the cursor line number and the current line as a deletion.
The current line replaced draft additionally supplies a newline and the addition marker, leaving the replacement text for the model to generate.
The current line completed draft, which we select for the 3-in-1 client, instead repeats the current line after the addition marker, anticipating an insertion that preserves the existing line.

\begin{figure}[htbp]
  \centering
  \begingroup
\footnotesize
\setlength{\fboxsep}{0pt}
\textbf{File content:} \texttt{/path/to/util.py}, cursor at the end of the signature.
\par\smallskip
{\ttfamily
def count\_words(text: str) -> int:\\
\colorbox{ghgreen!18}{\strut\ \ \ \ return len(text.split())}
}
\par\medskip
\hrule
\medskip
\begin{subfigure}[t]{0.49\linewidth}
  \caption{No draft}
  \smallskip
  \begin{tabular}{@{}l@{}}
    \texttt{/path/to/util.py:0} \\
    \texttt{-def count\_words(text: str) -> int:} \\
    \texttt{+def count\_words(text: str) -> int:} \\
    \texttt{+\ \ \ \ return len(text.split())}
  \end{tabular}
\end{subfigure}\hfill
\begin{subfigure}[t]{0.49\linewidth}
  \caption{File path (2-in-1 draft)}
  \smallskip
  \begin{tabular}{@{}l@{}}
    \texttt{\colorbox{ghpurple!15}{\strut/path/to/util.py:}0} \\
    \texttt{-def count\_words(text: str) -> int:} \\
    \texttt{+def count\_words(text: str) -> int:} \\
    \texttt{+\ \ \ \ return len(text.split())}
  \end{tabular}
\end{subfigure}
\par\medskip
\begin{subfigure}[t]{0.49\linewidth}
  \caption{Current line}
  \smallskip
  \begin{tabular}{@{}l@{}}
    \texttt{\colorbox{ghpurple!15}{\strut/path/to/util.py:0}} \\
    \texttt{\colorbox{ghpurple!15}{\strut-def count\_words(text: str) -> int:}} \\
    \texttt{+def count\_words(text: str) -> int:} \\
    \texttt{+\ \ \ \ return len(text.split())}
  \end{tabular}
\end{subfigure}\hfill
\begin{subfigure}[t]{0.49\linewidth}
  \caption{Current line replaced}
  \smallskip
  \begin{tabular}{@{}l@{}}
    \texttt{\colorbox{ghpurple!15}{\strut/path/to/util.py:0}} \\
    \texttt{\colorbox{ghpurple!15}{\strut-def count\_words(text: str) -> int:}} \\
    \texttt{\colorbox{ghpurple!15}{\strut+}def count\_words(text: str) -> int:} \\
    \texttt{+\ \ \ \ return len(text.split())}
  \end{tabular}
\end{subfigure}
\par\medskip
\begin{subfigure}[t]{0.49\linewidth}
  \caption{Current line completed (selected 3-in-1 draft)}
  \smallskip
  \begin{tabular}{@{}l@{}}
    \texttt{\colorbox{ghpurple!15}{\strut/path/to/util.py:0}} \\
    \texttt{\colorbox{ghpurple!15}{\strut-def count\_words(text: str) -> int:}} \\
    \texttt{\colorbox{ghpurple!15}{\strut+def count\_words(text: str) -> int:}} \\
    \texttt{+\ \ \ \ return len(text.split())}
  \end{tabular}
\end{subfigure}\hfill
\begin{minipage}[t]{0.49\linewidth}
  \smallskip
  \textbf{Legend}
  \par\smallskip
  \colorbox{ghgreen!18}{\strut Inserted text}
  \par\smallskip
  \colorbox{ghpurple!15}{\strut Speculative decoding draft}
\end{minipage}
\endgroup

  \caption{The five draft formats in the comparison.
  Each panel shows the same illustrative diff patch response; purple highlighting marks the supplied draft prefix, and green marks the suggested insertion in the file content.
  No draft prefix is supplied in panel (a).}
  \label{fig:speculative-decoding-draft-variants}
\end{figure}

\paragraph{Experimental Setup.}
All formats use the same ordered set of 4{,}792 recorded editing contexts and identical prompts; only the speculative decoding draft changes.
We set the sampling temperature to 0.
We test each format in a separate session of the Locust load-testing tool \citep{locust}, with successive concurrency levels of 1, 4, 8, 12, and 16 clients.
Each client sends its next request immediately after the previous request completes.
Each level has a 120-second measurement window following an unmeasured warmup of 30 to 60 seconds, with a request-rate stability check.
We reuse contexts cyclically when we exhaust the input set.
We assign requests to a measurement window by their start time, including those that complete after the window ends; we exclude warmup and cooldown requests.
We compute achieved requests per second (RPS) by dividing the number of requests that start in the measurement window by 120 seconds.
We measure latency at the load generator from request construction through receipt and parsing of the full response, not time to first token or client display time.
We report the arithmetic mean, median (p50), and 95th percentile (p95), computing percentiles by linear interpolation between ordered observations.
The current line series uses the rerun included in the comparison report.

\begin{figure}[htbp]
  \centering
  \resizebox{\linewidth}{!}{\input{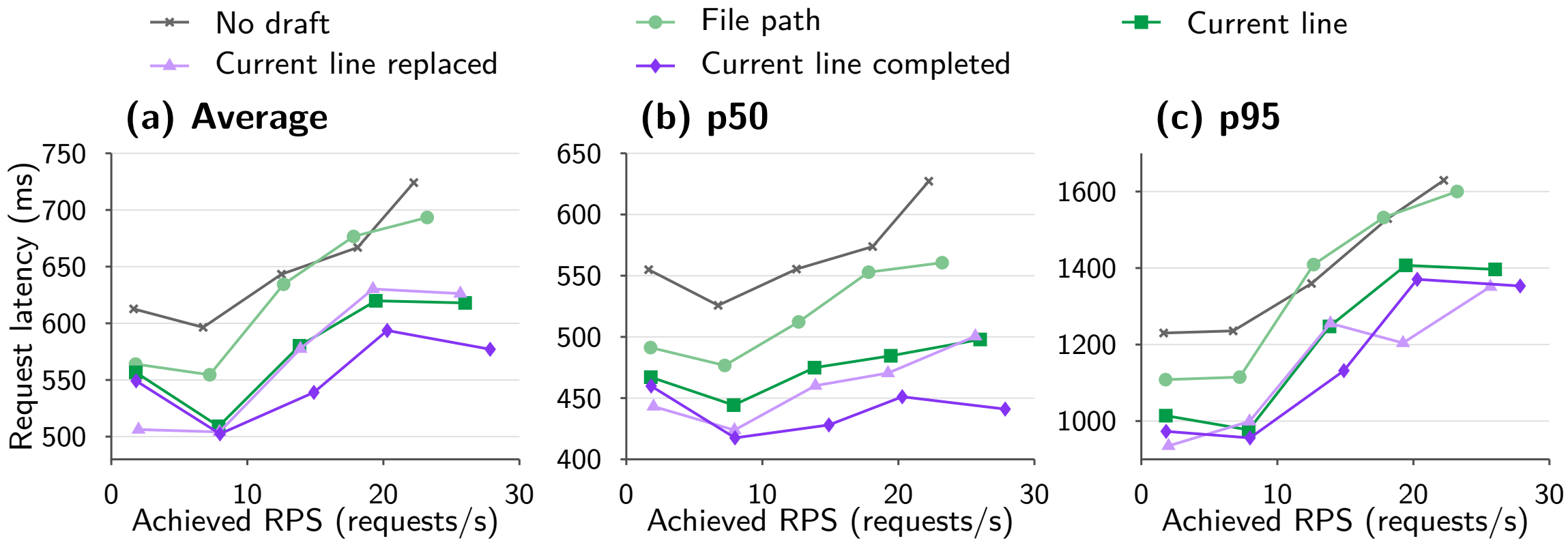}}
  \caption{Request latency versus achieved RPS for five draft formats on the fixed 2-in-1 model: average, p50, and p95.
  From left to right, the five points on each curve correspond to 1, 4, 8, 12, and 16 concurrent clients.
  Lines connect measured points and do not represent additional experiments.
  Statistics retain failed requests, matching the source comparison report.}
  \label{fig:speculative-decoding-latency}
\end{figure}

\paragraph{Latency Results.}
Figure~\ref{fig:speculative-decoding-latency} plots latency against achieved RPS.
At each tested concurrency level, all three current line drafts have lower mean latency than the file path draft.
At 16 concurrent clients, the current line completed draft has a mean latency of 576.8\,ms, compared with 693.0\,ms for the file path draft and 723.7\,ms for the no-draft baseline.
These measurements favor the current line completed draft among the five formats tested on the 2-in-1 model.
The 3-in-1 model's completion-focused training motivates our expectation that the current line completed draft matches its response prefixes more often than those of the 2-in-1 model.
We therefore select the current line completed draft for the 3-in-1 model.

\section{Patch Validity Taxonomy}
\label{app:patch-validity-taxonomy}

Tables~\ref{tab:patch-validity-critical} and~\ref{tab:patch-validity-suboptimal} show the patch validity taxonomy and corresponding examples for critical and suboptimal cases, respectively.

\begingroup
  \small
  \setlength{\tabcolsep}{4pt}
  \begin{longtable}{>{\raggedright\arraybackslash}p{8.4cm}ll}
  \caption{Critical patch validity taxonomy. These responses either cannot be correctly parsed and
  applied to the file or produce a visible defect. Each minimal example uses
  the reference file shown at the top, with line numbers as they appear in
  the prompt.}
  \label{tab:patch-validity-critical} \\
    \toprule
    \multicolumn{3}{l}{\textbf{Reference file} \texttt{test.py}, as it appears
      in the prompt} \\
    \multicolumn{3}{l}{%
      \begin{tabular}{@{}l@{}}
        \texttt{1|def hello():} \\
        \texttt{2|\ \ \ \ print("hello")} \\
        \texttt{3|\ \ \ \ x = 1} \\
        \texttt{4|\ \ \ \ return True} \\
        \texttt{5|} \\
        \texttt{6|def goodbye():} \\
        \texttt{7|\ \ \ \ print("goodbye")} \\
        \texttt{8|\ \ \ \ return False}
      \end{tabular}} \\
    \midrule
    Check & Scope & Example \\
    \midrule
    \endfirsthead
    \multicolumn{3}{l}{\tablename~\thetable: Critical patch validity taxonomy (continued).} \\
    \toprule
    Check & Scope & Example \\
    \midrule
    \endhead
    \midrule
    \multicolumn{3}{r}{Continued on next page.} \\
    \endfoot
    \bottomrule
    \endlastfoot
    Any line starts with a space                                  & Response
      & \texttt{\ test.py:2} \\
    Patch contains empty lines                                    & Patch
      & \begin{tabular}[t]{@{}l@{}}
          \texttt{test.py:2} \\
          \texttt{-\ \ \ \ print("hello")} \\
          \texttt{} \\
          \texttt{+\ \ \ \ print("world")}
        \end{tabular} \\
    \addlinespace
    Response does not contain \texttt{:}                          & Response
      & \texttt{print("hello")} \\
    Patch header does not contain \texttt{:}                      & Patch
      & \texttt{test.py-2} \\
    Patch header content after \texttt{:} is not a digit sequence & Patch
      & \texttt{test.py:line2} \\
    File named in the patch header is absent from the prompt      & Patch
      & \texttt{main.py:2} \\
    Line number in the patch header is absent from the prompt     & Patch
      & \texttt{test.py:17} \\
    All patches in the response are no-ops                        & Response
      & \begin{tabular}[t]{@{}l@{}}
          \texttt{test.py:2} \\
          \texttt{-\ \ \ \ print("hello")} \\
          \texttt{+\ \ \ \ print("hello")}
        \end{tabular} \\
    \addlinespace
    First line of a patch is not a deletion                       & Patch
      & \begin{tabular}[t]{@{}l@{}}
          \texttt{test.py:2} \\
          \texttt{+\ \ \ \ print("world")}
        \end{tabular} \\
    \addlinespace
    First deletion line does not match the prompt line it names   & Patch
      & \begin{tabular}[t]{@{}l@{}}
          \texttt{test.py:2} \\
          \texttt{-\ \ \ \ print("hello!")} \\
          \texttt{+\ \ \ \ print("world")}
        \end{tabular} \\
    \addlinespace
    Patch interleaves deletions and additions                     & Patch
      & \begin{tabular}[t]{@{}l@{}}
          \texttt{test.py:2} \\
          \texttt{-\ \ \ \ print("hello")} \\
          \texttt{+\ \ \ \ print("world")} \\
          \texttt{-\ \ \ \ x = 1} \\
          \texttt{+\ \ \ \ x = 2}
        \end{tabular} \\
    \addlinespace
    Added lines duplicate the code following the insertion point  & Patch
      & \begin{tabular}[t]{@{}l@{}}
          \texttt{test.py:2} \\
          \texttt{-\ \ \ \ print("hello")} \\
          \texttt{+\ \ \ \ print("world")} \\
          \texttt{+\ \ \ \ x = 1}
        \end{tabular} \\
    \addlinespace
    Added lines duplicate the code preceding the insertion point  & Patch
      & \begin{tabular}[t]{@{}l@{}}
          \texttt{test.py:3} \\
          \texttt{-\ \ \ \ x = 1} \\
          \texttt{+\ \ \ \ print("hello")} \\
          \texttt{+\ \ \ \ x = 2}
        \end{tabular} \\
  \end{longtable}
\endgroup

\begingroup
  \small
  \setlength{\tabcolsep}{4pt}
  \begin{longtable}{>{\raggedright\arraybackslash}p{8.4cm}ll}
  \caption{Suboptimal patch validity taxonomy. These responses are applicable
  and correct, but inefficient. Each minimal example uses the
  reference file of Table~\ref{tab:patch-validity-critical}.}
  \label{tab:patch-validity-suboptimal} \\
    \toprule
    Check & Scope & Example \\
    \midrule
    \endfirsthead
    \multicolumn{3}{l}{\tablename~\thetable: Suboptimal patch validity taxonomy (continued).} \\
    \toprule
    Check & Scope & Example \\
    \midrule
    \endhead
    \midrule
    \multicolumn{3}{r}{Continued on next page.} \\
    \endfoot
    \bottomrule
    \endlastfoot
    A no-op patch appears among non-no-op patches                 & Response
      & \begin{tabular}[t]{@{}l@{}}
          \texttt{test.py:2} \\
          \texttt{-\ \ \ \ print("hello")} \\
          \texttt{+\ \ \ \ print("hello")} \\
          \texttt{} \\
          \texttt{test.py:7} \\
          \texttt{-\ \ \ \ print("goodbye")} \\
          \texttt{+\ \ \ \ print("bye")}
        \end{tabular} \\
    \addlinespace
    Patch deletes and re-adds lines it does not change            & Patch
      & \begin{tabular}[t]{@{}l@{}}
          \texttt{test.py:2} \\
          \texttt{-\ \ \ \ print("hello")} \\
          \texttt{-\ \ \ \ x = 1} \\
          \texttt{+\ \ \ \ print("hello")} \\
          \texttt{+\ \ \ \ x = 2}
        \end{tabular} \\
    \addlinespace
    Too many unchanged lines separate the modified lines          & Patch
      & \begin{tabular}[t]{@{}l@{}}
          \texttt{test.py:2} \\
          \texttt{-\ \ \ \ print("hello")} \\
          \texttt{-\ \ \ \ x = 1} \\
          \texttt{-\ \ \ \ return True} \\
          \texttt{+\ \ \ \ print("world")} \\
          \texttt{+\ \ \ \ x = 1} \\
          \texttt{+\ \ \ \ return False}
        \end{tabular} \\
    \addlinespace
    Pure deletion carries unneeded additions and deletions        & Patch
      & \begin{tabular}[t]{@{}l@{}}
          \texttt{test.py:2} \\
          \texttt{-\ \ \ \ print("hello")} \\
          \texttt{-\ \ \ \ x = 1} \\
          \texttt{+\ \ \ \ print("hello")}
        \end{tabular} \\
  \end{longtable}
\endgroup

\section{Input Context Example}
\label{app:input-context}

Figure~\ref{fig:unified-model-prompt} illustrates the key components of the unified model input context: recently viewed code snippets, current file content, edit diff history, cursor position, and instructions.

\begin{figure}[H]
\begin{lstlisting}[style=promptstyle,aboveskip=0pt]
<|recently_viewed_code_snippets|>
<|recently_viewed_code_snippet|>
code_snippet_file_path: /path/to/antarctica.py
40|penguin2 = Penguin("adelie")
41|penguin2.swim()
42|feed_penguin(penguin2, fish="krill")
<|/recently_viewed_code_snippet|>
<|/recently_viewed_code_snippets|>

<|current_file_content|>
current_file_path: /path/to/penguins.py
0|def feed_penguin(penguin, fo):
1|    penguin.meals += 1
2|    penguin.hungry = False
3|    penguin.last_ate_time = datetime.now()
4|
5|    print(f"{penguin.name} ate {fish.type}")
6|
7|penguin = Penguin("emperor")
8|feed_penguin(penguin, fish="silverfish")
<|/current_file_content|>

<|edit_diff_history|>
--- /path/to/penguins.py
+++ /path/to/penguins.py
@@ -0,1 +0,1 @@
-def feed_penguin(penguin, fish):
+def feed_penguin(penguin, fo):

<|/edit_diff_history|>

<|cursor_position|>
0|def feed_penguin(penguin, fo<|cursor|>):
<|/cursor_position|>

The developer was working on a section of code within the `current_file_content` -
carefully note their `cursor_location` marked with `<|cursor|>`. Using the given
`recently_viewed_code_snippets`, `current_file_content`, `edit_diff_history`, and
`cursor_location`, please continue the developer's work. Output a modified diff
format with a sequence of intuitive next changes, where each patch must start with
`<filename>:<line number>`. Order changes by priority and flow; for instance, edits
adjacent to the user's cursor should always be prioritized, followed by lines near
the cursor, followed by lines farther away. If there are no good edit candidates,
output the empty string "". Avoid undoing or reverting the developer's last
change unless there are obvious typos or errors. Adhere meticulously to the diff
format.
\end{lstlisting}
  \caption{The unified prompt corresponding to the editing context in
  Figure~\ref{fig:diff-patch-format}. Recently viewed code, the current file, recent edit history, cursor position, and additional instructions guide the model to output
  a sequence of edit patches that best fit the editing context.}
  \label{fig:unified-model-prompt}
\end{figure}

\clearpage
\section{Additional Offline-Online Evaluation Metric Correlation Analysis}
\label{app:phase1-phase2-offline-online-correlation-details}
\begin{figure}[H]
  \centering
  \includegraphics[width=\linewidth]{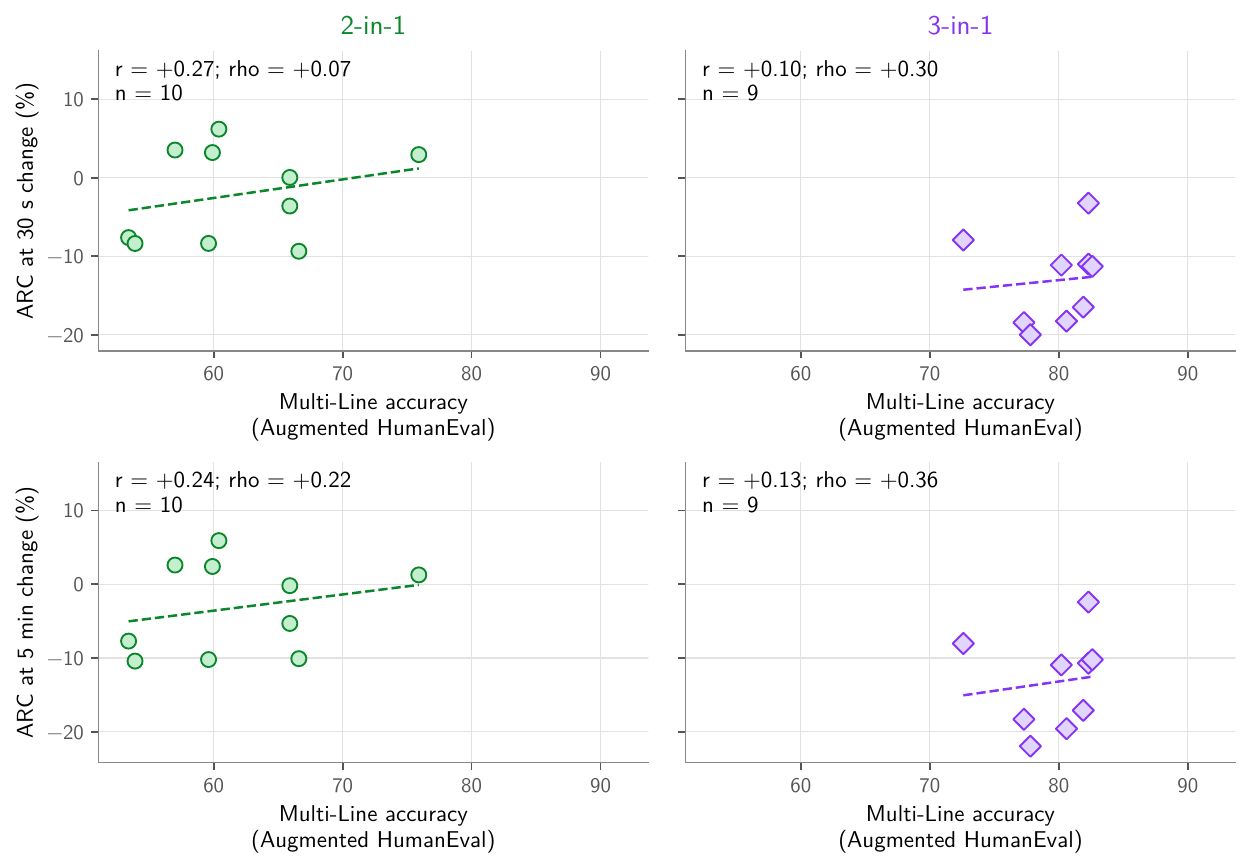}
  \caption{Augmented HumanEval Multi-Line accuracy versus ARC at 30 seconds
  and 5 minutes, relative to each experiment's control.
  Accuracy has only weak positive associations with retained characters
  in both phases.}
  \label{fig:phase1-phase2-human-eval-correlation}
\end{figure}

\begin{figure}[H]
  \centering
  \includegraphics[width=\linewidth,height=0.78\textheight,keepaspectratio]{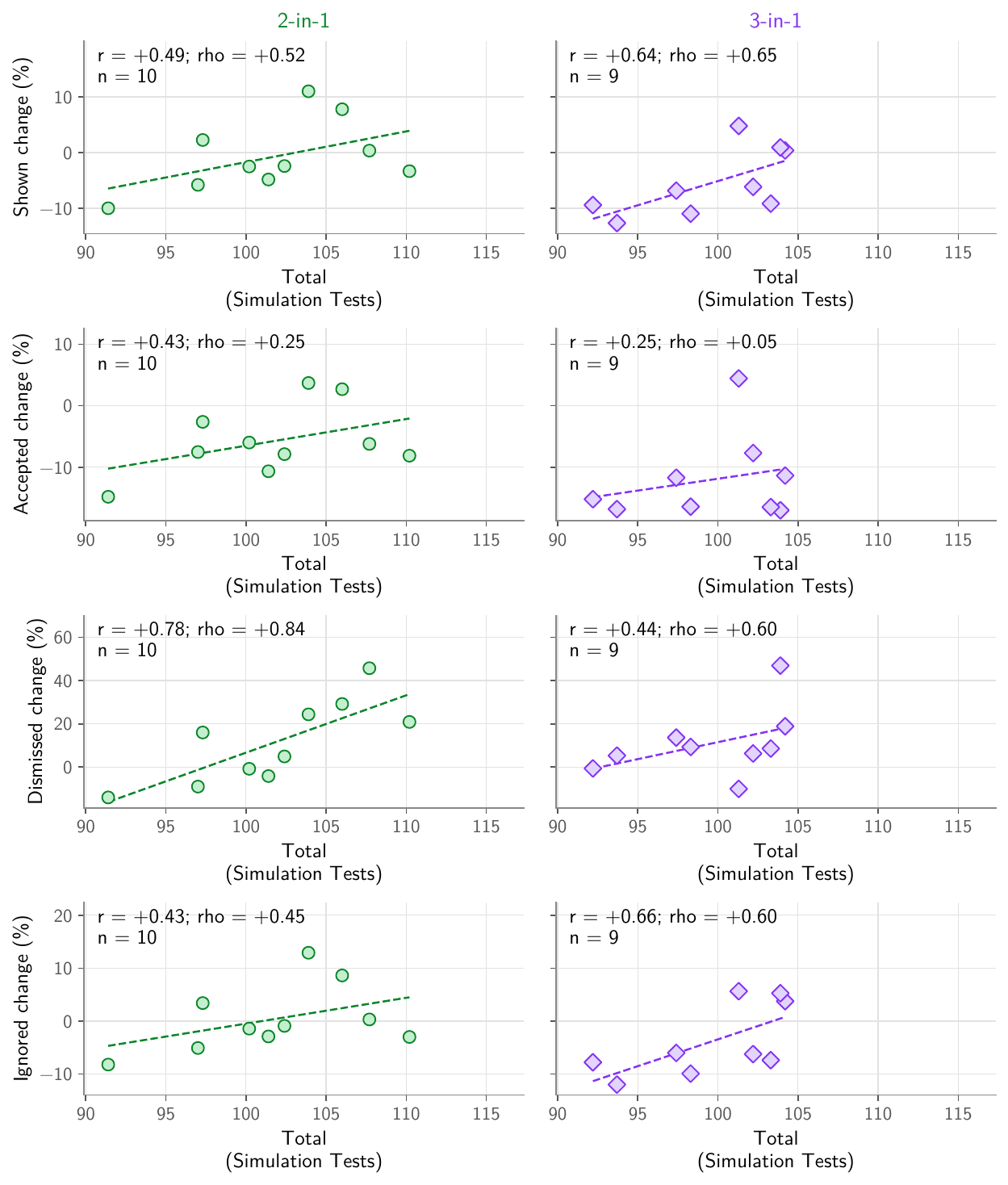}
  \caption{Simulation Tests Total versus online suggestion changes,
  relative to each experiment's control.
  Higher scores accompany more shown and dismissed suggestions in both phases,
  a pattern consistent with the benchmark capturing variation in suggestion frequency.}
  \label{fig:phase1-phase2-simulation-correlation}
\end{figure}

\begin{figure}[H]
  \centering
  \includegraphics[width=\linewidth]{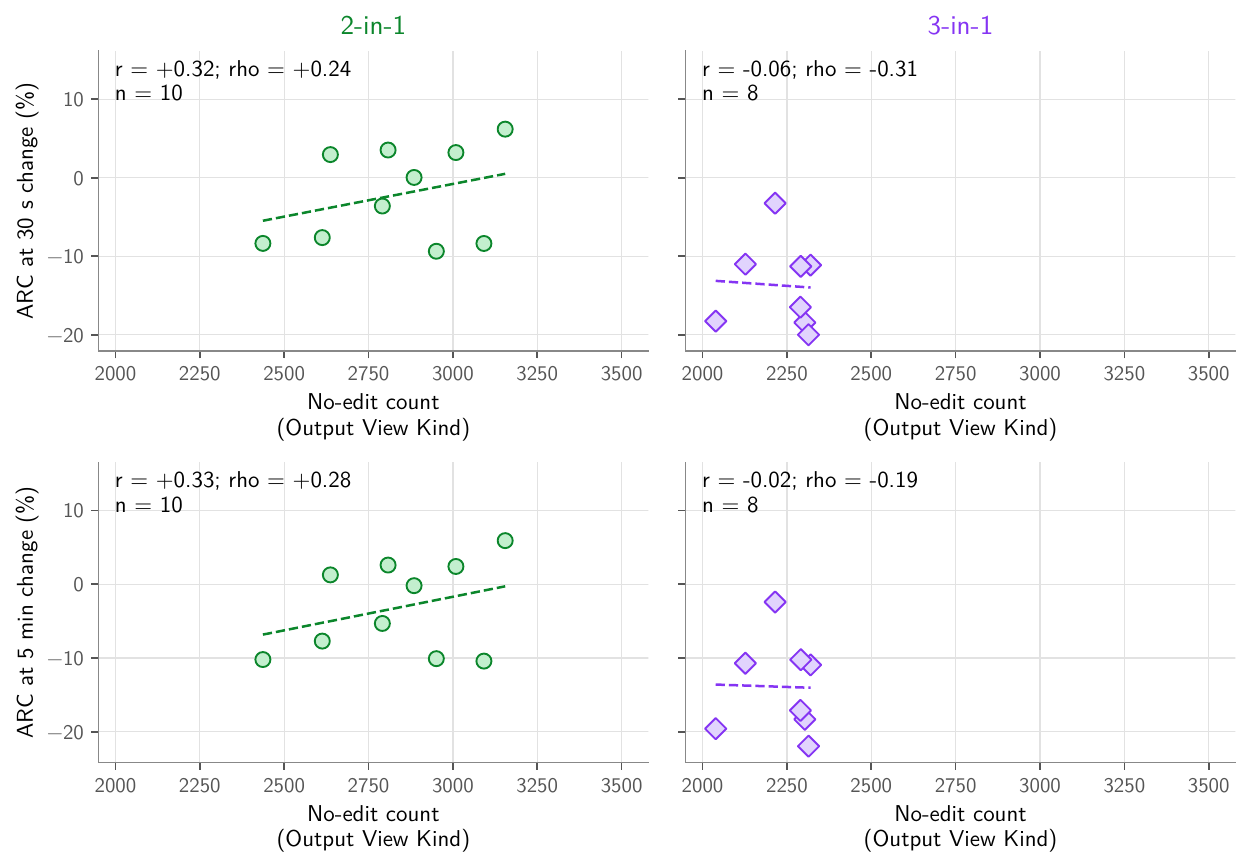}
  \caption{Output View Kind no-edit counts versus ARC at 30 seconds and
  5 minutes, relative to each experiment's control.
  Rank correlations are weakly positive in Phase~1 and negative in Phase~2,
  providing no consistent association between abstention and retained characters.}
  \label{fig:phase1-phase2-view-kind-correlation}
\end{figure}

\begin{figure}[H]
  \centering
  \includegraphics[width=\linewidth,height=0.78\textheight,keepaspectratio]{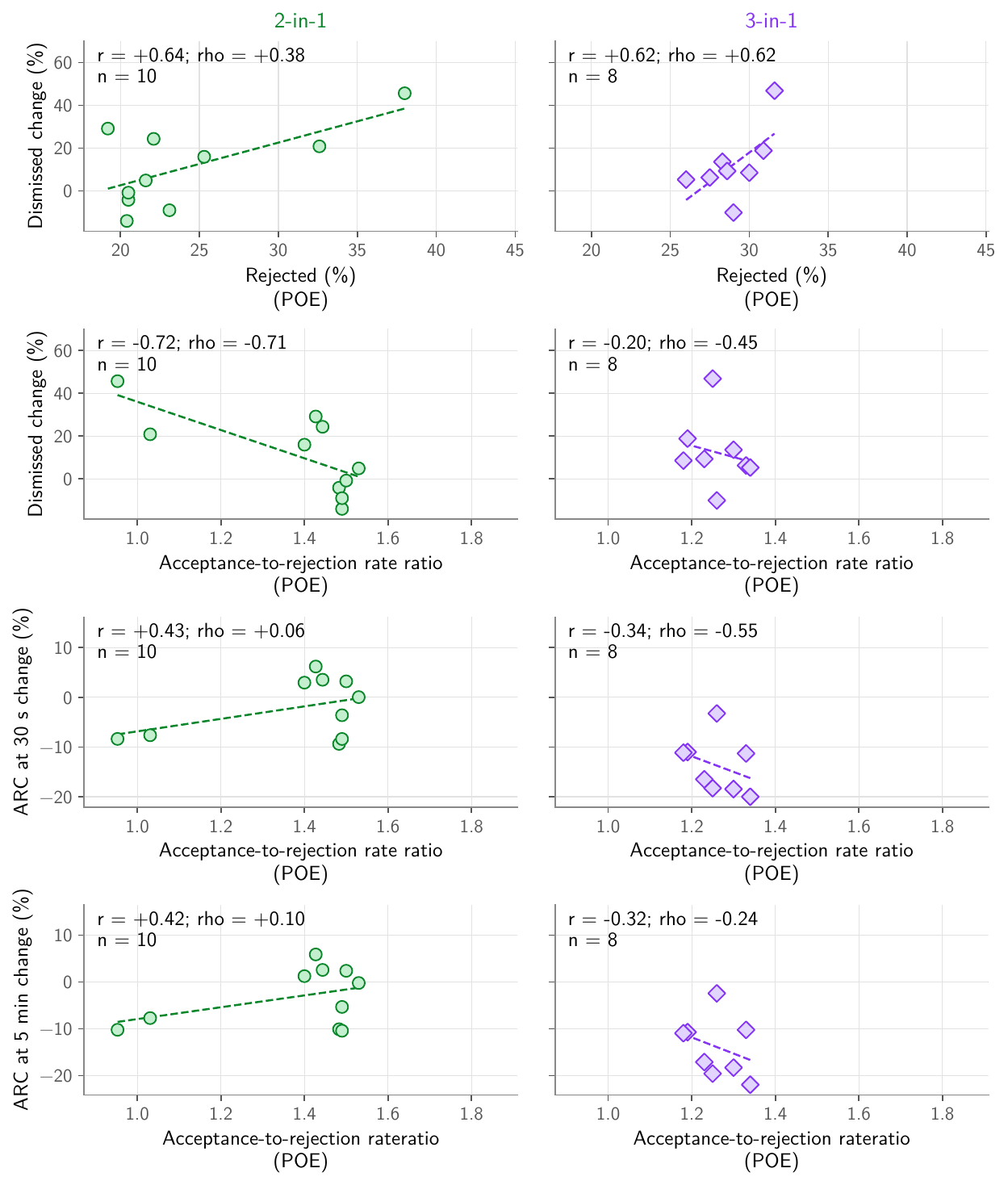}
  \caption{POE rejection rate and acceptance-to-rejection ratio versus
  dismissals and ARC, relative to each experiment's control.
  Higher rejection rates and lower acceptance-to-rejection ratios accompany more
  dismissals in both phases; associations with ARC are less consistent.}
  \label{fig:phase1-phase2-poe-outcomes-correlation}
\end{figure}

\begin{figure}[H]
  \centering
  \includegraphics[width=\linewidth]{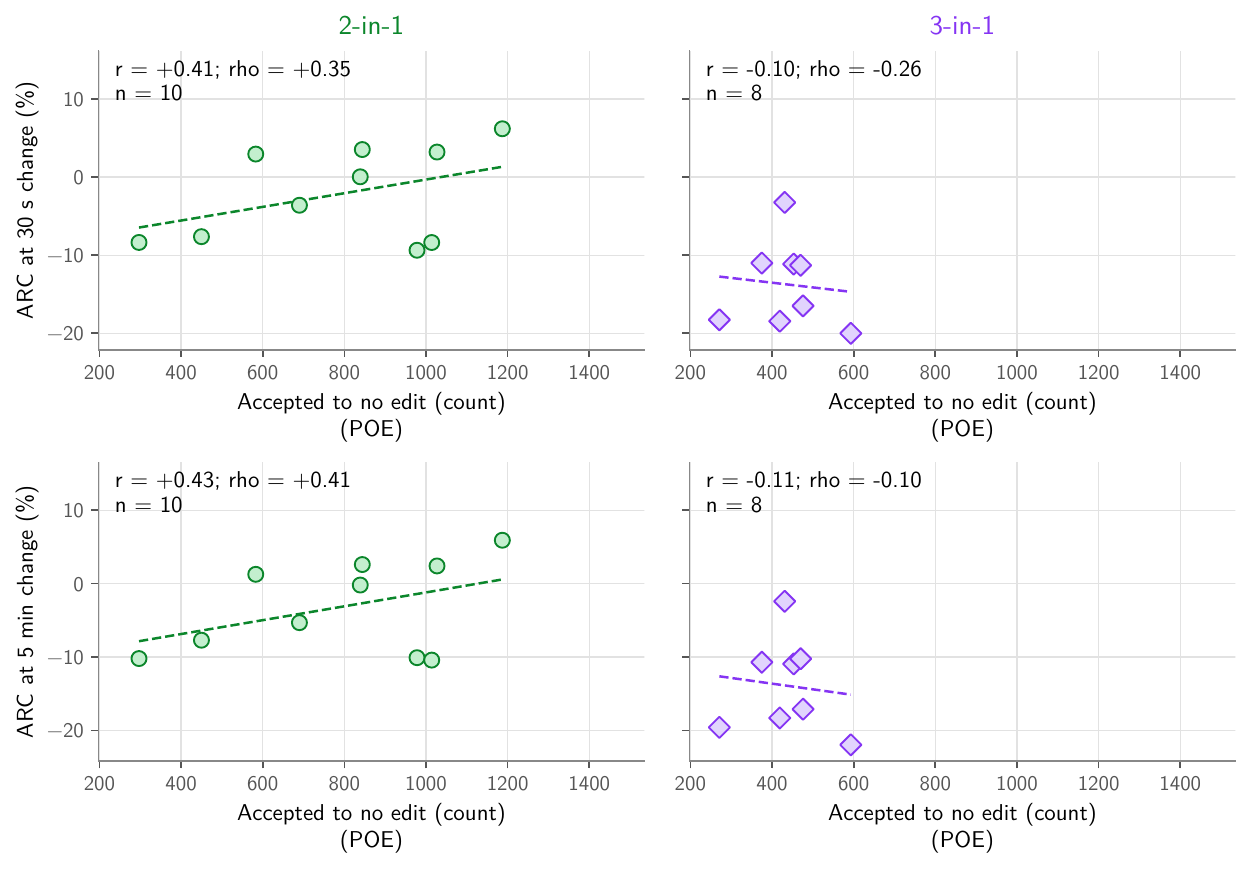}
  \caption{POE counts of previously accepted suggestions converted to no edit
  versus ARC at 30 seconds and 5 minutes, relative to each experiment's control.
  Rank correlations change sign between phases, providing no consistent
  association between these conversions and retained characters.}
  \label{fig:phase1-phase2-poe-abstention-correlation}
\end{figure}

\begin{figure}[H]
  \centering
  \includegraphics[width=\linewidth,height=0.78\textheight,keepaspectratio]{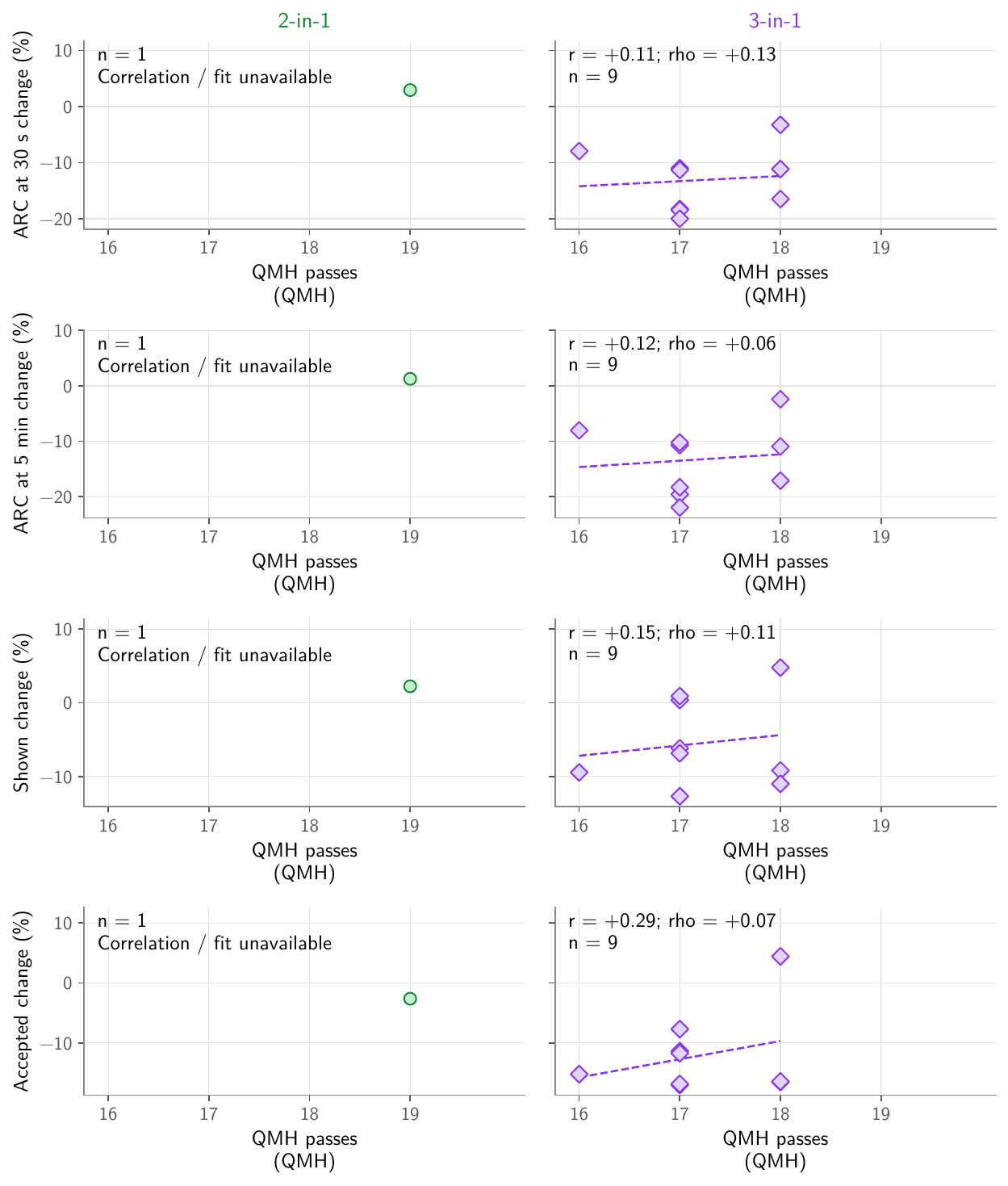}
  \caption{QMH passes versus ARC, shown, and accepted suggestions, relative to
  each experiment's control. Associations are weak among the nine 3-in-1
  candidates; the single 2-in-1 observation cannot support a correlation estimate.
  These results do not establish QMH as a predictor of online performance.}
  \label{fig:phase1-phase2-qmh-correlation}
\end{figure}

This appendix contains the 18 selected offline-online metric pairings we discuss in Section~\ref{sec:phase1-phase2-offline-online-correlation}, which we show separately for the 2-in-1 and 3-in-1 candidates in 36 phase-specific panels. We show the single 2-in-1 QMH observation descriptively, but it does not support a correlation analysis.
Figures~\ref{fig:phase1-phase2-human-eval-correlation}, \ref{fig:phase1-phase2-simulation-correlation}, \ref{fig:phase1-phase2-view-kind-correlation}, \ref{fig:phase1-phase2-poe-outcomes-correlation}, \ref{fig:phase1-phase2-poe-abstention-correlation}, and \ref{fig:phase1-phase2-qmh-correlation} report phase-specific Pearson $r$, Spearman $\rho$, sample sizes, and least-squares fits.
We use shared axes across the two phases for each pairing, but fit lines only over each phase's observed offline range.
All online values are relative changes from controls specific to each A/B experiment.
No-edit and conversion counts are not normalized abstention rates.
Repeated coordinates can overlap, particularly among the nine 3-in-1 QMH candidates in a narrow score range.
The uncertainty, selection, and dependence limitations we discuss in the main text apply to each of these analyses.
None of the displayed phase-specific associations meets the adjusted 0.05 significance threshold.

\end{document}